\documentclass[english]{iopart}
\pdfoutput=1  
  \usepackage{graphicx}
 \newcommand{\beq}{\begin{equation}}
\newcommand{\eeq}{\end{equation}}

\newcommand{\beqr}{\begin{eqnarray}}
\newcommand{\eeqr}{\end{eqnarray}}
\newcommand{\barr}{\begin{array}}
\newcommand{\earr}{\end{array}}
\newcommand{\bal}{\begin{align}}
\newcommand{\eal}{\end{align}} 
\newcommand{\bmu}{\begin{multline}}
\newcommand{\emu}{\end{multline}}

\begin{document}

 \title{ Hard hexagon partition function for complex fugacity}
\author{  M. Assis$^1$, J.L. Jacobsen$^{2,3}$, I. Jensen$^4$, J-M. Maillard$^5$ and B.M. McCoy$^1$ }
\address{$^1$ CN Yang Institute for Theoretical Physics, State
  University of New York, Stony Brook, NY, 11794, USA} 
\address{$^2$ Laboratoire de Physique Th{\'e}orique, {\'E}cole Normale
  Sup{\'e}rieure, 24 rue Lhomond, 75231 Paris Cedex, France}
\address{$3$ Universit{\'e} Pierre et Marie Curie, 4 Place Jussieu,
  75252 Paris, France}
\address{$^4$ ARC Center of Excellence for Mathematics and Statistics
of Complex Systems, Department of Mathematics and Statistics, The
University of Melbourne, VIC 3010, Australia}
\address{$^5$ LPTMC, UMR 7600 CNRS, 
Universit\'e de Paris, Tour 23,
 5\`eme \'etage, case 121, 
 4 Place Jussieu, 75252 Paris Cedex 05, France}

\begin{abstract}

We study the analyticity of the partition function of the hard hexagon
model in the complex fugacity plane by computing zeros and transfer
matrix eigenvalues for large finite size systems. We find that the
partition function per site computed by Baxter in the thermodynamic
limit for positive real values of the fugacity is not sufficient to
describe the analyticity in the full complex fugacity plane. We also
obtain a new algebraic equation for the low density partition function per
site.
 
\end{abstract}

\noindent {\bf AMS Classification scheme numbers}: 34M55, 
47E05, 81Qxx, 32G34, 34Lxx, 34Mxx, 14Kxx 
\vskip .5cm

 {\bf Key-words}:  Hard hexagon model, 
 partition function zeros, transfer matrix eigenvalues, Hauptmoduls, 
modular functions.

\vspace{.2in}

\section{\label{sec:intro} Introduction}

The hard hexagon model was solved by Baxter over 30 years 
ago~\cite{baxterhh,baxbook}. More precisely Baxter computed the 
thermodynamic limit of the grand  partition function per site for real
positive values of the fugacity $z$
\begin{eqnarray} 
\hspace{-0.9in}&& \quad \quad  \quad  \quad 
\lim_{L_v,L_h\rightarrow \infty}Z_{L_v,L_h}^{1/L_vL_h}(z)
\quad \quad \, \,  {\rm with}  \, \, \quad \quad
 0\,<\, L_v/L_h\,< \,\infty \quad \quad {\rm fixed}.
\end{eqnarray}
Even more precisely Baxter computed
the limit
\begin{eqnarray} 
\hspace{-0.9in}&& \qquad \quad  \quad  \quad 
\kappa(z)\,\,  =\,\, \, 
\lim_{L_h\rightarrow \infty}\, \lambda_{\rm max}(z;L_h)^{1/L_h}, 
\end{eqnarray}
where $\lambda_{\rm max}(z;L_h)$ is the largest eigenvalue of the transfer
matrix. Baxter found that there are two distinct regions of 
positive fugacity
\begin{eqnarray}
 \hspace{-0.9in}&& \qquad \quad  \quad  \quad 
0\,\leq \, z \,< \,z_c 
\quad \quad \, \, {\rm and} \,\,  \quad  \quad  \quad 
z_c\,<\,z\,<\,\infty, 
\end{eqnarray}
with
\begin{eqnarray}
\hspace{-0.9in}&& \qquad \quad \quad  \quad 
z_c\,= \,\,\frac{11+5{\sqrt 5}}{2}
\,\, = \,\,\, 11.0901699473 \,\, \cdots,
\end{eqnarray}
where in each separate region the partition function per site 
has separate analytic expressions, which we denote by $\kappa_{-}(z)$ 
for the low density, and by $\kappa_{+}(z)$ for the high density
intervals respectively. The low density function $\kappa_{-}(z)$ has
branch points at $z_c$ and 
\begin{eqnarray} 
\hspace{-0.9in}&& \qquad \quad  \quad 
z_d\,\,=\,\,\, -{{1} \over {z_c}} 
\,\, = \,\,\, \frac{11-5{\sqrt 5}}{2}
\,\, = \,\, -0.0901699473 \,\, \cdots,
\end{eqnarray}
is real and positive in the interval $z_d \leq z \leq z_c$ 
and is analytic in 
the plane  cut along the real axis 
from $z_d$ to $-\infty$ and $z_c$ to $+\infty$. Conversely the high
density function $\kappa_{+}(z)$ is real and positive for 
$z_c \, \leq\,  z\,  < \, +\infty$ and is analytic 
in the plane cut along the real axis
from $z_c $ to $-\infty$.  

For the purpose of thermodynamics it is sufficient to restrict
attention to positive values of the fugacity. However, it is of
considerable interest to investigate the behavior of the partition
function for complex values of $z$ as well. For finite size systems
the partition function is, of course, a polynomial and as such can be
specified by its zeros. 

In the thermodynamic limit the free energy
will be analytic in all regions which are the limit of the zero free
regions of the finite system~\cite{yl1952}. In general there will be
several such regions. One such example with three regions is given by
Baxter~\cite{baxter3}.  There appears to be no
general theorem stating when the free energy of a system can be continued
through the locus of zeros.
 
 The analytic structure of the free energy in the complex
fugacity plane is not in general determined by the the free
energy on the positive $z$ axis and for hard hexagons it is only for
the positive $z$ axis that a complete analysis has been carried
out. In this paper we address the problem of determining 
analyticity  in the complex $z$ plane by  computing
the partition function zeros on lattices as
large as $39\times 39$ and comparing these zeros with the locus
computed from  the limiting partition
functions per site computed by Baxter for real positive value for the
fugacity $0 \, \leq \, z \, \leq \, \infty$.  We will see 
that the two functions $\kappa_{\pm}(z)$ are not sufficient to 
describe the location of the zeros in the complex $z$ plane.
 There is, of course, no reason that $\kappa_{\pm}(z)$
  should be sufficient to represent the partition function in the
  entire complex $z$ plane.
We propose in  section~\ref{sec:discneck}
an extension of Baxter's methods which can explain our results in the
portion of the complex plane not covered by $\kappa_{\pm}(z)$.

In section~\ref{sec:pre} we present  the relation between partition function zeros 
and the eigenvalues of the transfer matrix with special attention to the
differences between cylindrical and toroidal boundary conditions.

In section~\ref{sec:pf} we begin by recalling the results of Baxter~\cite{baxterhh} 
for $\kappa_{\pm}(z)$ and the subsequent analysis of Joyce~\cite{joyce} 
for the high density regime for the polynomial relation between
$z$ and $\kappa_{+}(z)$. For the low density regime we derive a 
new polynomial relation between $z$ and $\kappa_{-}(z)$. Some details of 
the analysis of $\kappa_{\pm}(z)$ and the associated density $\rho_{-}(z)$ 
are presented in \ref{app:sing} and \ref{app:rhoexp}. 

In section~\ref{sec:ev}  we compute transfer matrix eigenvalues and equimodular
curves for the maximum eigenvalues for values of $L_h$ as large as 30.
We demonstrate the difference between the equimodular
curves of the full transfer matrix and the equimodular curves
for eigenvalues restricted to the sector $P\, =\, 0$. These equimodular
curves are compared with  the partition functions per site 
$\kappa_{\pm}(z)$.

In section~\ref{sec:zero}  we present the results for partition function zeros
for both toroidal and cylindrical boundary conditions for a variety
of $L_h\, \times\,  L_v$ lattices. For $L_h\, =\, L_v$ the largest sizes
are $39\, \times \, 39\, $ for cylindrical and $27\, \times \,  27\,$ 
for toroidal boundary conditions. We compare the zeros with $\kappa_{\pm}(z)$ 
and with the equimodular eigenvalue curves of section~\ref{sec:ev}  for both the 
cases $L_v\, =\, L_h$ and $L_v\, \gg\,  L_h$ and we analyze the density 
of zeros on the negative $z$ axis. We analyze the dependence of the
approach as $L\rightarrow \infty$ of the endpoints $z_d(L)$ and
$z_c(L)$ to $z_d$ and $z_c$ by means of finite size scaling and
identify several correction to scaling exponents.

In section~\ref{sec:disc}  we  use our results to discuss
the relation of the free energy on the positive real fugacity axis to the  
partition function in the full complex fugacity plane and some
concluding remarks are made in section~\ref{sec:conc}. A description of the
methods used for the numerical computations of eigenvalues and zeros
are given in \ref{app:TM} and  the numerical details of the finite size
scaling are given in \ref{app:FSS}.

\section{\label{sec:pre} Preliminaries}

In this section we review the concepts of partition function, transfer
matrix, free energy and partition function zeros and highlight the
properties we discuss in later sections. 

\subsection{\label{sec:prepf}  Partition function}

The hard hexagon model is defined on a triangular lattice, which is
conveniently viewed as a square lattice with an added diagonal on 
each face as shown in figure~\ref{fig:hhbweights}. Particles are placed on the
sites of the lattice with the restriction that if there is a 
particle at one site no particle is allowed at the six
nearest neighbor sites. The grand canonical partition function  
on the lattice with $L_v$ rows and $L_h$ columns is computed as
\begin{eqnarray} 
\hspace{-0.9in}&& \qquad \quad \quad  \quad 
Z_{L_v,L_h}(z)\, \, =\, \, \, \sum_{N=0}^{\infty}\, g(N) \cdot \, z^N, 
\end{eqnarray}
where $g(N)$ is the number of allowed configurations with $N$
particles. By definition on a finite lattice the partition function 
is a polynomial which can be described by its zeros $z_k$ as
$\, \, \prod\, (1-z/z_k)$.  For hard hexagons the order of
  the polynomial is bounded above by $L_vL_h/3$ which becomes an equality
when it is an integer.

\subsection{\label{sec:pretm} Transfer matrices}

\begin{figure}[h!]

\begin{center}
\begin{picture}(100,100)
\put(20,0){\includegraphics[width=3cm]{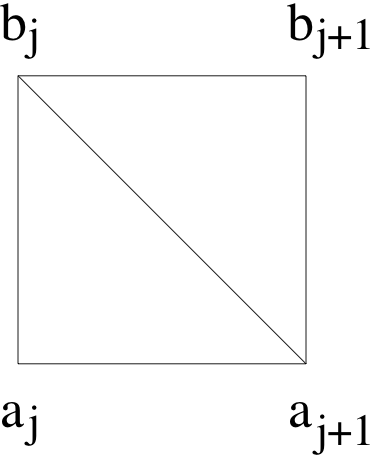}} 
\end{picture}
\end{center}

\caption{\label{fig:hhbweights} Boltzmann weights for the transfer matrix of hard hexagons}

\end{figure}

An alternative and quite different representation of the partition
function on the finite lattice is given in terms of a  transfer
matrix $T(z;L_h)$ computed in terms of the local Boltzmann weights in
figure~\ref{fig:hhbweights}  as
\begin{eqnarray}
 \hspace{-0.9in}&& \qquad \quad  \quad 
T_{\{b_1,\cdots b_{L_h}\},\{a_1,\cdots.a_{L_h}\}}
\,\,  =\,\,\, \,   \prod_{j=1}^{L_h}\, W(a_j,\, a_{j+1};\, b_j,\, b_{j+1}),  
\label{tmat}
\end{eqnarray}
where the occupation numbers  $a_j$ and $b_j$ take the 
values $0$ and $1$ and for
periodic boundary conditions in the horizontal
direction we use the convention that $L_h+1\equiv 1$. Then 
the hard hexagon  weights  $W(a_j,a_{j+1};b_j,j_{j+1})$ 
are written (see page 403 of~\cite{baxbook}) in the form
\begin{eqnarray}
\hspace{-0.9in}&& \qquad \quad  \quad 
W(a_j,a_{j+1};b_j,b_{j+1}) \,  \, = \, \,  \, 0
\nonumber\\
\label{hhwb}
\hspace{-0.9in}&& \qquad \quad  \quad 
 {\rm for} \quad  \quad 
a_ja_{j+1}\, =\,\, b_jb_{j+1}\, =\,\, a_jb_j\, =\,\, a_{j+1}b_{j+1}
\, =\,\, a_{j+1}b_{j}\, =\, 1, 
\end{eqnarray} 
and otherwise:
\begin{eqnarray}
\label{hhw}
 \hspace{-0.9in}&& \qquad \quad  \quad \quad 
W(a_j,a_{j+1};\, b_j,b_{j+1})\,\, = \,\,\,\, z^{(a_j+a_{j+1}+b_j+b_{j+1})/4}.
\end{eqnarray}
This transfer matrix does not satisfy $T \,= \,T^t$:   
thus there may be complex eigenvalues even for $z \,\geq \, 0$.
As far as real values of $\, z$
are concerned, the matrix elements are all non
negative for $z\, \geq\,  0$ and, thus, by the Perron-Frobenius 
theorem the maximum eigenvalue is real and positive. 

For lattices with toroidal boundary conditions where there 
are periodic boundary conditions in the vertical direction 
\begin{eqnarray}
\label{transper}
\hspace{-0.9in}&& \qquad \quad  \quad  \quad 
Z^P_{L_v,L_h}(z) \,\,\,\, = \,\,\,\,  {\rm Tr}\,\, T^{L_v}(z;L_h).
\end{eqnarray}
For lattices with cylindrical boundary conditions where there 
are free boundary condition in the vertical direction
\begin{eqnarray} 
\label{transfree}
\hspace{-0.9in}&& \qquad \quad  \quad   \quad 
Z^C_{L_v,L_h}(z)\,\,\,  =\,\, \,\,
 \langle {\bf v}_B|T^{L_v}(z;L_h)|{\bf v}'_B\rangle, 
\end{eqnarray}
where ${\bf v}_B$ and ${\bf v}'_B$ are suitable vectors 
for the boundary conditions on rows $1$ and $L_v$. 
For the transfer matrix (\ref{tmat}) with Boltzmann weights given 
by the symmetrical form (\ref{hhwb}) with (\ref{hhw}) 
the components of the vectors 
 ${\bf v}_B$ and ${\bf v}'_B$ for free boundary conditions are
\begin{eqnarray} 
\hspace{-0.9in}&& \qquad \quad  \quad 
{\bf v}_B(a_1,a_2,\cdots, a_{L_h})\,\, = \,\,\, 
{\bf v}'_B(a_1,a_2,\, \cdots,\, a_{L_h})
\, \, =\,\, \,\, \prod_{j=1}^{L_h}\, z^{a_j/2}. 
\label{vector}
\end{eqnarray}

When the transfer matrix is diagonalizable
(\ref{transper}) and (\ref{transfree}) may be written in terms of the
eigenvalues $\lambda_k$  and eigenvectors
 ${\bf v}_k$ of the transfer matrix
$T_{L_h}(z)$ as
\begin{eqnarray}
\label{eigen1}
\hspace{-0.9in}&& \quad  \,   \,  
Z^P_{L_v,L_h}(z)\, =\,\,\,  \sum_k \, \lambda_k^{L_v}(z;L_h)
\qquad \quad \quad \hbox{and}  \\ 
\label{eigen2}
\hspace{-0.9in}&& \quad  \, \,  
Z^C_{L_v,L_h}(z)\, =\,\,\,  \sum_k\lambda_k^{L_v}(z;L_h)\cdot \, c_k 
\quad \, \, 
{\rm where}  \quad \, \, \,   \, \,  
c_k\, =\,\,\,  (\bf{v}_B\cdot \bf{v}_k)(\bf{v}_k \cdot \bf{v}'_B).
\end{eqnarray}

\subsection{\label{sec:pretdl} The thermodynamic limit}

For finite size systems the hard hexagon partition function is a polynomial
and the transfer matrix eigenvalues are all algebraic
functions. However, for physics we must study the thermodynamic
limit where $L_v,~L_h \, \rightarrow \, \infty$ and, because both the
partition function and the transfer matrix eigenvalues diverge in this
limit we consider instead of the partition function the free energy
\begin{eqnarray} 
\label{free}
\hspace{-0.9in}&& \qquad \quad  \quad  \quad 
-F/k_BT \,\,\, =\, \,\, 
 \lim_{L_v,L_h\rightarrow \infty}(L_vL_h)^{-1} \cdot \, \ln Z_{L_v,L_h}(z).
\end{eqnarray} 
For real positive values of $z$ this limit must be independent of the
aspect ratio $0\, <\,  L_v/L_h\, <\, \infty$ for thermodynamics to be valid.

In terms of the transfer matrix representations of 
the partition function (\ref{eigen1}) and (\ref{eigen2}) 
we take the limit $L_v\, \rightarrow \,  \infty$
\begin{eqnarray} 
\hspace{-0.9in}&& \qquad \quad  \quad  \quad 
\lim_{L_v\rightarrow \infty}\, L_v^{-1} \cdot \, \ln Z_{L_v,L_h}(z)
\,\, \,  =\, \,\, \,  \ln \lambda_{\rm max}(z;L_h).
\end{eqnarray}

For the limiting free energy (\ref{free}) to exist and be non zero it is
required that 
\begin{eqnarray} 
\hspace{-0.9in}&& \qquad \quad \quad  \quad \quad 
0\,\,  <\, \,
 \lim_{L_h\rightarrow \infty}\, L_h^{-1} \cdot \, \ln \lambda_{\rm max}(z;L_h) 
\,\,  <\,\,  \infty,   
\end{eqnarray}
or, equivalently, that the partition function per site exists
\begin{eqnarray} 
\hspace{-0.9in}&& \qquad \quad  \quad \quad 
\kappa(z)\,\,  =\, \,\, \,  \,  
\lim_{L_h \rightarrow \infty}{\lambda_{\rm max}}\, (z;L_h)^{1/L_h} \,\, \,  <\,\,  \infty, 
\end{eqnarray}
(in other words the maximum eigenvalue must be exponential in $L_h$).
This exponential behavior will guarantee that for real positive $z$
\begin{eqnarray} 
\label{limitint}
\hspace{-0.9in}&&  \quad 
\lim_{L_h\rightarrow \infty}\,  \lim_{L_v\rightarrow
  \infty}(L_vL_h)^{-1}\ln Z_{L_v,L_h}(z) \,\, \,  =\, \, \, 
\lim_{L_v,L_h\rightarrow \infty}(L_vL_h)^{-1} \cdot \, \ln Z_{L_v,L_h}(z), 
\end{eqnarray} 
independent of the aspect ratio $L_v/L_h$.
However for complex ``nonphysical values'' of $z$ this
independence of the ratio  $L_v/L_h$ is not obvious. In particular for
hard squares at $z\, =\, -1$ {\it all} eigenvalues of the transfer matrix lie on
the unit circle and the partition function $Z_{L_v,L_h}(-1)$ depends on
number theoretic properties~\cite{fse}-\cite{baxneg} of $L_v$ and $L_h$.

\subsection{\label{sec:prepfev} Partition function zeros versus transfer matrix eigenvalues}

It remains in this section to relate partition function zeros to
transfer matrix eigenvalues and eigenvectors. For finite lattices the
partition function zeros can be obtained from
(\ref{eigen1}) and (\ref{eigen2}) if all eigenvalues and
eigenfunctions are known. We begin with the simplest case where
\begin{eqnarray}
\label{emod1}
\hspace{-0.9in}&& \qquad \quad  \quad \quad  \quad 
L_v \, \,\,  \rightarrow \,\,  \,\infty 
\quad \quad  \quad  {\rm with~ fixed} \quad  \quad  \quad  L_h, 
\end{eqnarray}
considered by Beraha, Kahane and Weiss~\cite{bkw1}-\cite{bkw3} as  
presented  by Salas and Sokal \cite{ss}. This is the case of a
cylinder of infinite length with $L_h$ sites in the finite
direction where the aspect ratio $L_v/L_h\, \rightarrow \, \infty$. 

Generically the eigenvalues have different 
moduli and in the limit (\ref{emod1}) the partition function will 
have zeros when two or more maximum eigenvalues of $T(z;L_h)$ have equal 
moduli
\begin{eqnarray} 
\label{emod2}
\hspace{-0.9in}&& \qquad \quad  \quad  \quad   \quad 
|\lambda_1(z;L_h)| \, \,  =\, \, \,  |\lambda_2(z;L_h)|. 
\end{eqnarray}
The locus in the complex plane $z$ is called an equimodular curve \cite{woodzero}-\cite{wood3}. 
On this curve
\begin{eqnarray} 
\hspace{-0.9in}&& \qquad \quad  \quad  \quad   \quad 
 {{ \lambda_1(z;L_h)} \over {\lambda_2(z;L_h) }} 
\,\,  =\, \,\,   e^{i\phi(z)}, 
\end{eqnarray}
where $\phi(z)$ is real and depends on $z$. The density of zeros on
this curve is proportional to $d\phi(z)/dz$. 

A simple example occurs for  hard hexagons where
on segments of the negative $z$-axis there is a complex conjugate pair
of eigenvalues which have the maximum modulus.

However, we will see that for the hard hexagon model 
there are points in the complex plane where more than two eigenvalues
values have equal moduli. Indeed, for hard squares at $z=\, -1$, we
have previously noted that all eigenvalues have modulus one.

For values of $z$ in the complex plane where the interchange of 
 (\ref{limitint}) holds the limiting locus of partition function
zeros for the square $L_v=L_h$ lattice will coincide with the transfer
matrix equimodular curves. However there is no guarantee that the
interchange (\ref{limitint}) holds in the entire complex $z$ plane.

Our considerations are somewhat different for toroidal and cylindrical
boundary conditions and we treat these two cases separately.

\subsubsection{\label{sec:precb} Cylindrical boundary conditions \label{app:fss} }

For cylindrical boundary conditions the partition function is given by
(\ref{eigen2}) which in addition to the eigenvalues of $T(L_h)$
depends on the boundary vector ${\bf v}_B$ (\ref{vector}).
Because  of the periodic boundary conditions in the $L_h$
direction there is a conserved momentum $P$. Consequently the transfer 
matrix and translation operator 
may be simultaneously diagonalized.  Therefore the transfer matrix may 
be block diagonalized by a transformation which is independent of $z$ 
and hence the characteristic equation will
factorize. Furthermore the boundary vector (\ref{vector}) 
for the cylindrical case satisfies 
\begin{eqnarray} 
\hspace{-0.9in}&& \qquad \quad  \quad 
{\bf v}_B(a_1,a_2,\, \cdots, \, a_{L_h})
\,\,\, =\, \, \, \,\,
{\bf v}_B(a_{L_h},a_1,\cdots,\,  a_{L_h-1})
\end{eqnarray}
and thus is also translationally invariant. 
Therefore  the   only eigenvectors which contribute to the partition
function in (\ref{eigen2}) lie in the translationally invariant
subspace where $P\, =\, 0$. Consequently we are able to restrict our 
attention to the reduced transfer matrix for this translationally
invariant sector where the momentum of the state is $P\, =\, 0$ 
because all of the scalar products $c_k$ in (\ref{eigen2}) for 
eigenvectors in sectors with $P\, \neq \, 0\, $ vanish.

\subsubsection{\label{sec:prepb} Toroidal boundary conditions \newline }

For toroidal boundary conditions the partition function in
(\ref{eigen1}) is the sum over all eigenvalues and a new feature
arises because for $P\neq 0, \pi$ the eigenvalues for $\pm P$ are
degenerate in modulus, but may have complex conjugate phases which are
independent of $z$. By grouping these two eigenvalues together we see
that the discussion leading to (\ref{emod2}) still applies. 
There are now three types of equimodular curves:
\vskip .1cm 
1)  Two eigenvalues are equal for crossings of eigenvectors 
with  $\, P\, =\, 0,\, \pi$,  
\vskip .1cm 
2) Three eigenvalues are equal for crossings of eigenvectors  of 
$\, P\, =\, 0,\, \pi$ with $ \, P\, \neq \,\,  0,\, \pi$, 
\vskip .1cm 
3) Four eigenvalues are equal for crossing of eigenvectors 
with $\,  P\, \neq \,\,  0, \, \pi$.

\subsubsection{\label{sec:preas} Nonzero finite aspect ratios $L_v/L_h$ \newline }

We are, of course, not really interested in the limit (\ref{emod1})
but rather in the case of finite nonzero aspect ratio $L_v/L_h$ and
particularly in the isotropic case $L_v=L_h$. There
is apparently no general theory for finite nonzero aspect ratio in
the literature and we will study this case in detail below.

\section{\label{sec:pf} The partition functions $\kappa_{\pm}(z)$ per site for hard hexagons}

Baxter \cite{baxterhh,baxbook} has computed the fugacity and 
the partition function per site in
terms of an auxiliary variable $x$ using the functions
\begin{eqnarray} 
\hspace{-0.9in}&& \quad \,   \quad 
G(x)\,\, = \,\, \,
\prod_{n=1}^{\infty}\, \frac{1}{(1-x^{5n-4})(1-x^{5n-1})},   
\\
\hspace{-0.9in}&& \quad \,   \quad 
H(x)\,\, =\,\,\, 
\prod_{n=1}^{\infty}\, \frac{1}{(1-x^{5n-3})(1-x^{5n-2})}, 
\quad \quad \, \,   
Q(x)\,\, =\,\,\, \prod_{n=1}^{\infty}\, (1-x^n).
\end{eqnarray} 
For high density where
$0 \, < \, z^{-1}\, < \,z_c^{-1}$ the results are
\begin{eqnarray} 
\label{zhi}
\hspace{-0.9in}&& \qquad \quad  \quad  
z\,\, =\,\,\,   
 {{1} \over {x}} \cdot \, \Bigl({{G(x)} \over {H(x)}}\Bigr)^5
\qquad \qquad \hbox{and} 
\end{eqnarray}
\begin{eqnarray} 
\label{khi}
\hspace{-0.9in}&& \qquad \quad  \quad 
\kappa_{+}\,\, =\,\, \,
  {{1} \over {x^{1/3}}} \cdot \,  \frac{ G^3(x) \, Q^2(x^5)}{H^2(x)}
 \cdot \, \prod_{n=1}^{\infty}
\frac{(1-x^{3n-2})(1-x^{3n-1})}{(1-x^{3n})^2}, 
\end{eqnarray}
where, as $x$ increases from $0$ to $1$, the value of $z^{-1}$ 
increases from  $0$ to $z_c^{-1}$.

For low density where $\, 0 \, \leq \, z \, < \, z_c$
\begin{eqnarray} 
\label{zlo}
\hspace{-0.9in}&& \quad \quad  \quad 
z\,\, =\,\,\,
 -x \cdot \, \Bigl({{H(x)} \over {G(x)}}\Bigr)^5
\qquad \qquad \hbox{and} 
\end{eqnarray}
\begin{eqnarray} 
\label{klo}
\hspace{-0.9in}&& \quad \quad  \quad 
\kappa_{-}\,\,  =\,\,  \, 
\frac{H^3(x) \, Q^2(x^5)}{G^2(x)} \cdot \, \prod_{n=1}^{\infty} \, 
\frac{(1-x^{6n-4})(1-x^{6n-3})^2(1-x^{6n-2})}
{(1-x^{6n-5})(1-x^{6n-1})(1-x^{6n})^2}, 
\end{eqnarray}
where, as $x$ decreases from $0$ to $-1$, the value of $z$ increases
from $0$ to $z_c$.

\subsection{\label{sec:pfalg} Algebraic equations for $\kappa_{\pm}(z)$}

The auxiliary variable $x$ can be eliminated between the expressions
for $z$ and $\kappa$ (\ref{zhi})-(\ref{klo}) and the resulting 
functions $\kappa_{\pm}(z)$ are in fact algebraic
functions of $z$. 
To give these algebraic equations we follow Joyce \cite{joyce}
 and introduce the functions
\begin{eqnarray}
\hspace{-0.9in}&& \qquad \quad  \quad 
\Omega_1(z)\, =\,\,\,\, 1 +11 z -z^2,  
\\
\hspace{-0.9in}&& \qquad \quad  \quad 
\Omega_2(z)\, =\,\,\,\,  z^4+228 z^3 +494 z^2 -228 z +1,  
\\
\hspace{-0.9in}&& \qquad \quad  \quad 
\Omega_3(z)\, =\,\,\,\,  (z^2+1) \cdot \, (z^4 -522 z^3 -10006 z^2 +522 z +1). 
\end{eqnarray}

For the  high density Joyce (see eqn. (7.9) in~\cite{joyce})
 showed that the function $\kappa_{+}(z)$ satisfies 
a polynomial relation of degree 24 in the variable  $\kappa_{+}(z)$ 
\begin{eqnarray}
\label{alghi} 
\hspace{-0.9in}&& \qquad \quad  \quad 
f_{+}(z,\kappa_{+})\,\,  = \, \,\, \,
 \sum_{k=0}^{4}\, C_{k}^{+}(z) \cdot  \, \kappa^{6k}_{+}
\,\, \, =\,\,\, \,  0, \qquad \quad \hbox{where} 
\end{eqnarray} 
\begin{eqnarray}
&&    C_0^{+}(z)\, =\, \, -3^{27}\,z^{22}\nonumber\\ 
&&C_1^{+}(z)\,=\,\,  -3^{19}\, z^{16} \cdot \, \Omega_3(z), 
 \nonumber \\
&&   C_2^{+}(z)\, =\, \,
 -3^{10} \, z^{10} \cdot \,
 [\Omega_3^2(z)\, -2430  \, z \cdot \,  \Omega^5_1(z)], 
 \nonumber\\
&&  C_3^{+}(z)\,=\,\, 
 -z^4 \cdot \, \Omega_3(z) \cdot \,
 [\Omega_3^2(z)\, -1458  \, z \cdot \, \Omega_1^5(z)]\nonumber\\ 
&&C_4^{+}(z) \,=\, \, \Omega_1^{10}(z).
\end{eqnarray}
Joyce has also derived an algebraic equation for 
the density (see eqn. (8.28) in~\cite{joyce})
 which follows from (\ref{alghi}).

For low density we have obtained by means of a Maple computation the
substantially more complicated polynomial relation which was not
obtained in~\cite{joyce} 
\begin{eqnarray} 
\label{alglo}
\hspace{-0.9in}&& \qquad \quad  \quad 
f_{-}(z,\kappa_{-}) 
\,\,= \,\,\,
\sum_{k=0}^{12} \,C_k^{-}(z) \cdot \, \kappa_{-}^{2k}
\, \,=\,\,\, 0,    \qquad \qquad \hbox{where} 
\end{eqnarray} 
\begin{eqnarray}
\hspace{-0.95in}&&  
\quad  C_0^{-}(z)\, =\,\, -2^{32} \cdot 3^{27} \cdot  z^{22}, \nonumber\\ 
\hspace{-0.95in}&&\quad C_1^{-}(z)\, =\, \, 0\nonumber\\ 
\hspace{-0.95in}&&  \quad C_2^{-}(z)\, =\, \, 2^{26} \cdot 3^{23} \cdot 31 \cdot
z^{18} \cdot \, \Omega_2(z), 
\nonumber\\
\hspace{-0.95in}&&  \quad 
C_3^{-}(z)\, = \, \, 
2^{26} \cdot 3^{19} \cdot 47 \cdot z^{16} \cdot \, \Omega_3(z), \nonumber\\
\hspace{-0.95in}&&  \quad C_4^{-}(z)\, =\, \, 
 -2^{17} \cdot 3^{18} \cdot 5701 \cdot z^{14} \cdot \, \Omega_2^2(z), 
 \nonumber \\
\hspace{-0.95in}&&  \quad  
C_5^{-}(z)\, =\, \, -2^{16} \cdot 3^{14} \cdot 7^2 \cdot 19 \cdot 37 \cdot 
z^{12} \cdot \, \Omega_2(z) \, \Omega_3(z),
\nonumber\\
\hspace{-0.95in}&&  \quad  
C_6^{-}(z)\, =\, \, 
-2^{10} \cdot 3^{10} \cdot 7 \cdot z^{10} \cdot \, [273001
\cdot\Omega_3^2(z) \, +2^6\cdot 3^5 \cdot 5 \cdot 4933 \cdot
z \cdot \, \Omega_1^5(z)], 
\nonumber\\
\hspace{-0.95in}&& \quad  
C_7^{-}(z)\, =\, \, -2^9 \cdot 3^{10} \cdot 11 \cdot 13 \cdot 139 \cdot 
z^{8} \cdot \, \Omega_3(z) \, \Omega_2^2(z), 
 \nonumber\\
\hspace{-0.95in}&&  \quad  
C_8^{-}(z)\, =\, \,
 -3^5 \cdot z^{6} \cdot \, \Omega_2(z) \cdot \, 
[7\cdot 1028327 \cdot  \, \Omega_3^2(z)\, 
 -2^6 \cdot 3^4 \cdot 11 \cdot 419 \cdot 
16811 \cdot z \cdot \, \Omega_1^5(z)],
\nonumber\\
\hspace{-0.95in}&&  \quad  
C_9^{-}(z)\, = \, \, 
-z^4  \cdot \, \Omega_3(z) \cdot \, [37 \cdot 79087 \, \Omega_3^2(z)
+2^6 \cdot 3^6 \cdot 5150251\cdot z \cdot \, \Omega_1^5(z)], 
 \nonumber\\
\hspace{-0.95in}&&  \quad   
C_{10}^{-}(z)\, =\, \, 
-z^2 \cdot \, \Omega_2^2(z) \, \cdot \,  [19\cdot 139\Omega_3^2(z)
 \,-2 \cdot 3^6 \cdot 151 \cdot 317 \cdot z \cdot \, \Omega^5_1(z)]
 \nonumber\\
\hspace{-0.95in}&& \quad  
C_{11}^{-}(z)\, =\, \, -\Omega_2(z)\, \Omega_3(z) \cdot \, 
[\Omega_3^2(z)\, -2\cdot 613 \cdot z \cdot\, \Omega_1^5(z)],\nonumber\\
\hspace{-0.95in}&& \quad C_{12}^{-}(z)\, = \,\,  \Omega_1^{10}(z).
\end{eqnarray}
We have verified that Joyce's algebraic equation  
for the density (see eqn. (12.10) in~\cite{joyce})
 follows from (\ref{alglo}).

We note the symmetry
\begin{eqnarray} 
\hspace{-0.9in}&& \qquad \quad  \quad   \quad  \quad 
z^{44} \cdot \, 
f_{\pm}\Bigl(-{{1} \over {z}},\, {{\kappa_{\pm}} \over {z}}\Bigr)
\,\, \, \,  = \,\,\, \,  \, 
f_{\pm}(z,\,\kappa).
\end{eqnarray}

In \ref{app:sing} we discuss the behavior $\kappa_{\pm}(z)$ at 
the singular points $z_c,~z_d$. 

\subsection{\label{sec:pfcomp} Partition function for complex $z$}

\begin{figure}[h!]

\begin{center}
\begin{picture}(160,150)
\put(0,-15){\includegraphics[width=6cm]{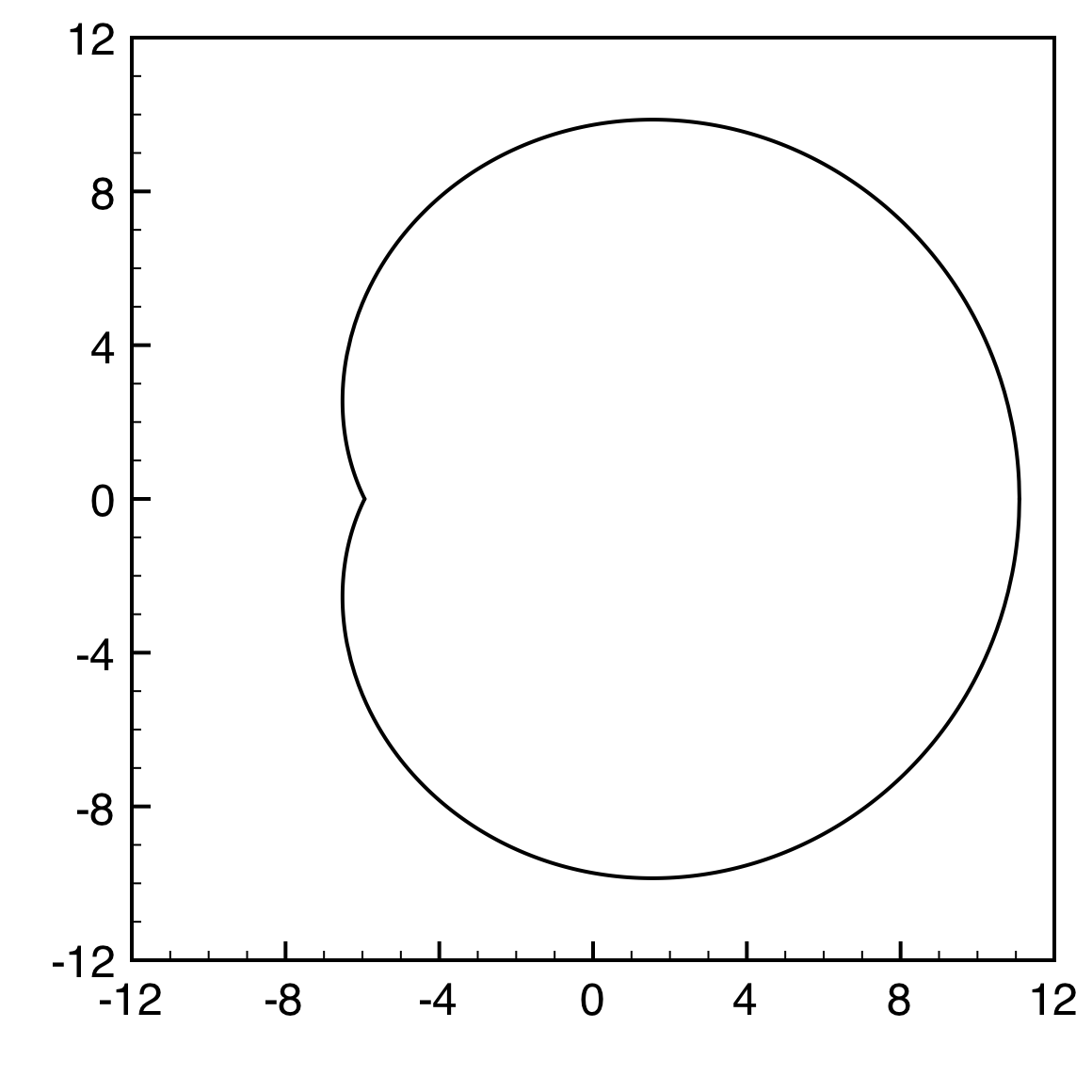}} 
\end{picture}
\end{center}

\caption{\label{fig:baxter} The equimodular curve for $|\kappa_{-}(z)|\,=\,|\kappa_{+}(z)|$
in the complex $z$ plane. The crossing of the positive $z$ axis is at
$z_c$ and the crossing of the negative $z$ axis is at $z=\,-5.925104\cdots$} 

\end{figure}

 From section~\ref{sec:prepfev}  we see that
the simplest construction of the partition function of hard hexagons 
in the complex $z$ plane would be if the low and high density
eigenvalues in the thermodynamic limit were the only two eigenvalues
of maximum modulus and that the interchange of limits (\ref{limitint})
holds for $z$ in the entire complex plane. 
Then the zeros would be given by  the
equimodular curve $|\kappa_{+}(z)|\, =\,|\kappa_{-}(z)|$. 
Because the partition functions per site
$\kappa_{\pm}(z)$ satisfy algebraic equations this curve will satisfy
an algebraic equation which can be found by setting 
$\kappa_{+}(z)\, =\, r\kappa_{-}(z)$ 
in the equation (\ref{alghi}) for $\kappa_{+}(z)$ and
computing the resultant between equations (\ref{alglo}) and
(\ref{alghi}). The solutions of this equation for $r$ on the unit circle 
will give all the locations where $\kappa_{\pm}(z)$ have equimodular
solutions. We have produced this equation using Maple but
unfortunately is it too large to print out. However, we are
only interested in the crossings of the maximum modulus eigenvalues. 
Consequently we have computed this curve not from its algebraic
equation but directly from the parametric representations 
(\ref{zhi})-(\ref{klo}). We plot this curve in figure~\ref{fig:baxter}. The curve 
crosses the positive real axis at $z_c$ and the negative 
real axis at $z\,=\, -5.9425104 \,\,\cdots $ which is exactly determined 
from the algebraic equation of the equimodular curve given in \ref{app:haupt}. 
The tangent to the equimodular curve is discontinuous at this negative
value of $z$. 

We will see  in the next section that this two eigenvalue assumption
is insufficient to  account for our finite size computations in some
regions of the plane.

\section{\label{sec:ev} Transfer matrix eigenvalues}

To obtain further information on the partition function in the complex
$z$ plane we compute, in this section, the eigenvalues for finite sizes of the 
transfer matrix $T(z; \,L_h)$ for the case of periodic boundary conditions 
in the $L_h$ direction.

There are two ways to study the eigenvalues of the transfer matrix;
analytically and numerically. Numerical computations can be carried 
out on matrices which are too large for symbolic computer programs to
handle. However, analytic computations reveal properties which cannot
be seen in numerical computations. Consequently we begin our
presentation with analytic results before we present our numerical
results.

\subsection{\label{sec:evana} Analytic results}

The eigenvalues of a matrix are obtained as the solutions of its
characteristic equation. For  the transfer matrices of the 
hard hexagon model this characteristic equation is a polynomial
in the parameter $z$ and the eigenvalue $\lambda$ with integer
coefficients. Consequently the eigenvalues $\lambda$ are algebraic
functions of $z$. In general such characteristic polynomials will be
irreducible (i.e. they will not factorize into products of polynomials
with integer coefficients). 

There are two important analytic non-generic features of the
hard hexagon eigenvalues: factorization of the characteristic
equation and the multiplicity of the roots of the resultant.

\subsubsection{\label{sec:evfac} Factorization of the characteristic equation \newline}

For a transfer matrix with cylindrical boundary conditions the
characteristic equation factorizes into subspaces characterized by a
momentum eigenvalue $P$. In general the characteristic polynomial 
in the translationally invariant $P\,=\, 0$ subspace  will be
irreducible. We have found that this is indeed the case for hard 
squares. However, for hard hexagons we find that for $L_h\,=\,12, \,15,\,18$,
the characteristic polynomial, for $P\, =\, 0$, factors into the product of two
irreducible polynomials with integer coefficients. We have not been
able to study the factorization for larger values of $L_h$ but we
presume that factorization always occurs and is a result of the 
integrability of hard hexagons. What is unclear is if for larger lattices 
a factorization into more than two factors can occur.

\subsubsection{\label{sec:evmul} Multiplicity of the roots of the resultant \newline}

An even more striking non-generic property of hard
hexagons is seen in the computation of the resultant of the
characteristic polynomial in the translationally invariant sector. The
zeros of the resultant locate the positions of all potential
singularities of  the solutions of the polynomials. 

We have been able to compute the resultant for $L_h\,=\,12,\, 15,\, 18$, and
find that almost all zeros of the resultant have multiplicity two which 
indicates that there is in fact no singularity at those points and that the 
two eigenvalues cross. This very dramatic property will almost certainly 
hold for all $L_h$ and must be a consequence of the integrability (although 
to our knowledge no such theorem is in the literature). 

\subsection{\label{sec:evcb} Numerical results in the sector $P=\, 0$}

For the partition function with cylindrical boundary conditions
only the transfer matrix eigenvalues with $P\,=\,0$ contribute. 
In this sector we have numerically computed eigenvalues of 
the transfer matrix, in the $P\,=\,0$ sector, for
systems of size as large as $L_h\,=\,30$ which has dimension 31836. 
For such large matrices brute force computations will obviously not be 
sufficient and we have developed algorithms specific to this problem 
which we sketch in \ref{app:TM}. We restrict our attention to
values of $L_h$ being a multiple of three, to minimize boundary effects
which will occur  when the circumference $L_h$ is incompatible with
the three-sublattice structure of the triangular lattice.

In figure~\ref{fig:hhcflocus} we plot the equimodular curves for the crossing of the
largest transfer matrix eigenvalues in the $P\, =\, 0$ sector 
for $L_h\, = \, 12,\,  15,\, 18,\, 21,  \, 24, \, 27$,
 and in figure~\ref{fig:comp1} we plot $L_h\, =\, 30$.  
It is obvious from these curves that more than two
eigenvalues of the transfer matrix contribute to the partition
function because of the increasing number of regions in the left half
plane which we refer to as the ``necklace''.
A striking feature is that there is a  pronounced mod 6 effect where for
$L_h\,\equiv\, 3 \, \,  ({\rm mod} \, 6)$ there is a level crossing curve in the
necklace on the negative real $z$ axis which is not present for 
$L_h\,  \equiv \,  0 \, \,  ({\rm mod} \, 6)$. The level crossing curves
separate the necklace into well defined regions. The number of these
regions is $\, L_h/3-4 \, $  for $L\,\leq \, 27$. The number of regions for
$L_h\,=\,30$ is the same as for $L_h\, =\, 24$. For $L_h\,=\,21, \, 27$ all 
the branch points of the necklace are given in table~\ref{tab:necklace1}
and in table~\ref{tab:necklace2} for $L_h\,=\,18,\, 24, \, 30$.

There are further features in figures~\ref{fig:hhcflocus}  and \ref{fig:comp1} which
deserve a more detailed discussion.

\subsubsection{\label{sec:evcomp} Comparison with the equimodular curve of $\kappa_{\pm}(z)$ \newline}

\begin{figure}[h!]

\begin{center}
\begin{picture}(360,480)
\put(0,300){\includegraphics[width=6cm]{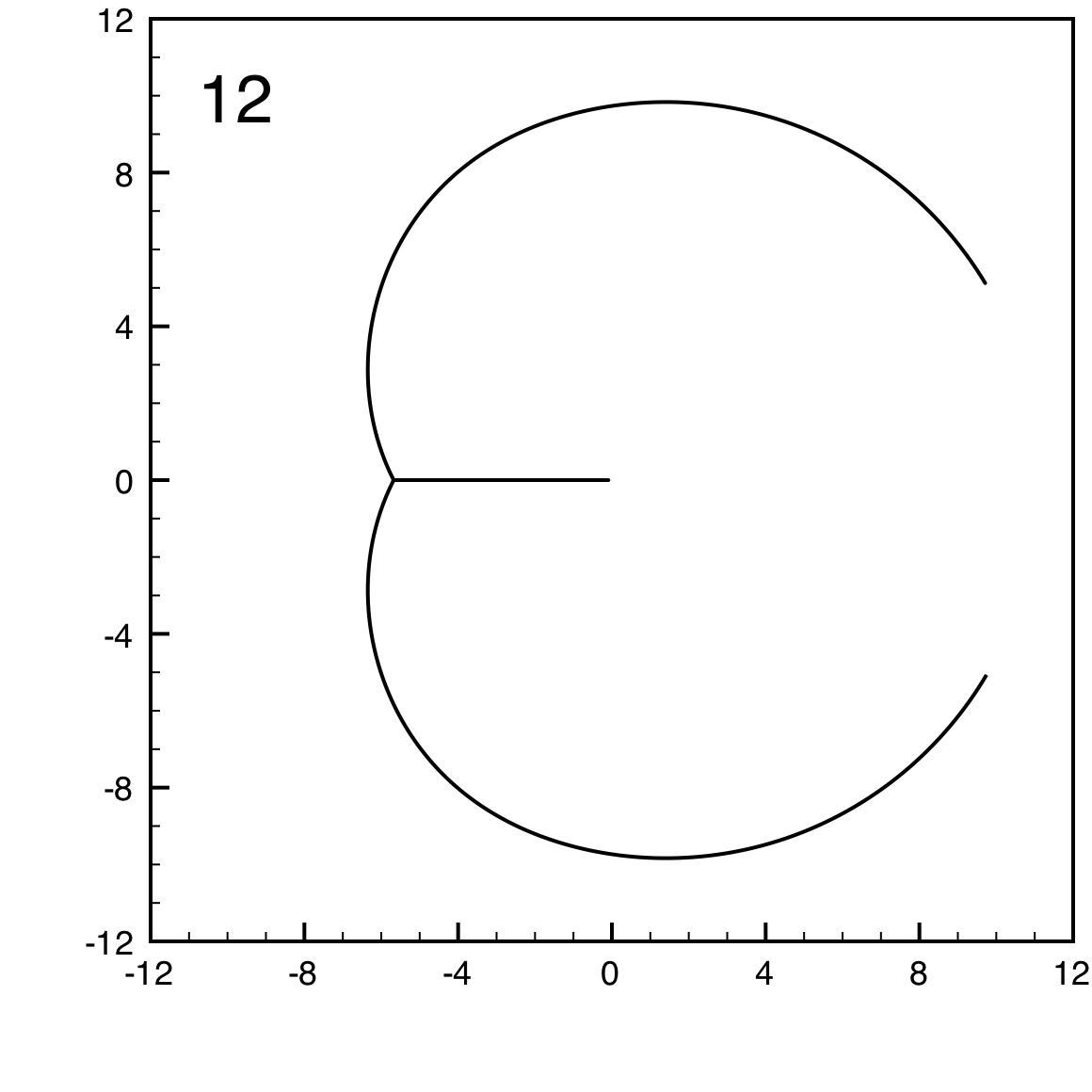}} 
\put(180,300){\includegraphics[width=6cm]{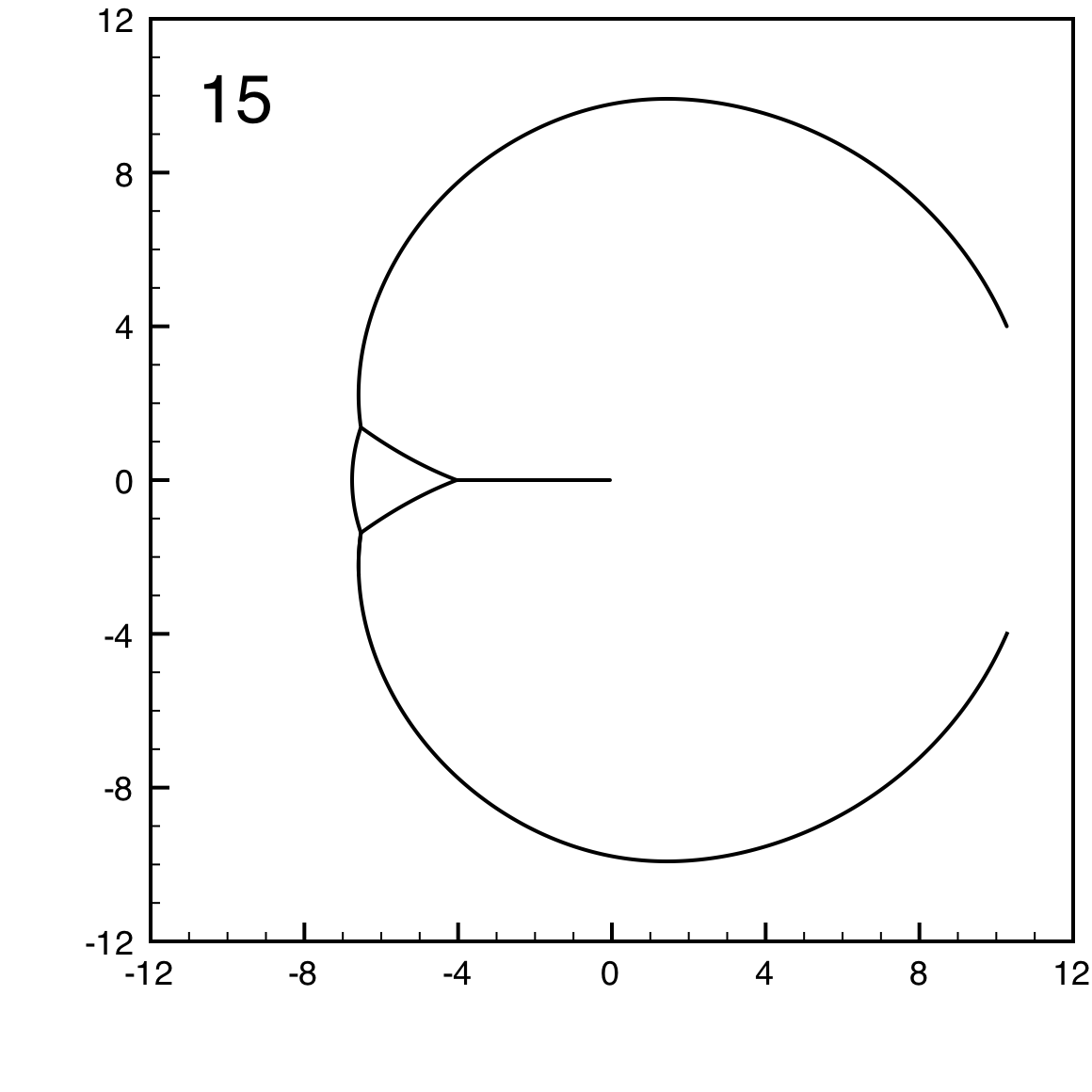}}
\put(0,140){\includegraphics[width=6cm]{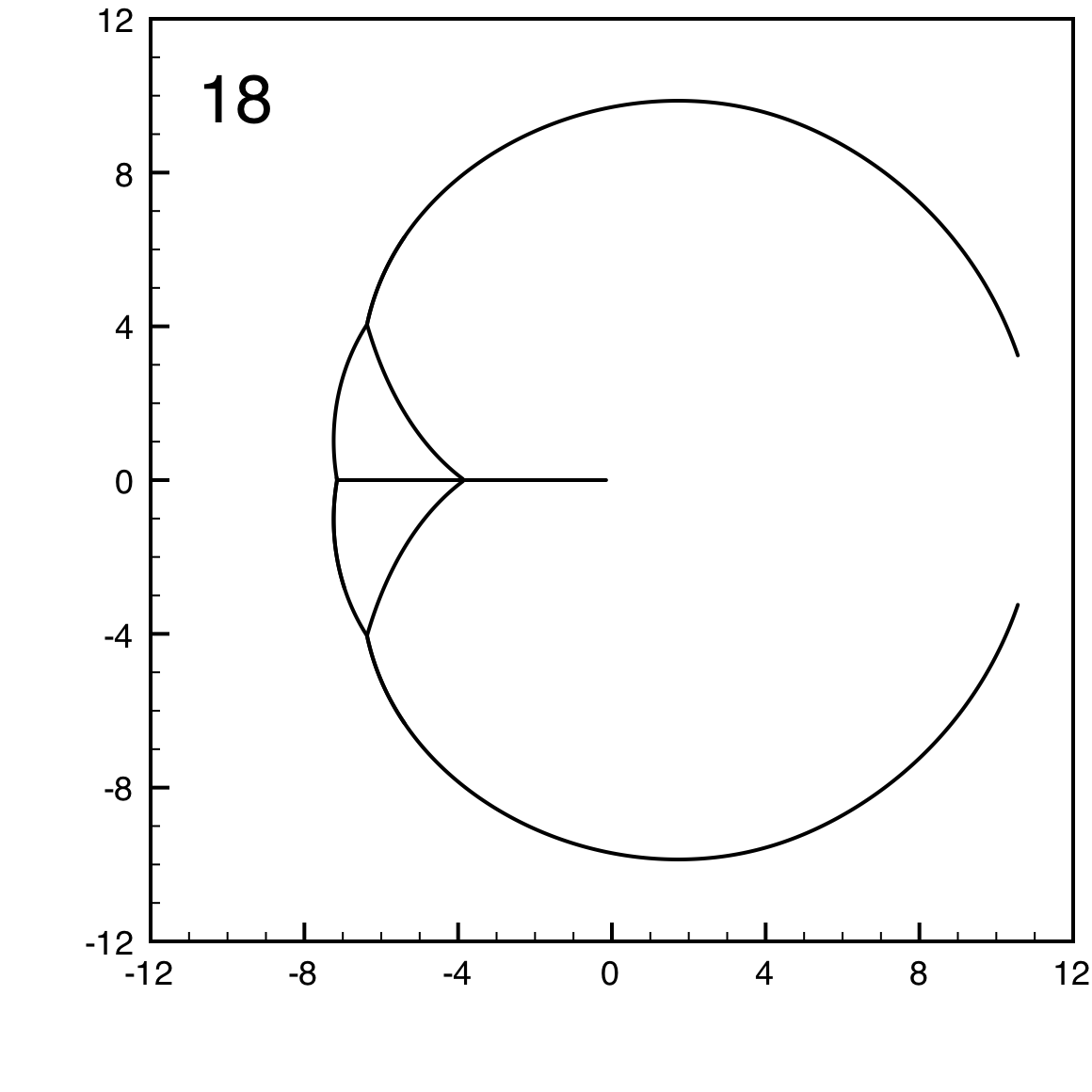}}
\put(180,140){\includegraphics[width=6cm]{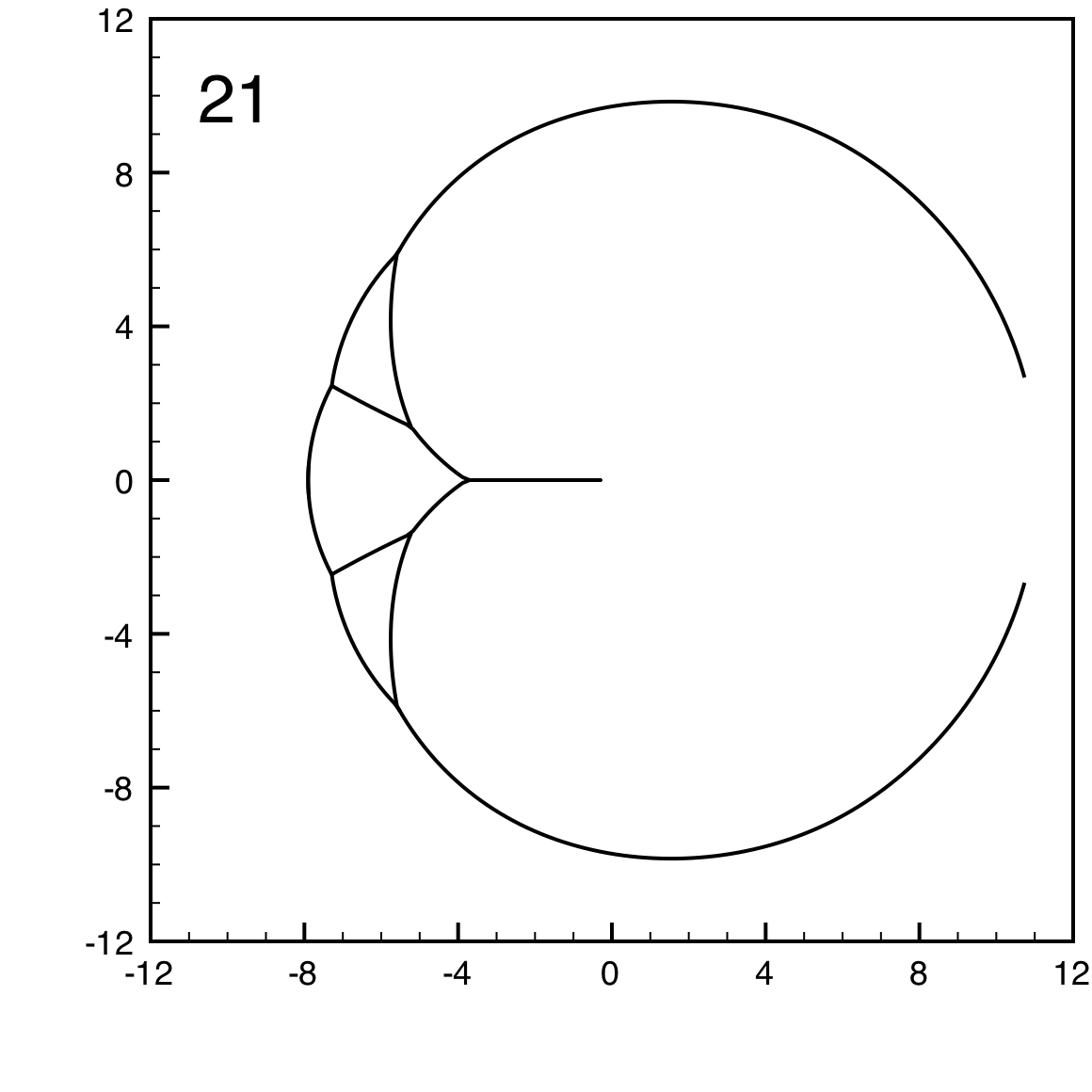}}
\put(0,-20){\includegraphics[width=6cm]{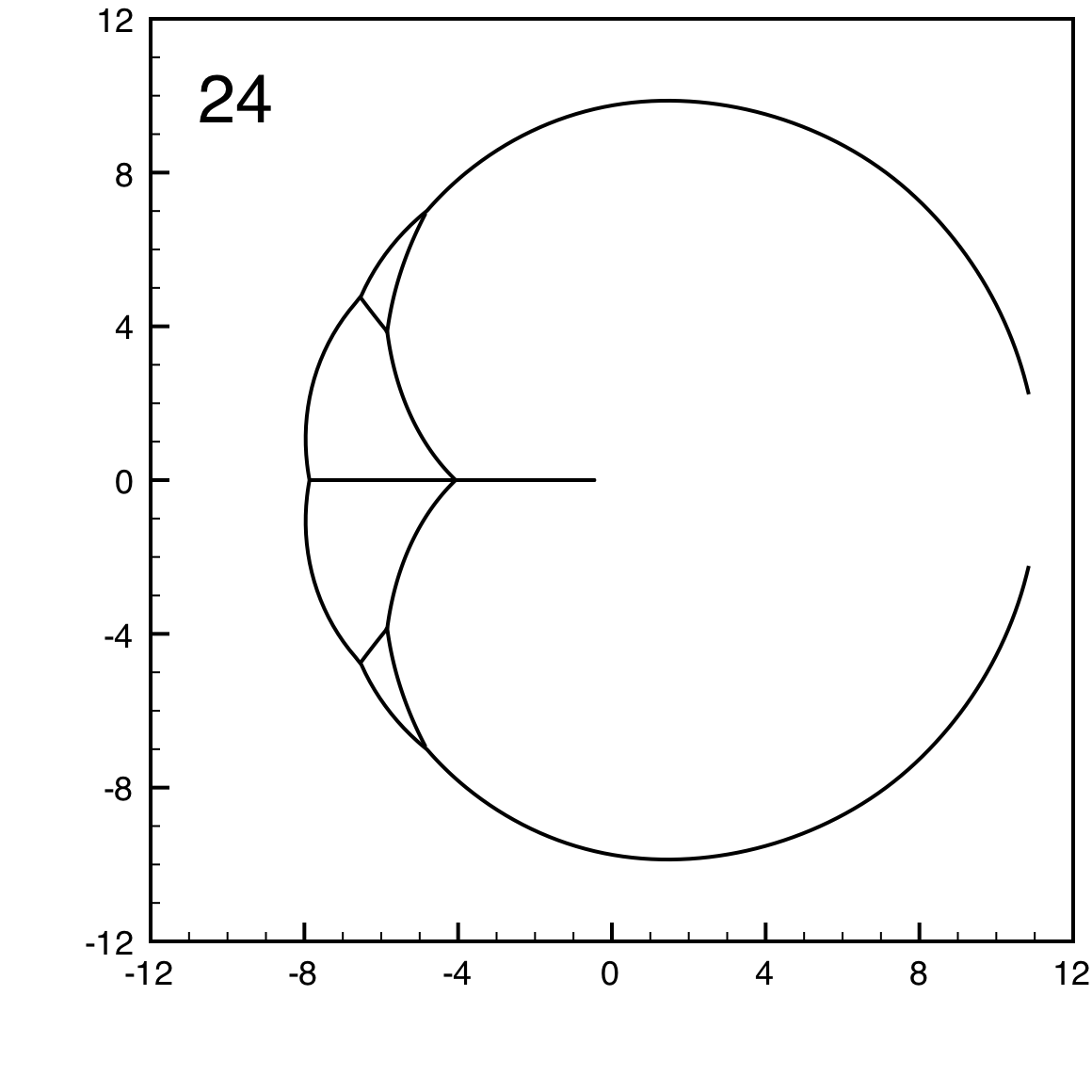}}
\put(180,-20){\includegraphics[width=6cm]{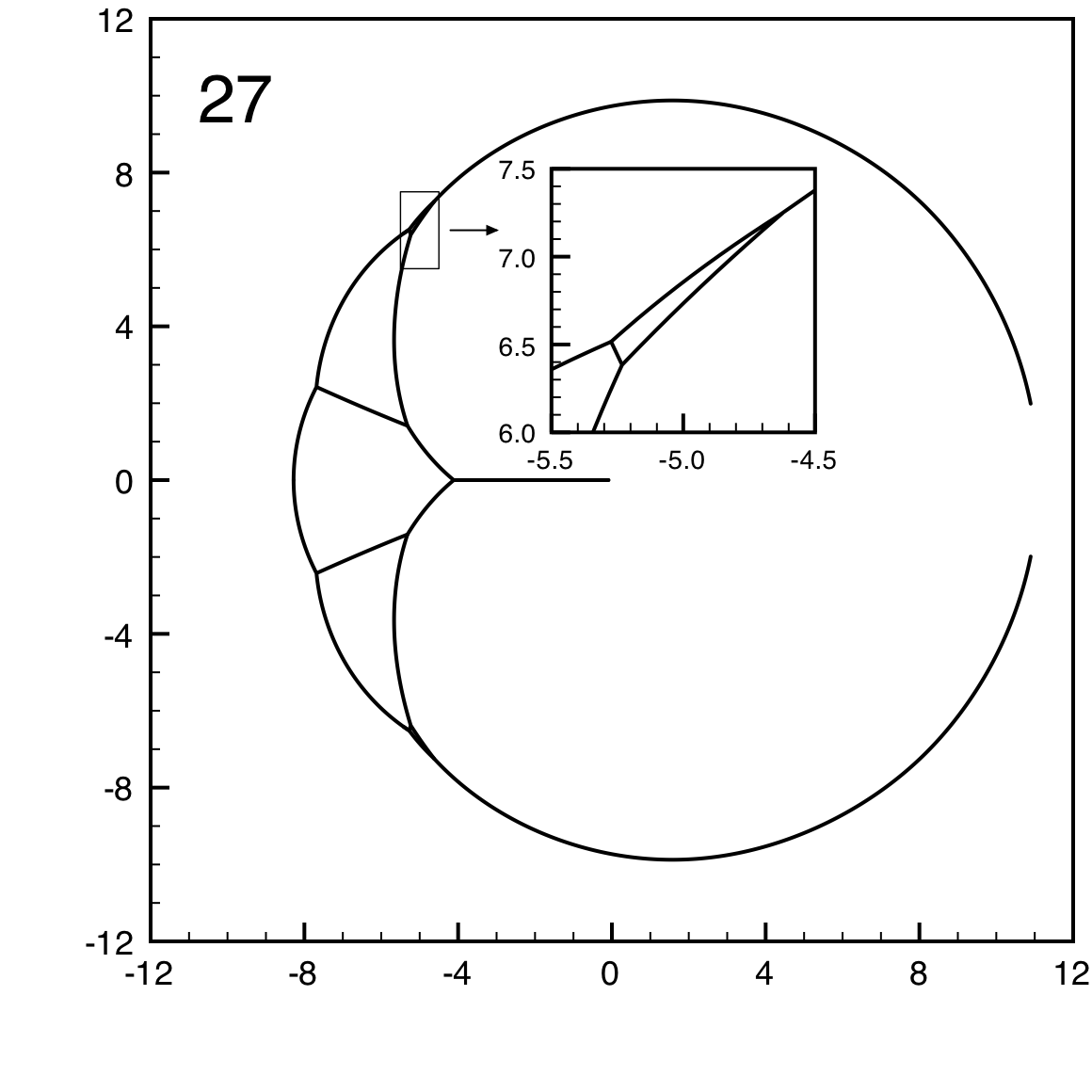}}
\end{picture}
\end{center}
\caption{\label{fig:hhcflocus} Plots in the complex fugacity plane $z$ of the 
equimodular curves of hard hexagon eigenvalues with cylindrical boundary
conditions of size $L_h\,=\,12,\,15,\,18,\,21,\,24,\,27$. The value of $L_h$
is given in the upper left hand corner of the plots.}
\end{figure}

\begin{figure}[ht]

\begin{center}
\begin{picture}(170,170)
\put(0,-5){\includegraphics[width=6cm]{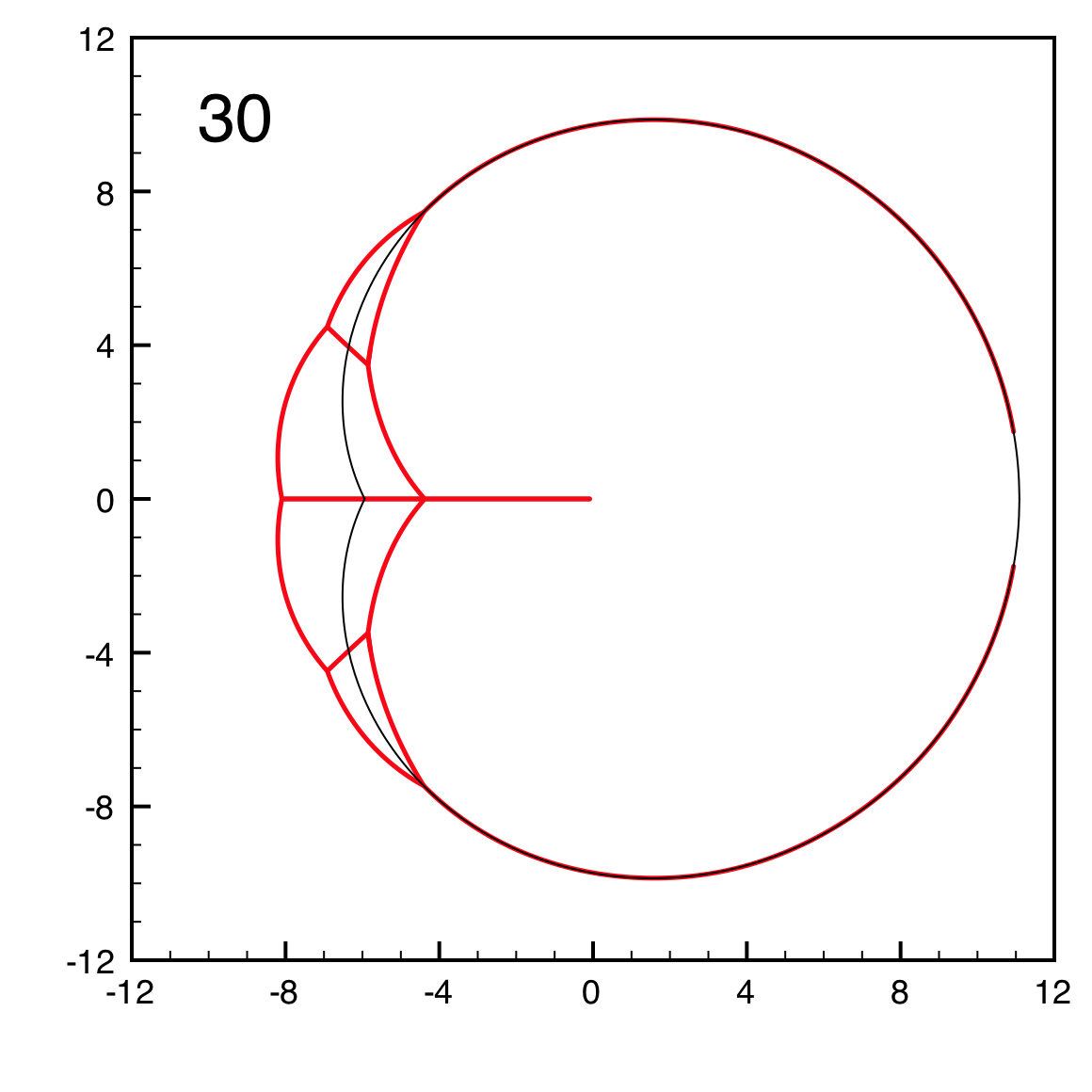}}
\end{picture}
\caption{ \label{fig:comp1} Comparison of the dominant eigenvalue crossings $L_h=30$
  shown in red with the equimodular curve
  $|\kappa_{+}(z)|\,=\,|\kappa_{-}(z)|$ of figure~\ref{fig:baxter} shown in black. Color
  online.}
\end{center}
\end{figure}

If the two eigenvalues $\kappa_{\pm}(z)$ computed in~\cite{baxterhh}
were sufficient to describe the $L_h\, \rightarrow \,\infty$
thermodynamic limit of
these finite size computations then the equimodular curves of figures~\ref{fig:hhcflocus}
 and \ref{fig:comp1} must approach the equimodular curve of $\kappa_{\pm}(z)$ of 
figure~\ref{fig:baxter}. We make this comparison for $L_h\,=\,30$ in figure~\ref{fig:comp1}.

In figure~\ref{fig:comp1} the agreement of the $\kappa_{\pm}(z)$ level crossing
curve with the eigenvalue equimodular curve for $L_h\,=\,30$ is
exceedingly  good in the entire portion of the plane which does not
include the necklace. However, in the necklace region the
$\kappa_{\pm}(z)$ curve does not agree with either the inner or outer
boundaries of the necklace but rather splits the necklace region into two
parts.  

A more quantitative argument follows from the values of the leftmost 
crossing with the negative real axis 
of the necklace given in table~\ref{tab:necklace1} for 
$L_h\, \equiv\, 0 \, \, ({\rm  mod} \,6)$ and in table~\ref{tab:necklace2}
 for $L_h\, \equiv\, 3 \, \, ({\rm mod} \,6)$. In both cases  
the left most crossing moves to the left to a value which if
extrapolate in terms of $1/L_h$ lies between 9 and 10. The Y branching
in tables~\ref{tab:necklace1} and~\ref{tab:necklace2} also moves to the left but does not extrapolate to a
value to the left of $z\,=\,-5.9425104\,\cdots$ where the $\kappa_{\pm}(z)$
equimodular curve crosses the negative real axis. We interpret this 
as implying that the necklace persists in the thermodynamic limit and that 
at least one more transfer matrix eigenvalue is needed  to explain 
the analyticity of the free energy in the complex $z$ plane.

\begin{table}[h]
\begin{center}
\begin{tabular}{|c|c|c|}\hline
$L_h=21$& $L_h=27$&comment\\ \hline
$-3.7731$&$-4.1138$&Y branching\\
$-5.5898\pm 5.8764i$&$-4.6228\pm 7.2480i$&necklace end\\
&$-5.2737\pm 6.5159i$&\\
&$-5.2321\pm 6.3840i$&\\
$-5.2264\pm 1.3949i$&$-5.3175\pm 1.4134i$&\\
$-7.2883\pm 2.4533i$&$-7.6848\pm 2.4225i$&\\
$-7.9020$&$-8.2803$&leftmost crossing \\ 
\hline
\end{tabular}
\end{center}
\caption{ \label{tab:necklace1} The branch points of the necklace of the equimodular curves 
of hard hexagons with cylindrical
boundary conditions for $L_h=\,21, \,27$.}

\end{table}

\begin{table}[h]
\begin{center}
\begin{tabular}{|c|c|c|c|}\hline
$L_h=18$& $L_h=\,24$&$L_h=30$&comment\\ \hline
$-3.8370$&$-4.0637$&$-4.3794$&Y branching\\
$-6.3703\pm 4.0485i$&$-4.8079\pm 7.0090i$&$-4.4043\pm 7.4623i$&necklace end\\
&$-6.5389\pm 4.7519i$&$-6.9134\pm 4.4771i$&\\
&$-5.8477\pm 3.8460i$&$-5.8526\pm 3.4864i$&\\
$-7.1499$&$-7.8663$&$-8.0937$&leftmost crossing \\ \hline
\end{tabular}
\end{center}
\caption{\label{tab:necklace2} The branch points of the necklace of the equimodular curves 
of hard hexagons with cylindrical
boundary conditions for $L_h=\,18,\, 24,\,30$.}

\end{table}

\subsubsection{\label{sec:evend} The endpoints $z_d(L_h)$ and $z_c(L_h)$ \newline}
 
In table~\ref{tab:endpoints} we give the endpoints which approach the
 unphysical and the physical singular points of the free energy
 $z_d$ and $z_c$. 
We also give in this table the ratio of the largest to the next
 largest eigenvalue at $z_d(L_h)$ and $z_c(L_h)$ 
as determined from
 eigenvalue crossings. In the limit $L_h \, \rightarrow \,  \infty$ this 
ratio must go to unity so the deviation from one is a measure of how 
far the  finite size $L_h$ is from the thermodynamic limit.

\begin{table}[h!]
\begin{center}
\begin{tabular}{|c|c|l||l|l|}\hline
$L_h$&$z_d(L_h)$ &$\lambda_1/\lambda_{\rm
      max}$&$z_c(L_h)$&$\lambda_1/\lambda_{\rm max}$\\ \hline
$12$&$-0.09051765$&$0.45085$&$~9.7432\pm 5.0712i$&$0.55487$\\
$15$&$-0.09037303$&$0.53048$&$10.2971\pm 3.9465i$&$0.54278$\\
$18$&$-0.09030007$&$0.59046$&$10.5753\pm 3.2016i$&$0.58463$\\
$21$&$-0.09026034$&$0.63709$&$10.7340\pm 2.6730i$&$0.62006$\\
$24$&$-0.09023555$&$0.67431$&$10.8310\pm 2.2852i$&$0.65030$\\
$27$&$-0.09021968$&$0.70467$&$10.8955\pm 1.9834i$&$0.67582$\\ 
$30$&$-0.09020833$&$0.72989$&$10.9389\pm 1.7499i$&$0.69827$\\
$\infty$&$-0.09016994$&$1.00000$&$11.09016994$&$1.00000$\\ \hline
\end{tabular}
\end{center}
\caption{\label{tab:endpoints} The values of the endpoints  
$z_d(L),\, z_c(L)$ for hard hexagons on the cylindrical
  lattice with length $L_h$ as determined from the equimodular
  eigenvalue curves 
and the ratios of the first excited state $\lambda_1$ to the largest
  eigenvalue $\lambda_{\rm max}$ at $z_d(L_h)$ and $z_c(L_h)$.}
\end{table}

\subsection{\label{sec:evpb} Eigenvalues for the toroidal lattice partition function}

\begin{figure}[ht]

\begin{center}
 
\begin{picture}(400,375)
\put(0,190){\includegraphics[width=6cm]{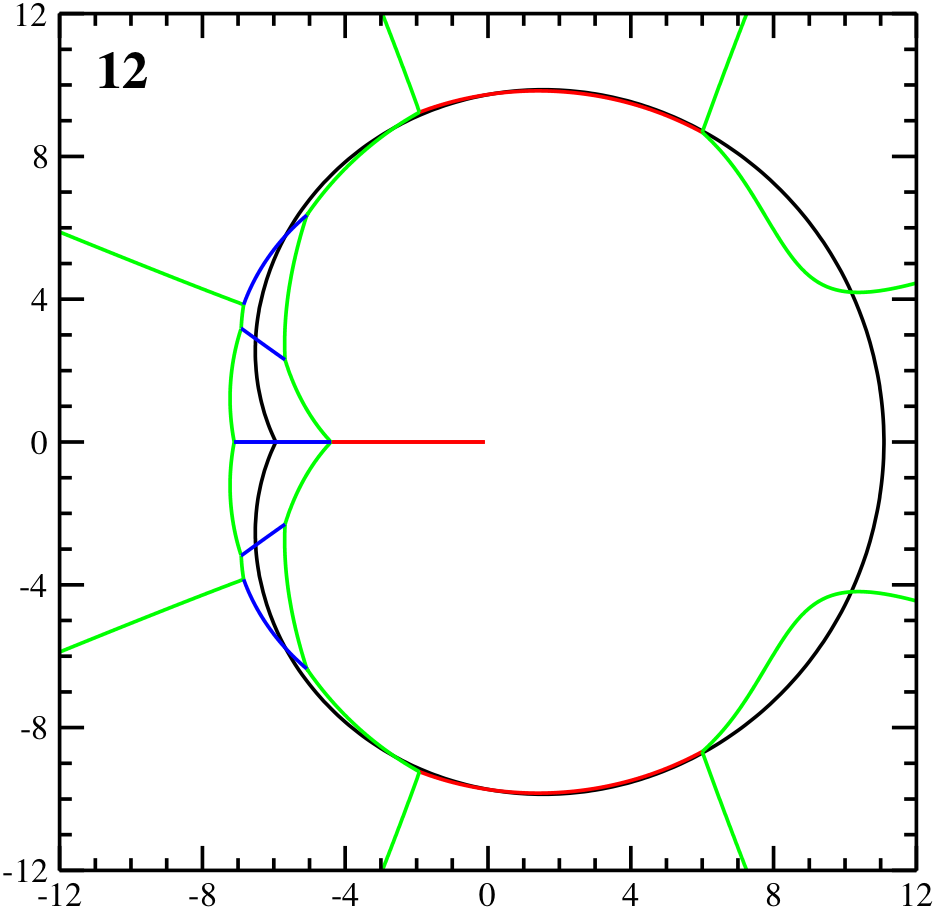}}
\put(200,190){\includegraphics[width=6cm]{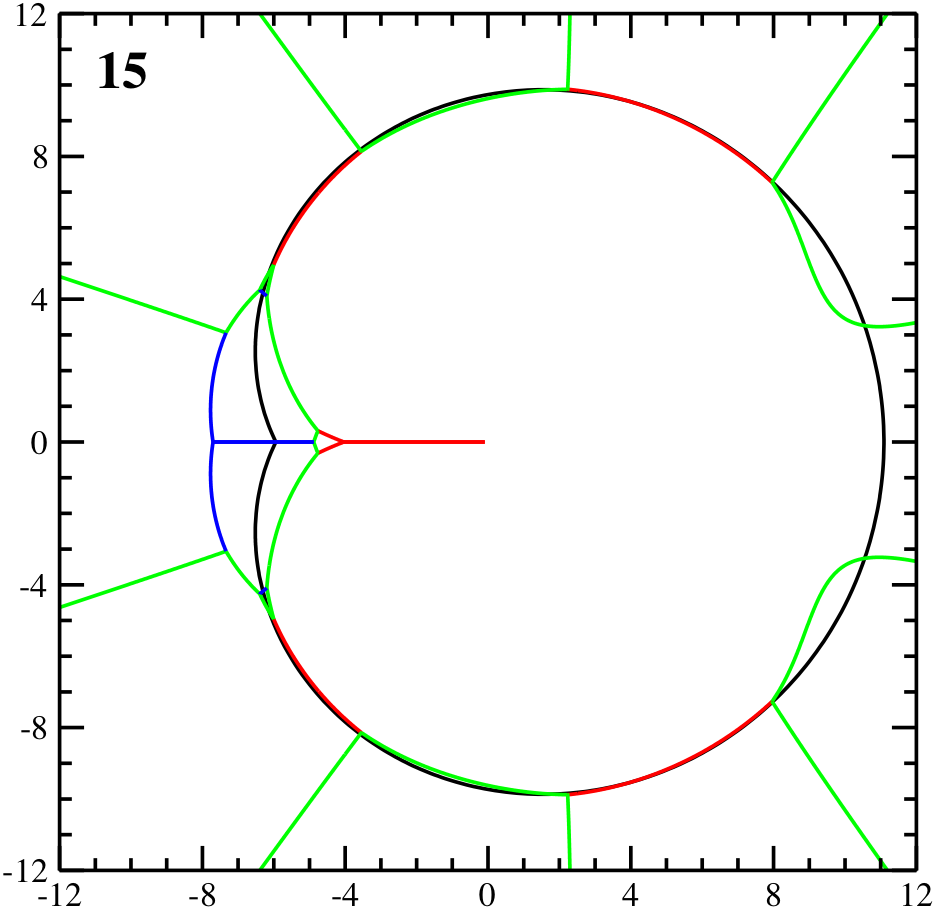}}
\put(0,0){\includegraphics[width=6cm]{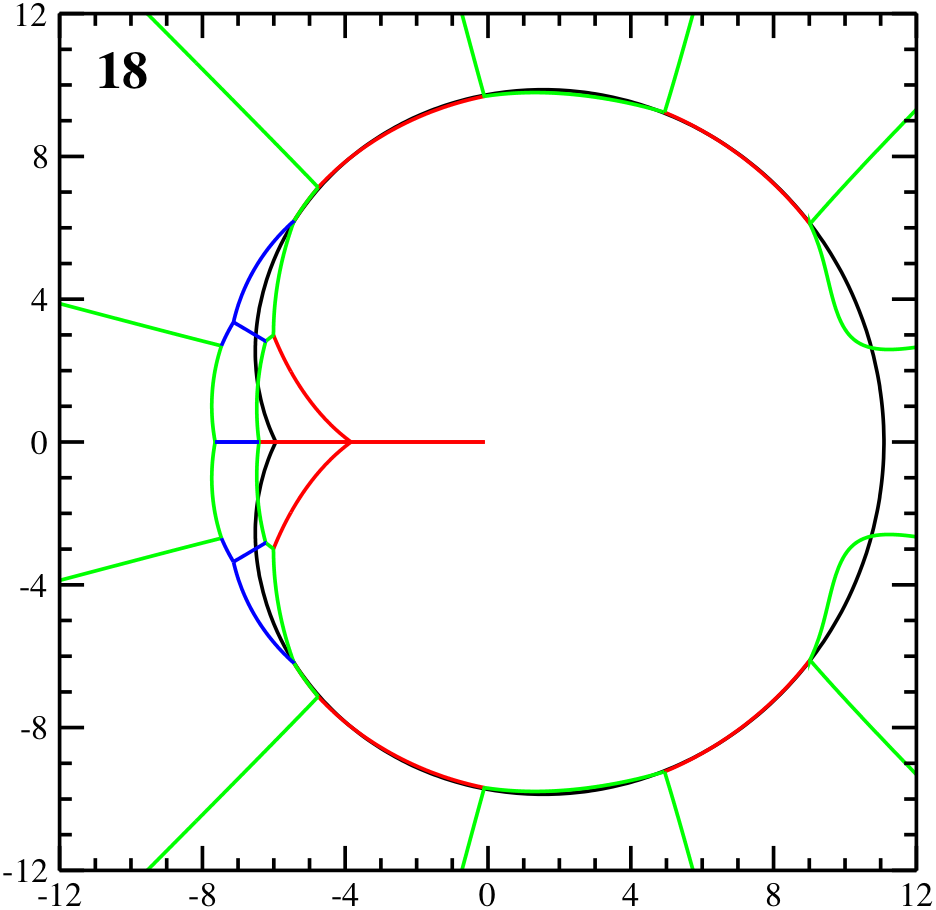}}
\put(200,0){\includegraphics[width=6cm]{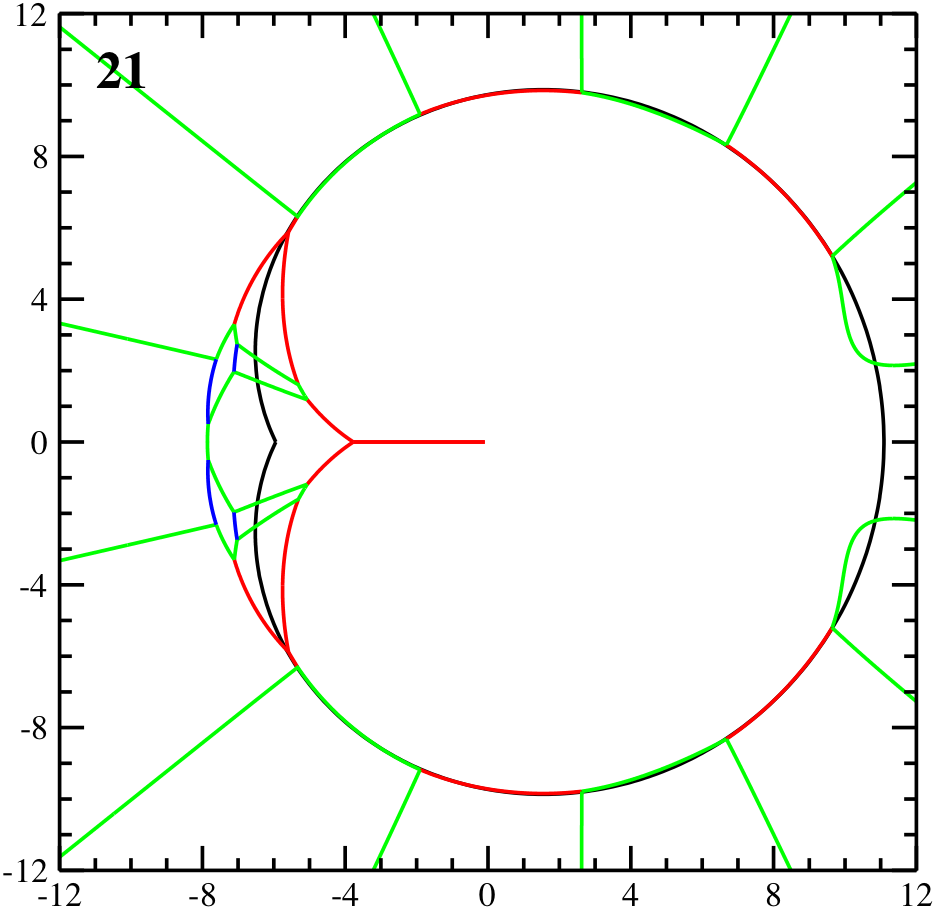}}
\end{picture}
\end{center}

\caption{\label{fig:hhpblocus} Plots in the complex fugacity plane $z$ of the equimodular
curves of hard hexagon eigenvalues for toroidal lattices for
$L=12,\, 15, \,18, \,21.$ On the red lines 2 eigenvalues are equimodular,
on the green lines 3 eigenvalues are equimodular and on the blue lines 4
eigenvalues are equimodular. The equimodular curve
$|\kappa_{-}(z)|\,=\,\,|\kappa_{+}(z)|$ is given in black for
comparison. Color online. }
\end{figure}

\begin{figure}[ht]

\begin{center}
\begin{picture}(495,400)
\put(0,200){\includegraphics[width=4cm]{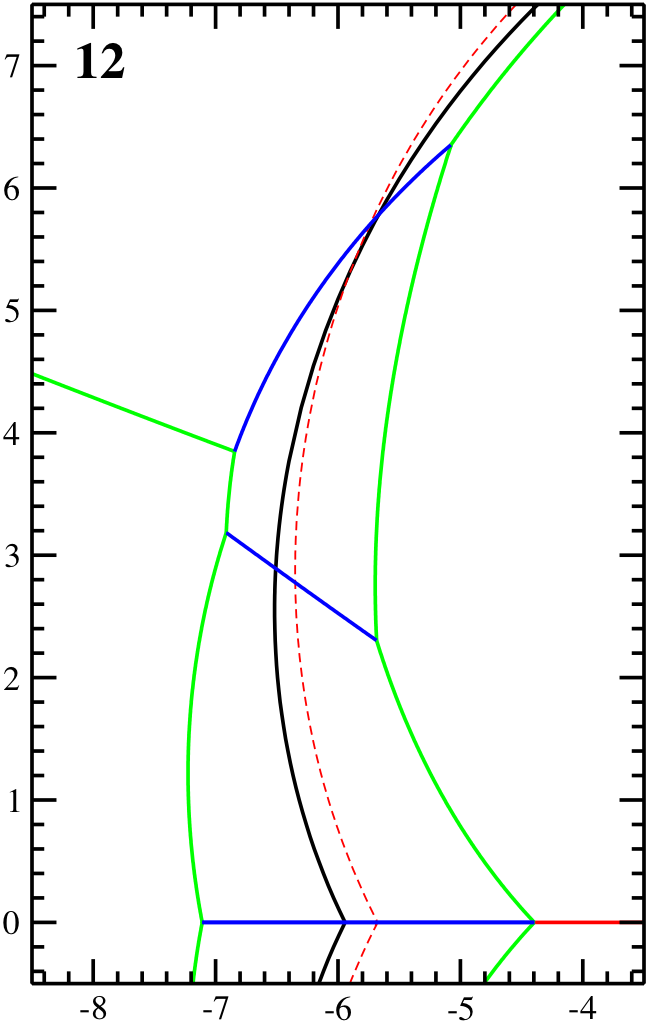}}
\put(165,200){\includegraphics[width=4cm]{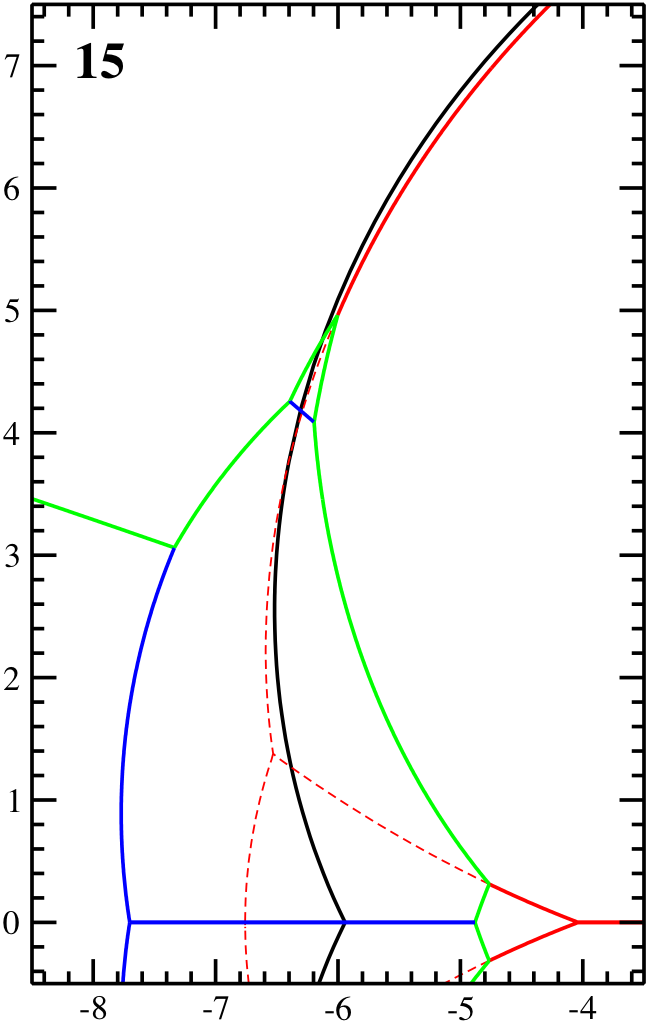}}
\put(335,200){\includegraphics[width=4cm]{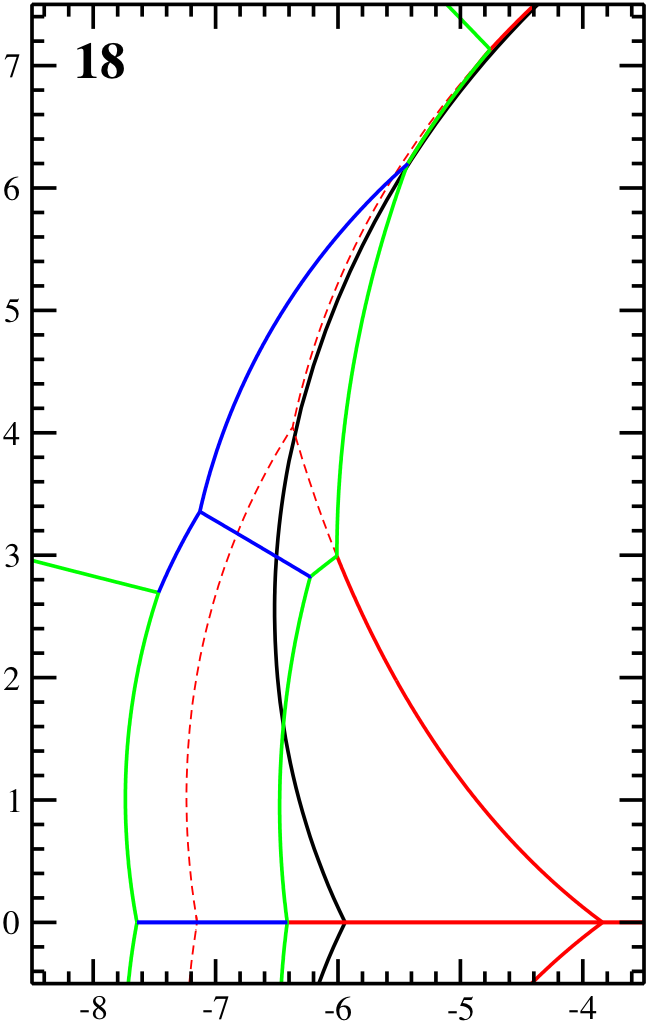}}
\put(65,0){\includegraphics[width=4cm]{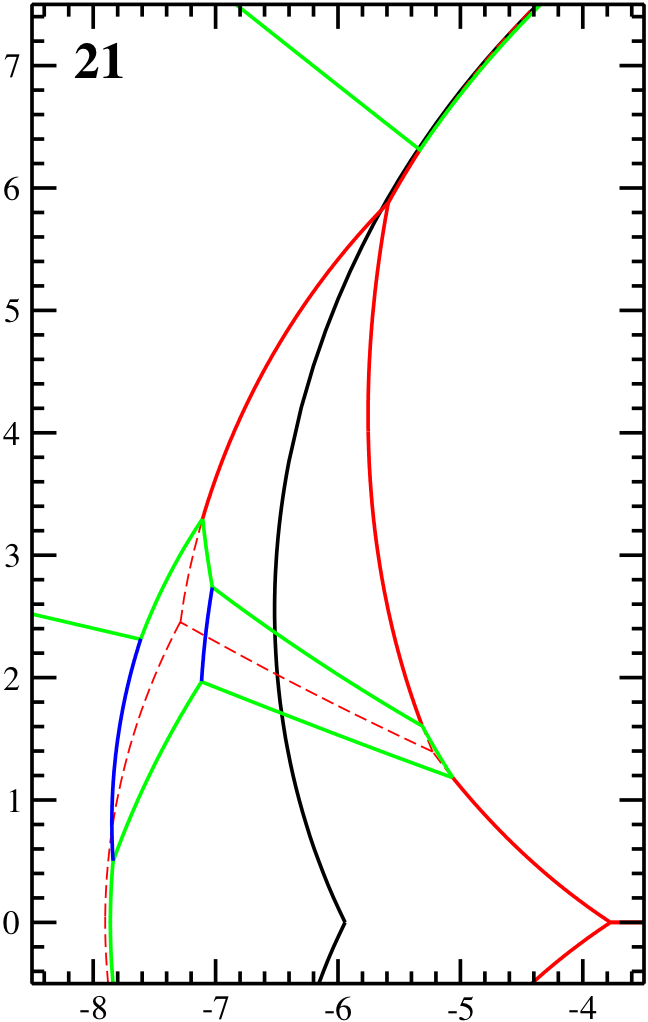}}
\put(230,0){\includegraphics[width=4cm]{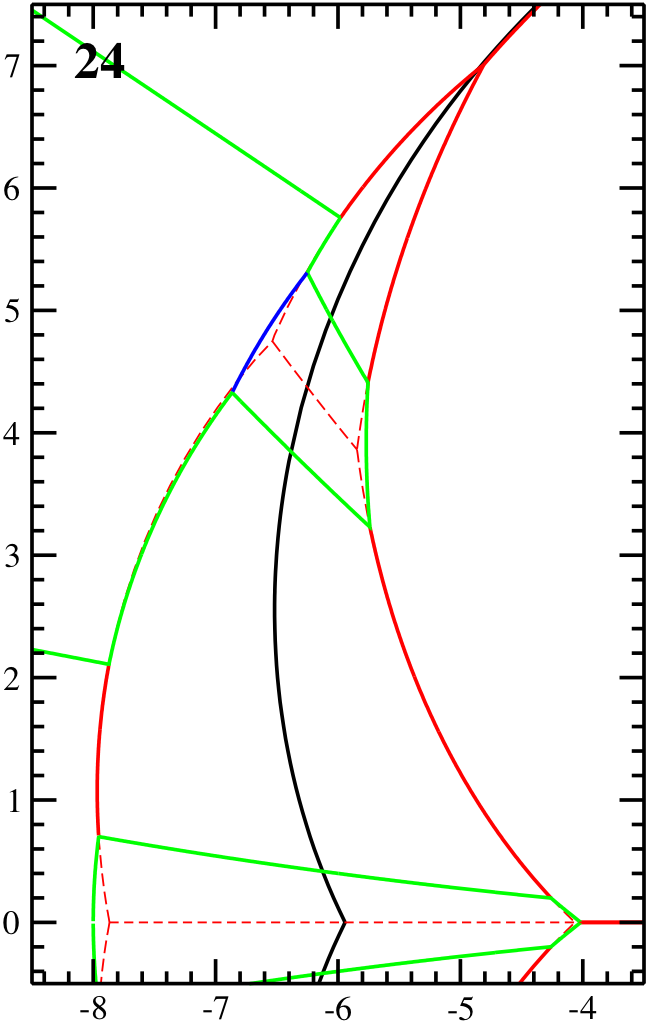}}
\end{picture}
\end{center}

\caption{\label{fig:hhpbnecklace} Comparison in the complex fugacity plane $z$ of the necklace
  region of the equimodular
curves of hard hexagon maximal eigenvalues for toroidal lattices for
$L_h=12,~15,~18,~21,~24$ with the eigenvalue crossing in the $P\,=\,0$ sector of
figure~\ref{fig:hhcflocus}. On the red lines 2 eigenvalues are equimodular,
on the green lines 3 eigenvalues are equimodular and on the blue lines 4
eigenvalues are equimodular. The dotted red curves are the additional
crossings in the $P=0$ sector from figure~\ref{fig:hhcflocus} which are now sub-dominant. 
The equimodular curve
$|\kappa_{-}(z)|\,=\,\,|\kappa_{+}(z)|$ is given in black for
  comparison. Color online. }
\end{figure}

For lattices with toroidal boundary conditions the eigenvalues of all
momentum sectors, not just $P\,=\,0$, contribute to the partition function.
In particular, in the thermodynamic limit   for 
$P\,=\,\pm 2\pi/3$, it is  shown in~\cite{baxterhh2} 
that there is an eigenvalue
$\lambda_{\pm 2\pi/3}(z;L_h)$ such that for $z\,\geq\, z_c$
\begin{eqnarray} 
\hspace{-0.9in}&& \qquad \quad  \quad  \quad 
 \lim_{L_h \rightarrow \infty }\, 
{{ \lambda_{\pm 2\pi/3}(z;L_h) } \over { \lambda_{\rm max}(z;L_h) }} 
\,\,\, = \,\,\,  e^{\pm 2\pi i/3}. 
\end{eqnarray}
These two eigenvalues with $P\, =\, \pm 2\pi/3$ cause significant
differences  from the equimodular curves for $P\, =\, 0$ for finite 
values of $L_h$. We illustrate this in figures~\ref{fig:hhpblocus} and \ref{fig:hhpbnecklace}. 
In figure~\ref{fig:hhpblocus} we plot the equimodular curves for toroidal boundary conditions for 
$L_h\,=\,9, \,12, \,15, \,18, \,21$. 

In these figures level crossings of 2 eigenvalues are
shown in red, of 3 eigenvalues in green and 4 eigenvalues in blue. For sectors 
separated by a red boundary, both sectors have momentum 
$P \, = \, 0$.  For sectors separated by a green boundary, one sector 
has momentum $P\,=\,0$, and the other has two eigenvalues of equal modulus and fixed
phases of $e^{\pm 2\pi i/3}$.  For sectors separated by a blue boundary, 
each sector has two eigenvalues of equal modulus, and fixed phases 
of $\, e^{\pm 2\pi i/3}$. The equimodular curve
 $|\kappa_{-}(z)|\,=\,\,|\kappa_{+}(z)|\,$ 
is plotted in black for comparison.

It is instructive to compare the necklace regions of the plots of 
figure~\ref{fig:hhpblocus} with the corresponding plots of figure 3 where the momentum
of the eigenvalues is restricted to $P=0$. We do this in figure 6
where in the necklace region we have added in dotted red lines the
$P\,=\, \,0$ level crossing of figure~\ref{fig:hhcflocus} which are, now,
 crossings of sub-dominant eigenvalues.

There are two important observations to make concerning these plots.

\subsubsection{\label{sec:evray} The rays out to infinity \newline}

The most striking difference between the equimodular curves for
cylindrical and toroidal boundary conditions is that 
there are ``rays'' of equimodular curves which go to infinity. These
rays all have three equimodular eigenvalues which separate a sector

\clearpage

\noindent with $\,P=\,0$ from a sector with $P\,=\,\pm 2\pi/3$. In the limit
$L_h\,\rightarrow \,\infty$ these three eigenvalues on the rays become 
equimodular independent of $z$, and thus there will be no zeros on these 
rays in the thermodynamic limit. 

\subsubsection{\label{sec:evdom} Dominance of $P=\, 0$ as $\, L_h\, \rightarrow \, \infty$ \newline}
\label{dominance}
We see in figure~\ref{fig:hhpbnecklace} that for the smaller values of $L_h$, such as $12$
and $15$, a sizable
portion the region in the necklace has momentum $P\,=\,\pm 2 \pi/3$. However, 
as seen in the plots for $L_h\,=\,18,~21$ and $24$ as $L_h$
increases the regions with $P\,= \, 0$ grow  and squeeze the
regions with $P\,=\,\,\pm 2 \pi/3$ down to a very small area. It is thus
most natural to conjecture that, in the limit $L_h\,\rightarrow\, \infty$,
only momentum $P\,=\,0$ survives, except possibly on the equimodular
curves themselves.

\clearpage

\section{\label{sec:zero} Partition function zeros}

We now turn to zeros of the partition function $Z_{L,L}(z)$ on the
lattices of size $L\times L$. Just as we required the creation of
specialized algorithms to compute the eigenvalues of the transfer
matrix so we need specialized algorithms to compute the polynomials.
We have studied both  cylindrical and toroidal boundary conditions.

\subsection{\label{sec:zerocb} Cylindrical boundary conditions} 

We have computed partition function zeros for lattices with
cylindrical boundary conditions for sizes up to $39\,\times\, 39$. We plot
these zeros in figure~\ref{fig:hhzero}. These plots share with the  equimodular
$P\,=\,0$ eigenvalue curves of figure~\ref{fig:hhcflocus}  the feature of having a necklace in the
left half plane beyond the Y branching.   These plots also have the 
feature that as the size increases the number of zeros inside the 
necklace region increases. However, in contrast with the
 equimodular $P\,=\,0$ eigenvalue curves there is no necklace for $L\,=\,15$.

\subsubsection{\label{sec:zerocbneck} Branching of the necklace \newline }

We give the left most crossing of the necklace, the Y branching point
and the necklace endpoint in table~\ref{tab:zeroneck}. We note that the left most
crossing is to the left of the corresponding left most crossing of the
transfer matrix eigenvalue equimodular crossings given in tables~\ref{tab:necklace1} 
and~\ref{tab:necklace2}. These crossings are moving to the right with increasing
$L$ for $L\,\geq \,27$. The Y branchings are moving to the right for
$L\,\geq \,30$. These trends are the opposite of what was found for the
transfer matrix eigenvalue curves which only went up to $L_h\,=\,27$. 
We note that for $15\times 15$ through $27\times 27$ there is only one
region in the necklace. However. for $30\times 30$ there are two
regions, for $33\times 33$ three, for $36\times 36 $ five and for
$39\times 39$ seven. It is unknown if the number of regions increases
for larger lattices.

\begin{table}[h!]
\begin{center}
\begin{tabular}{|c|c|l|c|}\hline
$L$&leftmost crossing& Y branching& necklace endpoints\\ \hline
$15$&no necklace&$-6.8311$&\\ 
$18$&$-8.666$&$-5.6655$&\\
$21$&$-9.1957$&$-4.5411~({\rm min})$&\\
$24$&$-8.8963\pm 0.264i$&$-4.7137$&\\
$27$&$-9.4969$&$-4.8031$&$-6.292287\pm 7.325196i$\\
$30$&$-9.2717\pm0.541i$&$-5.0851~({\rm max})$&$-5.515958\pm 8.174231i$\\
$33$&$-9.4610$&$-4.8875$&$-4.728011\pm 8.742729i$\\
$36$&$-9.213\pm 0.527i$&$-4.6972$&$-4.797746\pm 8.473961i$\\
$39$&$-9.3221$&$-4.5687$&$-4.270164\pm 8.792602i$\\ \hline
\end{tabular}
\end{center}
\caption{\label{tab:zeroneck} The necklace crossing and endpoints as a function of $L$ for the 
$L\times L$ lattice with cylindrical boundary conditions. There is a
  mod 6 phenomenon apparent in both the location of the necklace
  crossings and the endpoint. The necklace endpoints at $L=27,~33,~39$
and at $L\,=\,30,~36$ are moving to the right.}
\end{table}

\subsubsection{\label{sec:zerocbcomp} Comparison with the eqimodular curves \newline }

In figure~\ref{fig:hhzerolong} we compare the equimodular curves for $L_h=27$ with the
partition function zeros of the $27\times 27$ lattice by plotting the
partition function zeros for the lattices $27\times 27,~~27\times
54,~~27\times 135$ and $27\times 270$. This comparison clearly shows
how slight kinks for $27\times 27$ grow  into an equimodular
curve with  5 separate regions. 

 
\begin{figure}[h!]
\begin{center}
 \begin{picture}(4500,450)
  \put(0,300){\includegraphics[width=5cm]{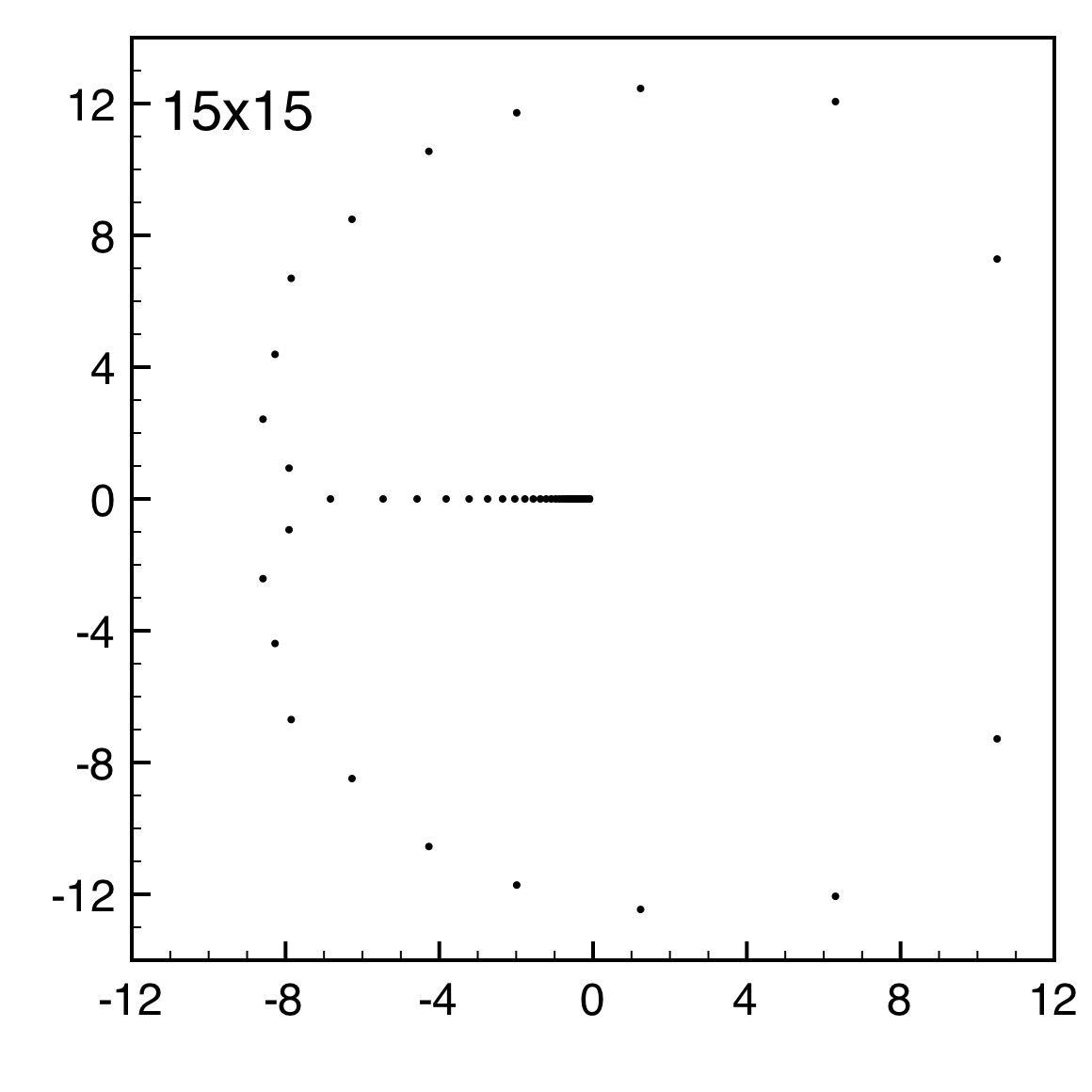}}
\put(150,300){\includegraphics[width=5cm]{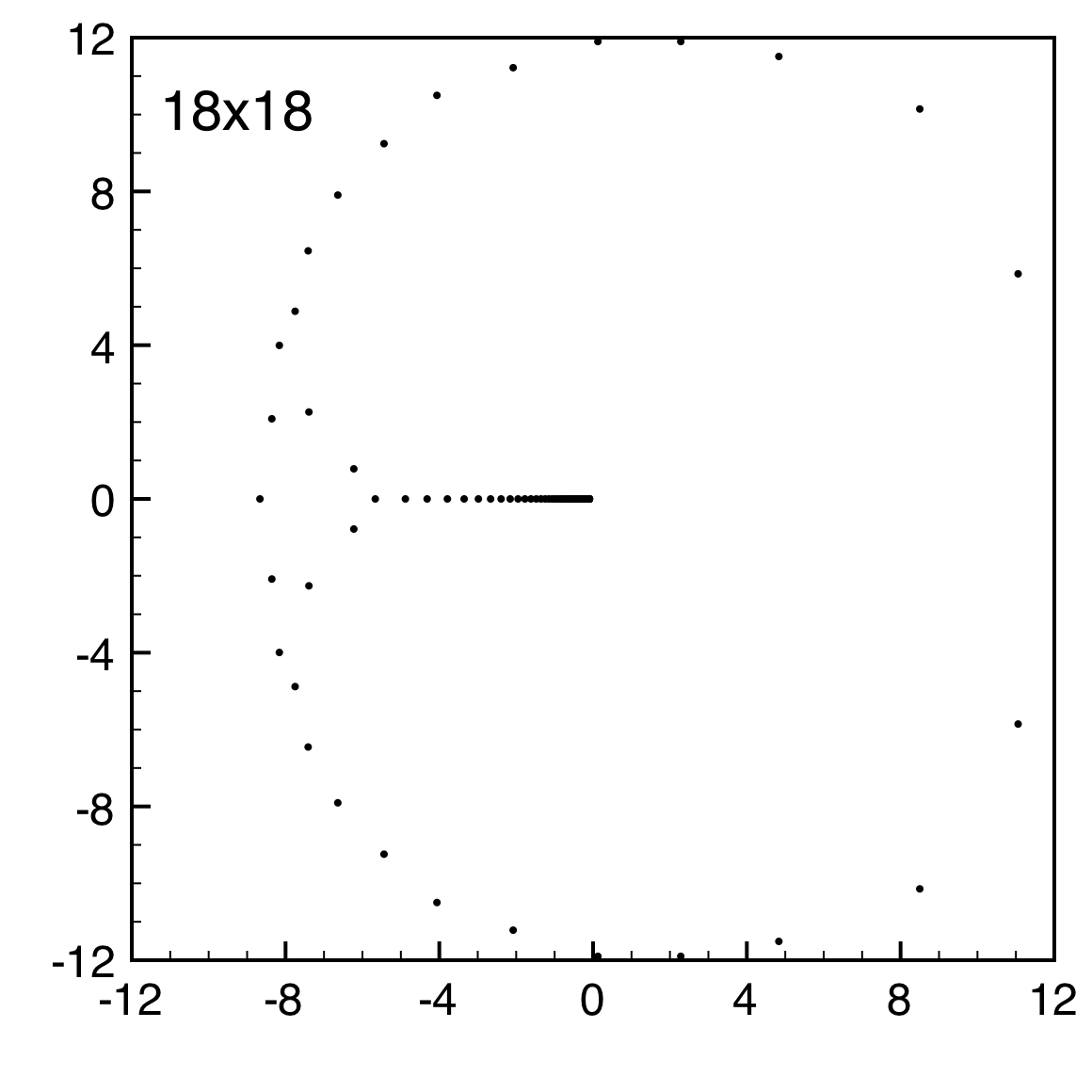}}
\put(300,300){\includegraphics[width=5cm]{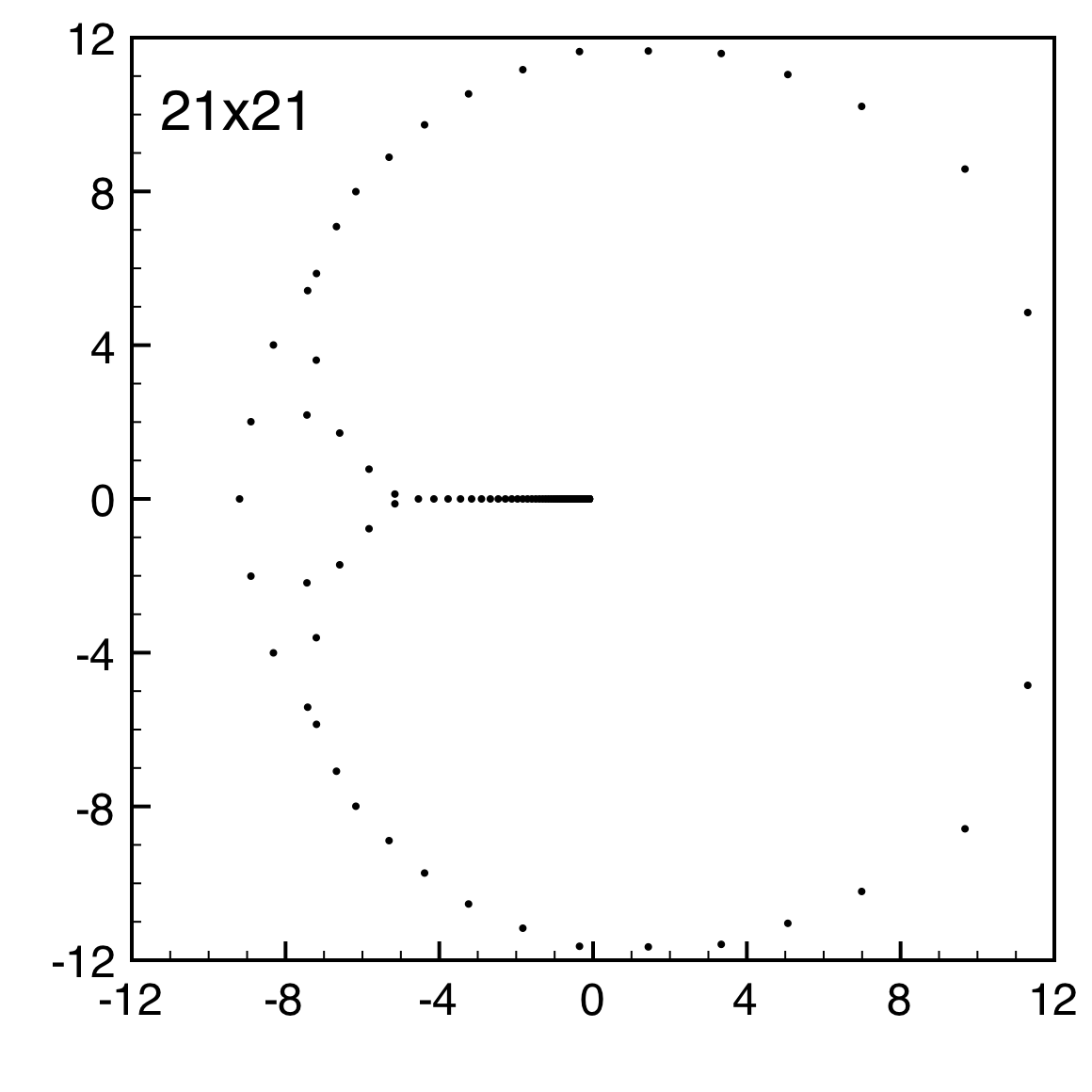}}
\put(0,150){\includegraphics[width=5cm]{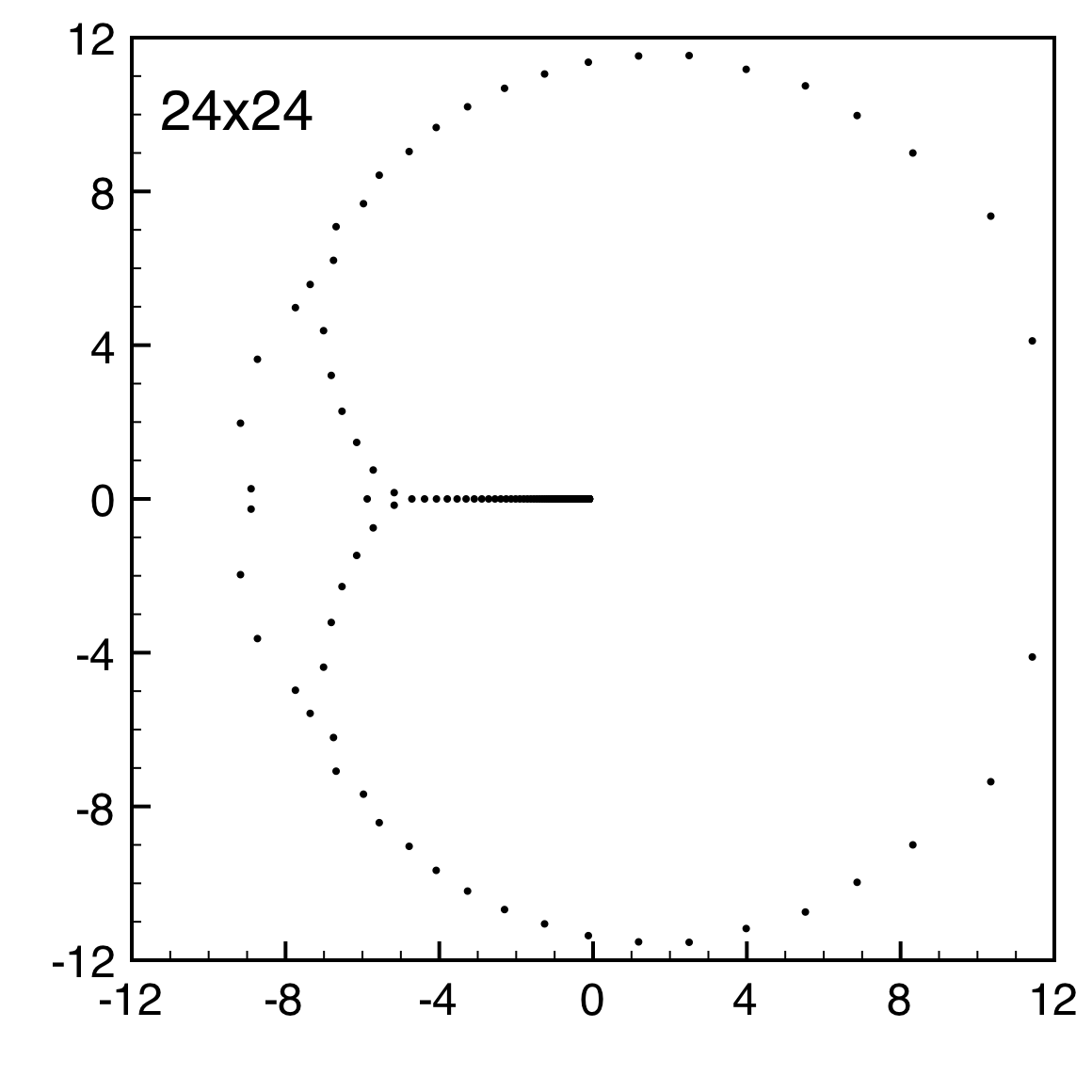}}
\put(150,150){\includegraphics[width=5cm]{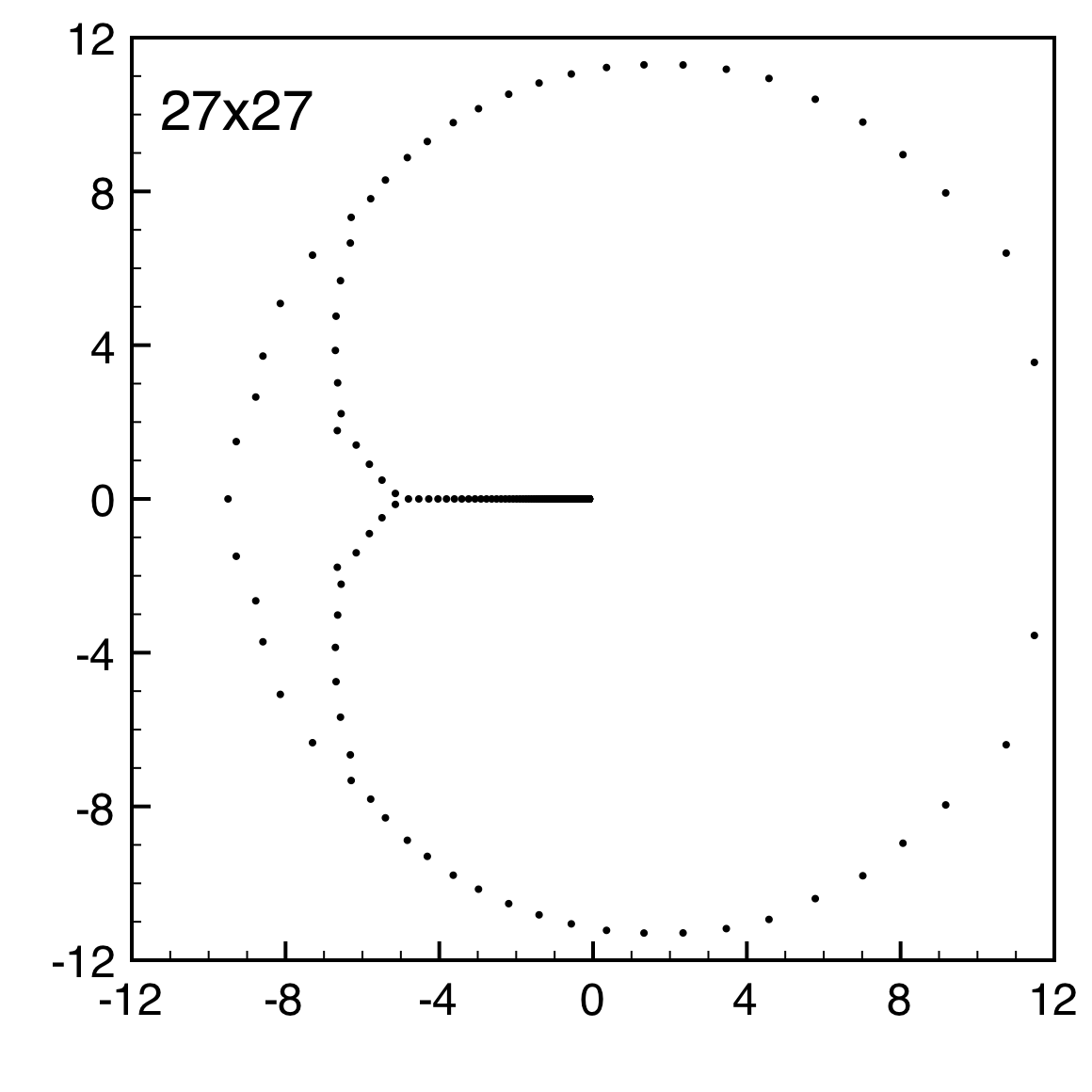}}
\put(300,150){\includegraphics[width=5cm]{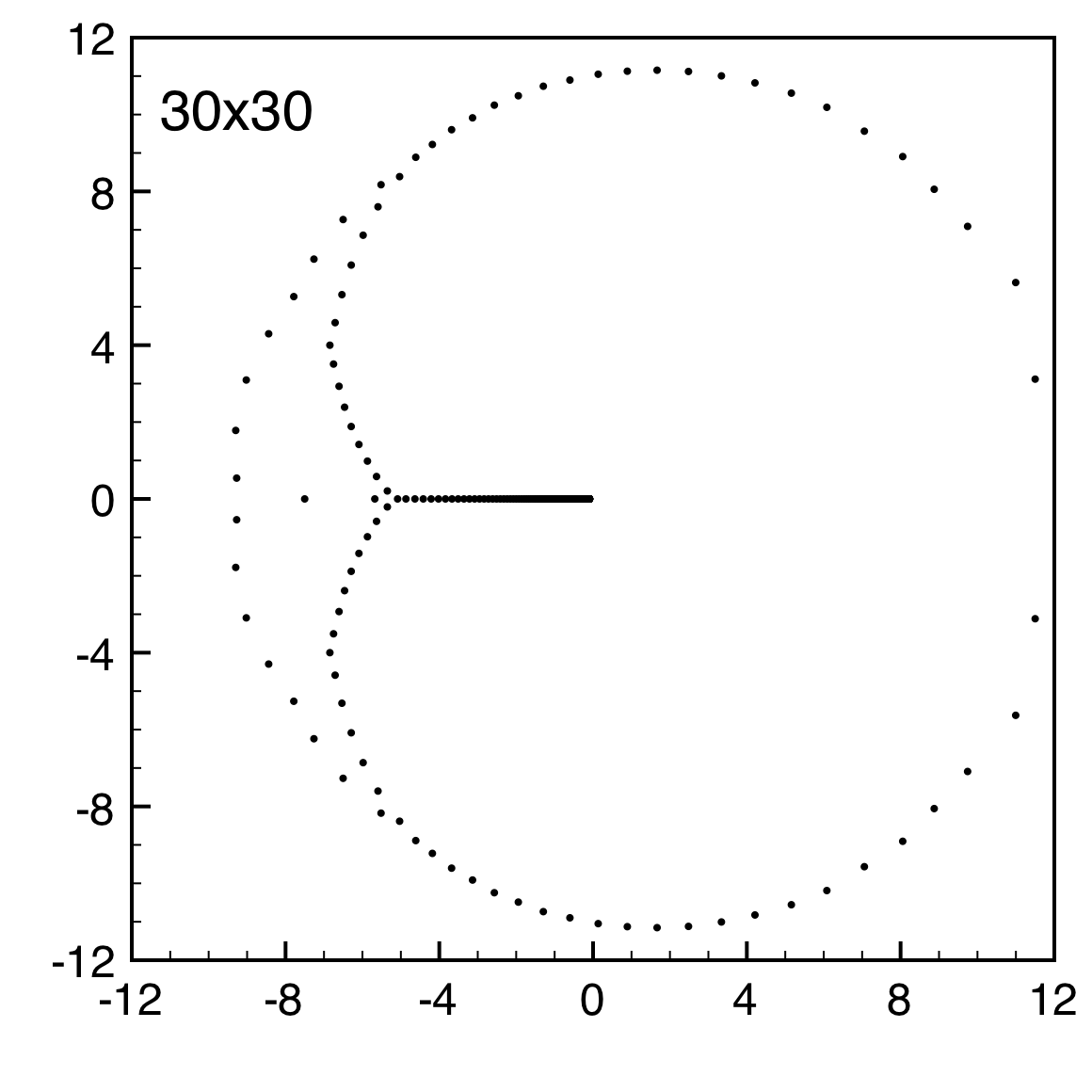}}
\put(0,0){\includegraphics[width=5cm]{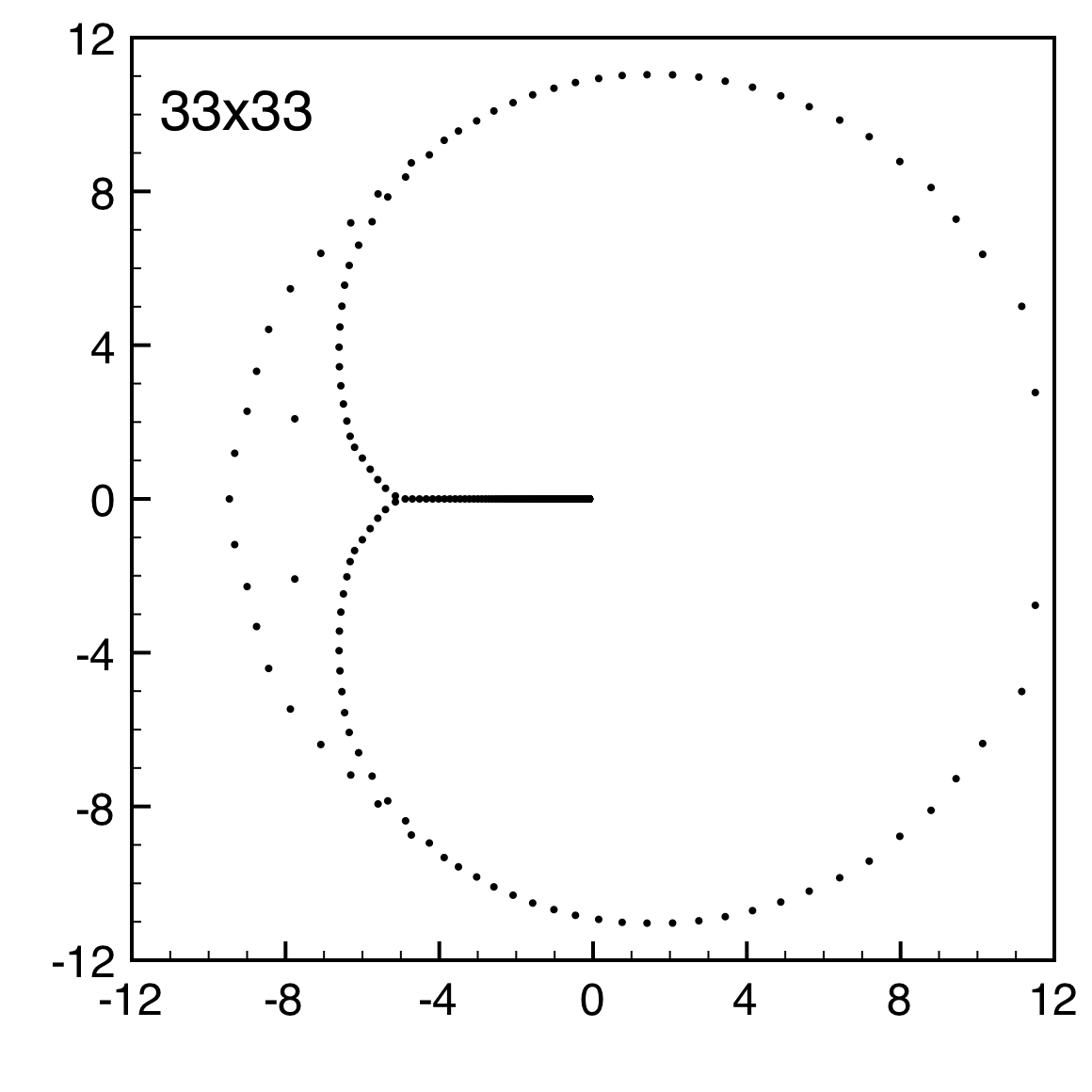}}
\put(150,0){\includegraphics[width=5cm]{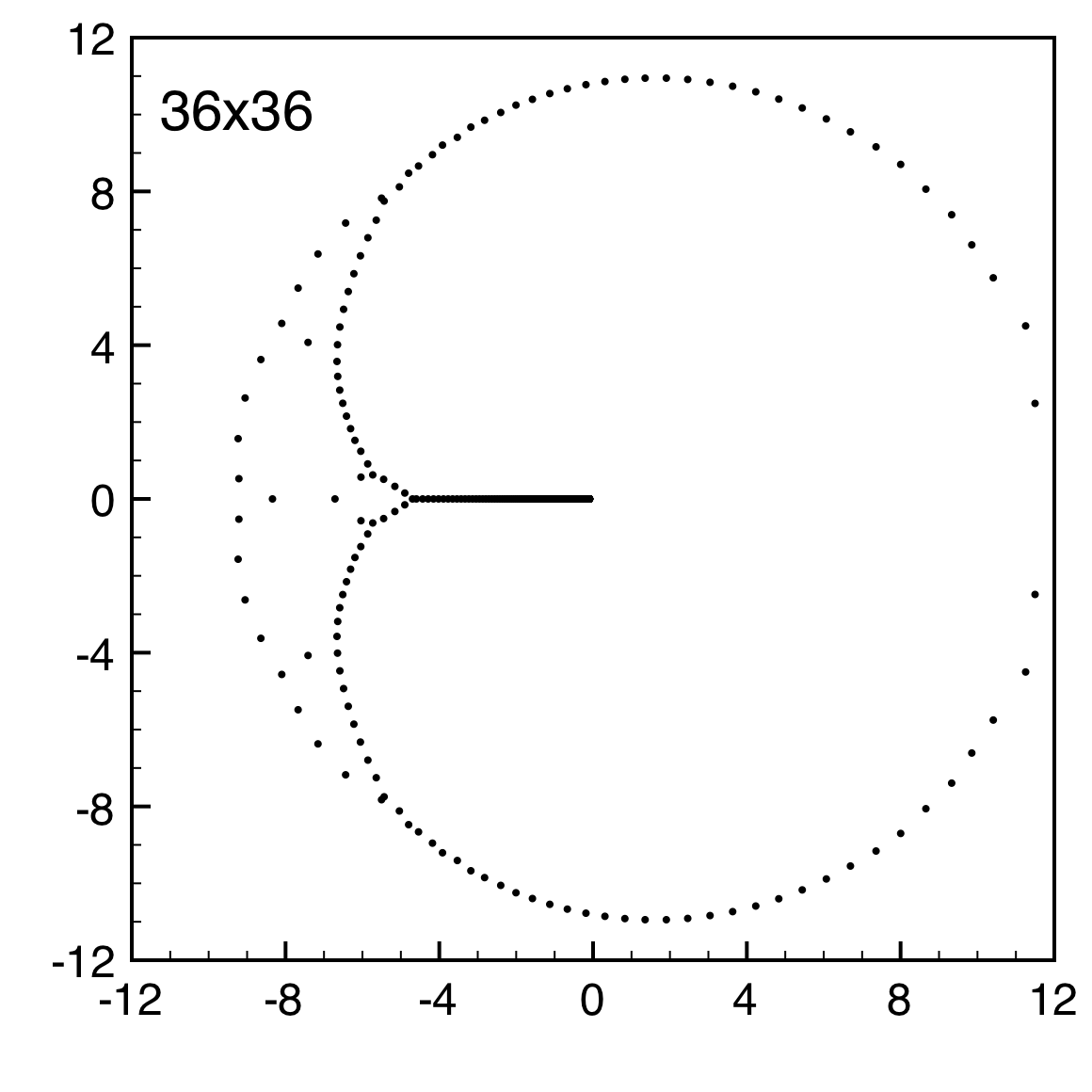}}
\put(300,0){\includegraphics[width=5cm]{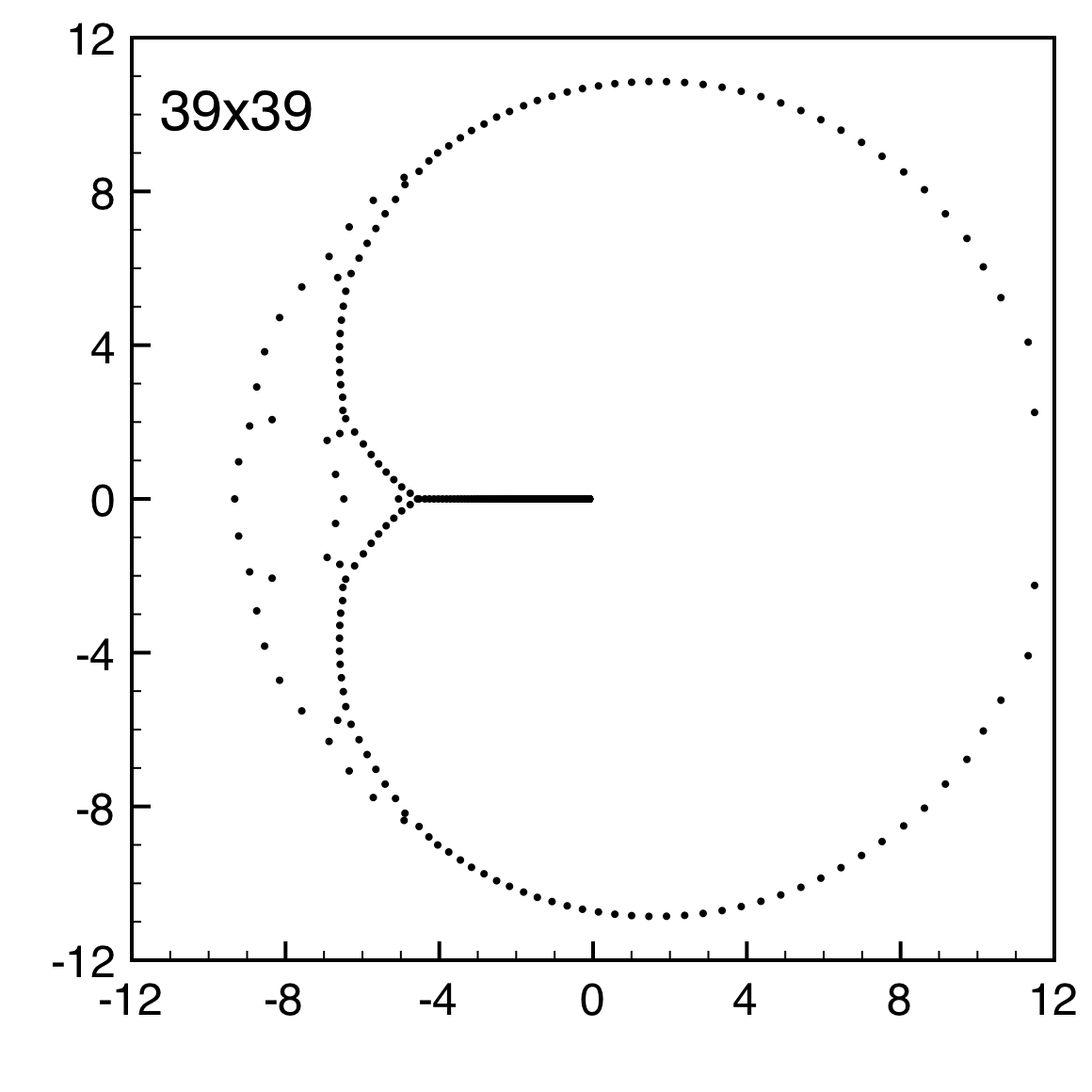}}
 \end{picture}
\end{center}
\caption{\label{fig:hhzero} Plots of partition function zeros in the complex fugacity 
plane $z$ of hard hexagon model for lattices with cylindrical boundary 
conditions of size $15\times 15,~18\times 18,~21\times 21,~24\times 24,
~27\times 27,~30\times 30,~33\times 33,~36\times 36,~39\times 39$.}

\end{figure}


\begin{figure}[h!]
\begin{center}
\begin{picture}(300,300)
\put(0,0){\includegraphics[width=10cm]{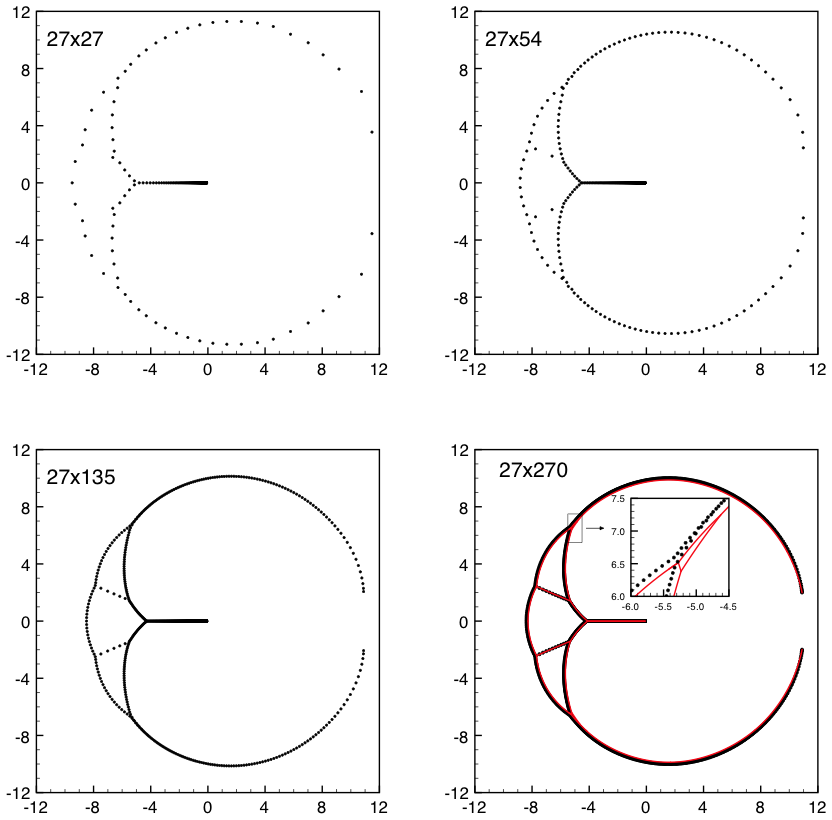}}
\end{picture}
\end{center}
\caption{ \label{fig:hhzerolong} The partition function zeros for the lattices $27\times
  27~,27\times 54~27\times 135$ and $27\times 270$. For $27\times 270$
  the equimodular curve is shown in red. Color online}
\end{figure}

\subsubsection{\label{sec:zerocbend} The endpoints $z_d(L)$ and $z_c(L)$ \newline }

In table~\ref{tab:zeroendpoints} we give the values of the endpoints which approach $z_d$
and $z_c$. We note that the values of $z_d(L_h)$ and $z_c(L_h)$
of table~\ref{tab:endpoints}  as determined from the equimodular curves are significantly 
closer to the limiting values $z_d$ and $z_c$ than the corresponding
values of table~\ref{tab:zeroendpoints}. We also note that, in
table~\ref{tab:endpoints}, ${\rm Re}(z_c(L_h))$ is monotonic and approaches $z_c$ from
below, while in table~\ref{tab:zeroendpoints} ${\rm Re}(z_c(L_h))$ is not monotonic 
and approaches $z_c$ from above.

\begin{table}[h!]
\begin{center}
\begin{tabular}{|c|c|l|}\hline
$L$&$z_d(L)$& \multicolumn{1}{c}{$z_c(L)$}\\ \hline
$  9$ & $  -0.0957417573$ & $   5.9002937473 \pm   12.2312152474i$ \\ 
$ 12$ & $  -0.0932266680$ & $   9.2335210855 \pm    9.3476347389i$ \\ 
$ 15$ & $  -0.0920714392$ & $  10.5114514245 \pm    7.2812520022i$ \\ 
$ 18$ & $  -0.0914523473$ & $  11.0571925423 \pm    5.8559364459i$ \\ 
$ 21$ & $  -0.0910853230$ & $  11.3084528958 \pm    4.8492670401i$ \\ 
$ 24$ & $  -0.0908515103$ & $  11.4268383658 \pm    4.1113758041i$ \\ 
$ 27$ & $  -0.0906942824$ & $  11.4806273673 \pm    3.5521968857i$ \\ 
$ 30$ & $  -0.0905839894$ & $  11.5012919280 \pm    3.1162734906i$ \\ 
$ 33$ & $  -0.0905039451$ & $  11.5044258314 \pm    2.7682753249i$ \\ 
$ 36$ & $  -0.0904442058$ & $  11.4981796564 \pm    2.4848695493i$ \\ 
$ 39$ & $  -0.0903985638$ & $  11.4869896404 \pm    2.2501329582i$ \\ 
$\infty$&$ -0.0901699437$&$11.0901699437$\\\hline
\end{tabular}
\end{center}
\caption{\label{tab:zeroendpoints} The values of $z_d(L)$, and 
$z_c(L)$ as a function of $L$ for the $L\times L$ lattice with
  cylindrical boundary conditions as determined from the zeros of the
  partition function.}

\end{table}

It is clear in table~\ref{tab:zeroendpoints} that $z_d(L)$  is converging rapidly to $z_d$
and a careful quantitative analysis well fits the data with the form 
\begin{equation}
\hspace{-0.95in}\quad \quad z_d(L)-z_d = b_0L^{-12/5}+b_1L^{-17/5}+b_2L^{-22/5}+a_3L^{-27/5}+\cdots
\label{zdfit}
\end{equation}
with
\begin{equation}
\hspace{-0.95in}\quad \quad b_0=1.7147(1), \quad b_1=-9.30(2), \quad b_2=48(2), \quad b_3=-180(30).
\end{equation}
The exponent $12/5$ is the leading exponent of the energy operator of
the Lee-Yang edge as
is seen from analysis of \cite{zuber} and \cite{cardy}. It is expected to
be the inverse of the correlation exponent $\nu$ at $z=z_d$ but a
computation of this correlation length is not in the literature. 

For $z_c(L)$ the data of table~\ref{tab:zeroendpoints} is well fit by
\begin{equation}
|z_c(L)|-z_c=a_0L^{-6/5}+a_1L^{-2}+a_2L^{-14/5}+\cdots 
\label{zcfit}
\end{equation}
where 
\begin{equation}
a_0=53.0(1), \quad a_1=-50(5), \quad a_2=-200(50)
\end{equation}
where the exponent $y=6/5$ is the inverse of the correlation length
exponent $\nu$ of the hard hexagon model at $z=z_c$ \cite{baxterhh2}.
The exponent $-2$  is consistent with $-y-|y'|$ where $y'=-4/5$ is the
exponent for the subdominant energy operator $\phi_{(3.1)}$ for the
three state Potts model \cite{dot} and the exponent $-14/5$ follows
  from $-y-2|y'|$. We note that the potential exponents $-y-1$ and
  $-2y$ do not appear in (\ref{zcfit}).

The analysis leading to (\ref{zdfit}) and (\ref{zcfit}) and the
relation with conformal field theory is given in appendix F.

\subsubsection{\label{sec:zerocbbax} Comparison with the equimodular curve of $\kappa_{\pm}(z)$ \newline }

\begin{figure}[b]

\begin{center}
 \begin{picture}(140,130)
\put(0,0){\includegraphics[width=5cm]{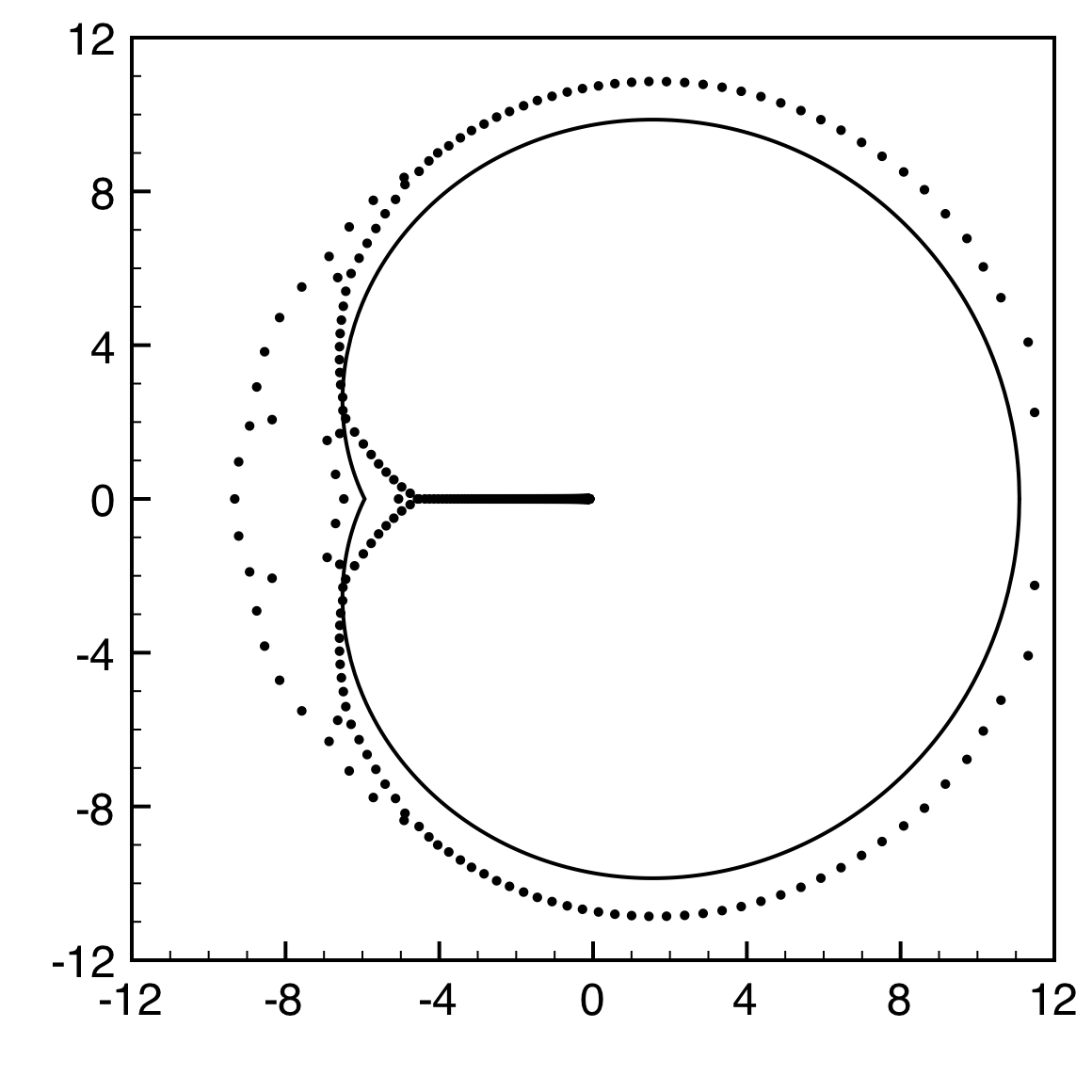}} 
\end{picture}
\end{center}
\caption{\label{fig:baxterw39} The equimodular curves for $|\kappa_{-}(z)|\,=\,|\kappa_{+}(z)|$
in the complex $z$ plane and the partition function zeros for
cylindrical boundary conditions on the $39\,\times \, 39$ lattice } 

\end{figure}

In figure~\ref{fig:baxterw39} we compare the zeros for the $39\, \times\, 39$ lattice 
with the equimodular curve of the $\kappa_{\pm}(z)$
of figure~\ref{fig:baxter}. Unlike the comparisons of figure~\ref{fig:comp1} the
$\kappa_{\pm}(z)$ equimodular curve does not have a region of overlap
with the zeros of the $39 \,\times\, 39$ lattice. However, in the region to
the right of the necklace, if 
\begin{eqnarray} 
\label{ind}
\hspace{-0.9in}&& \qquad \quad  \quad 
\lim_{L_h\rightarrow \infty,L_v\rightarrow \infty}
\, Z_{L_v,L_h}(z)^{1/L_vL_h}
\quad \quad {\rm is~independent~of} \quad L_v/L_h, 
 \nonumber 
\end{eqnarray}
then, for this region, the limiting locus of zeros will agree with the
$\kappa_{\pm}$ equimodular curve. We have examined this possibility
and find that we can well fit this portion of the zero locations of
figure~\ref{fig:hhzero} by a shifted cardioid 
\begin{eqnarray}
\label{cardioid}
\hspace{-0.9in}&& \qquad \quad  \quad 
{\rm Re}(z)\,\,=\,\,\,\, 
{{a} \over {2}} \, +c \,\,\,  +a\cos \theta\,\,
 +{{a} \over {2}} \cdot \, \cos 2\theta, 
\nonumber\\
\hspace{-0.9in}&& \qquad \quad  \quad 
{\rm Im}(z)\,\,=\,\,\, \, 
a\sin\theta \,\,\,  + {{a} \over {2}} \cdot \, \sin2\theta. 
\end{eqnarray}
The fitting parameters $a$ and $c$ depend on $L$,
and, when plotted versus $1/L$, these values fall very closely on a straight 
line which extrapolated to $L\, \rightarrow\, \infty$ gives a curve which 
is virtually indistinguishable from the $\kappa_{\pm}$ equimodular curve 
outside of the necklace regions. We take this to be evidence that in this 
non necklace region the limit (\ref{ind}) for cylindrical boundary conditions 
is independent of the ratio $L_v/L_h$. Further numerical details are
given in \ref{app:card}.

\subsection{\label{sec:zeropb} Toroidal boundary conditions}

\begin{figure}[h!]

\begin{center}
\begin{picture}(400,540)
\put(0,360){\includegraphics[width=6cm]{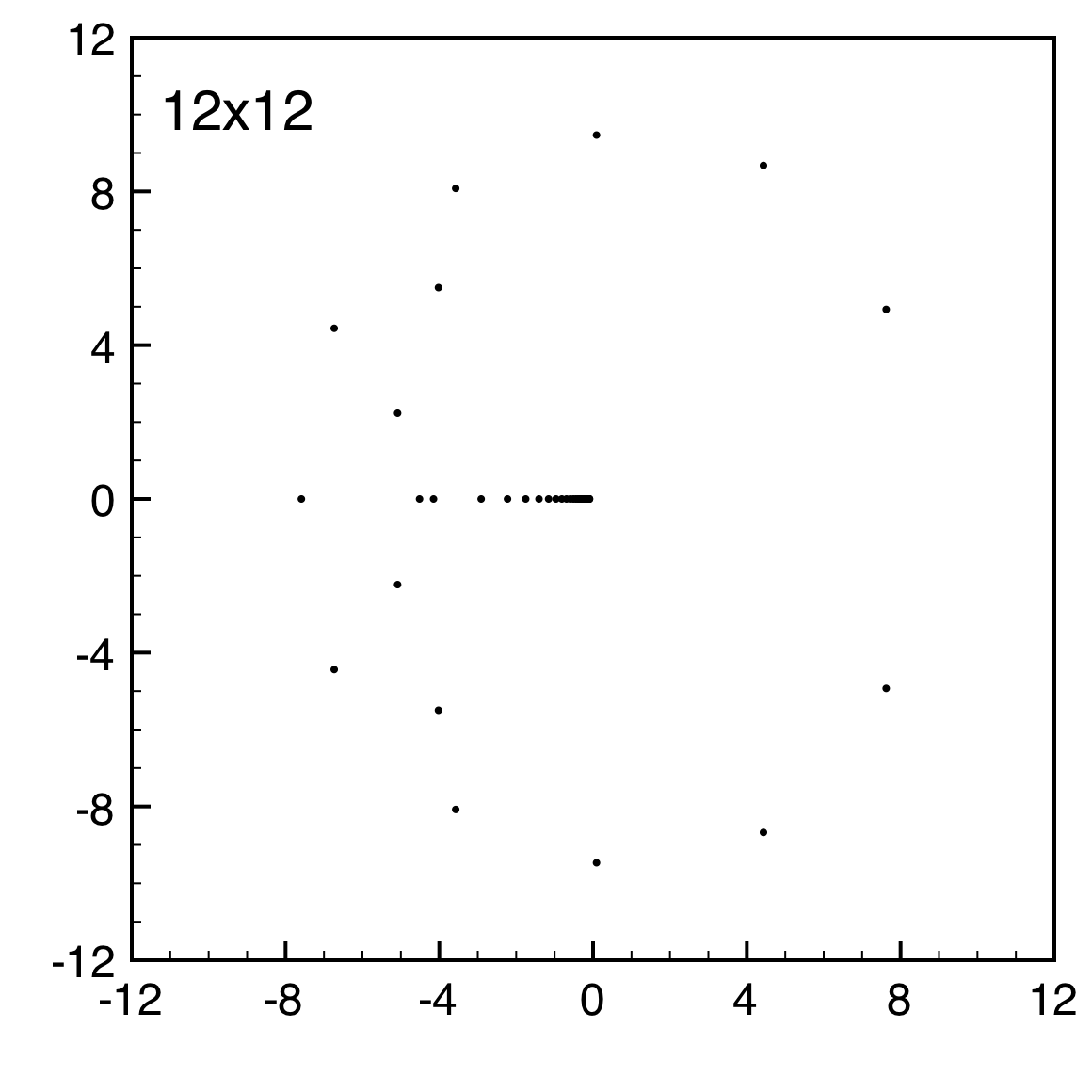}} 
\put(200,360){\includegraphics[width=6cm]{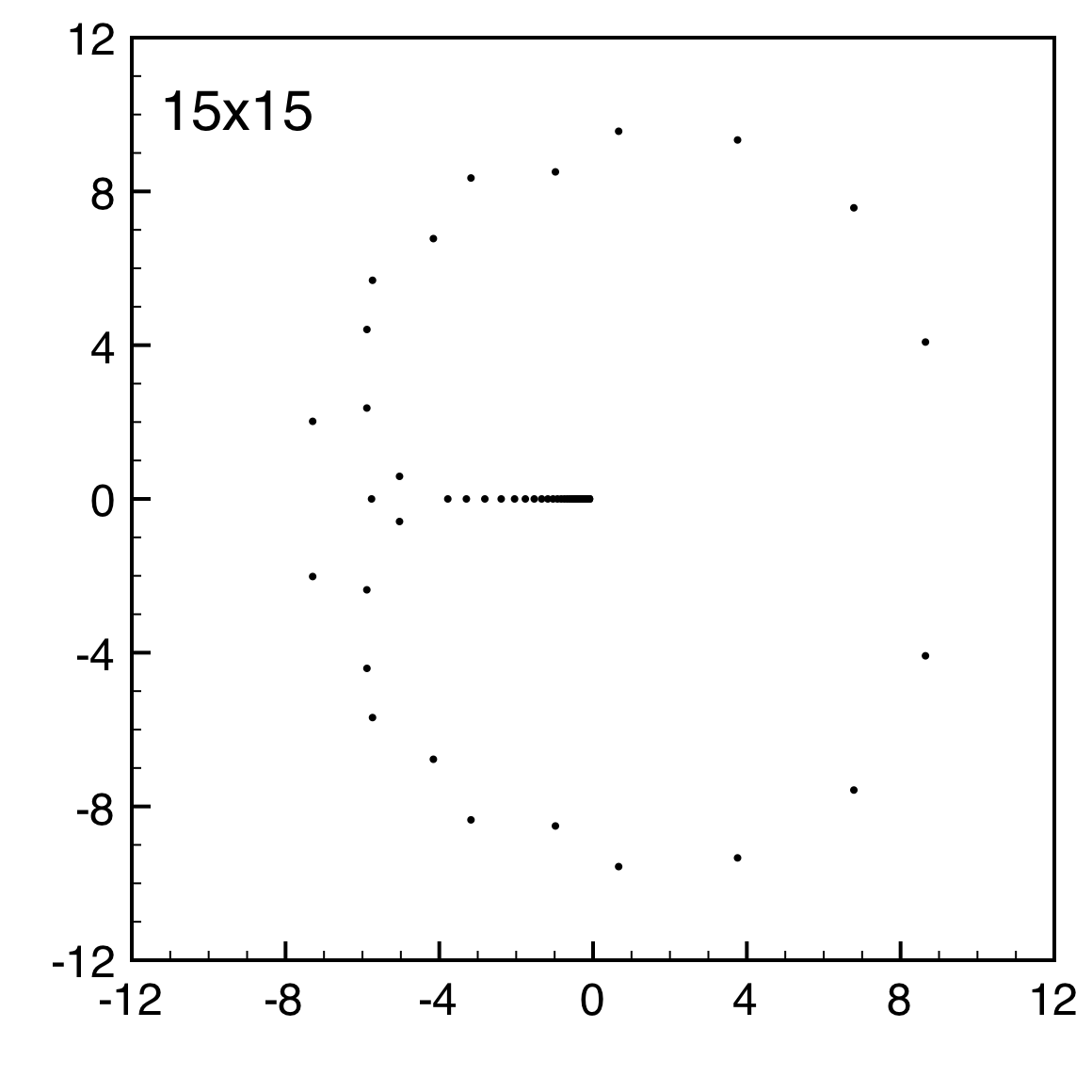}}
\put(0,180){\includegraphics[width=6cm]{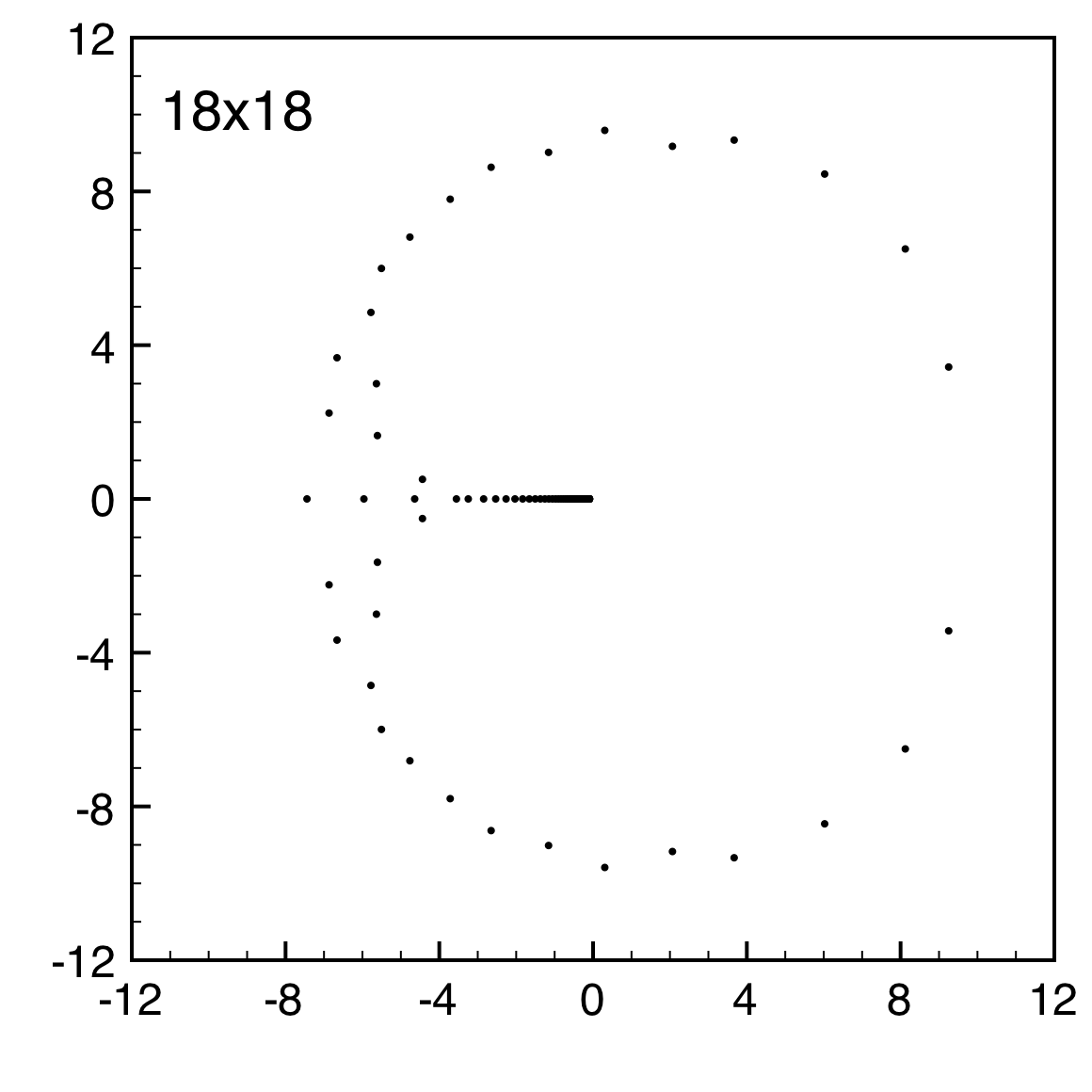}}
\put(200,180){\includegraphics[width=6cm]{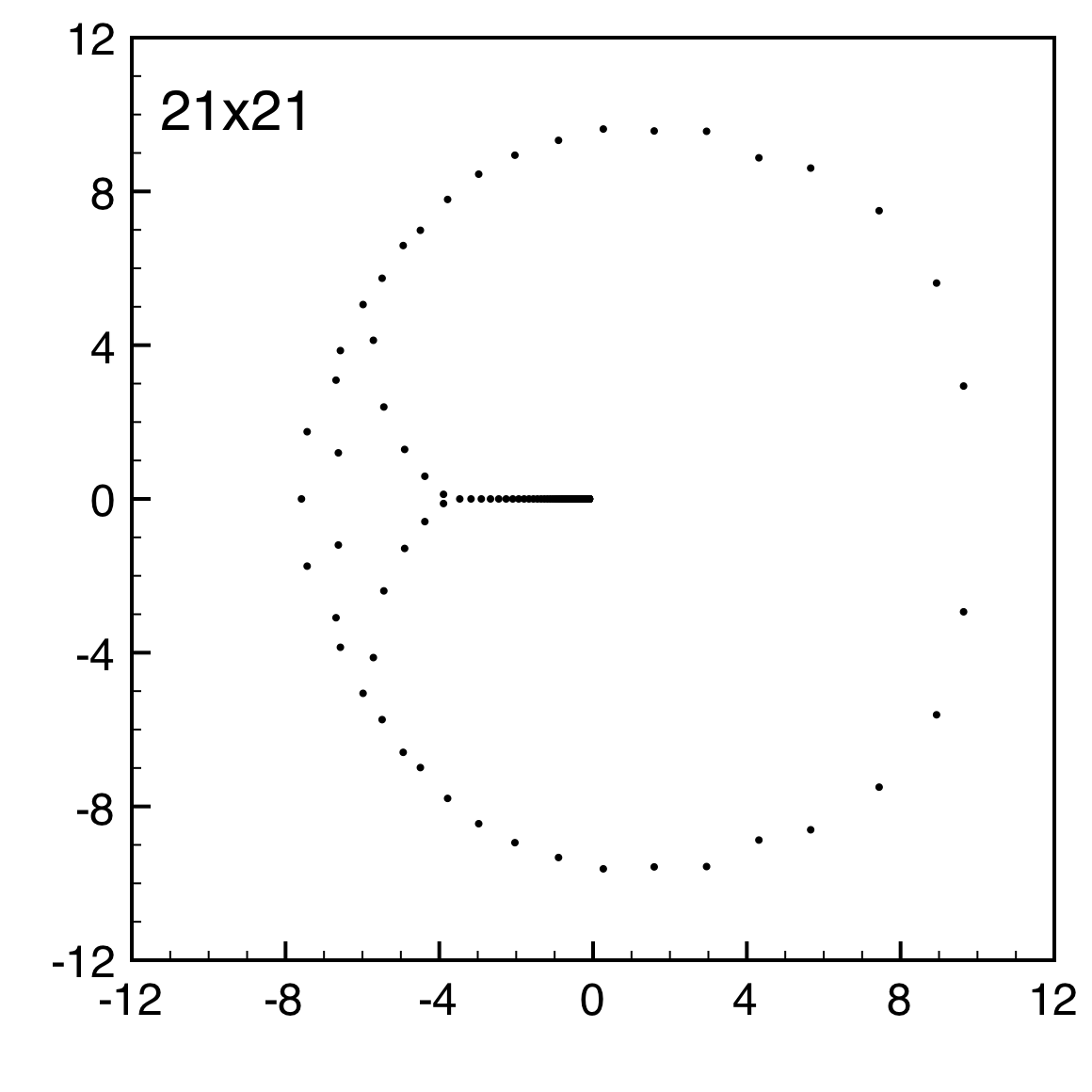}}
\put(0,0){\includegraphics[width=6cm]{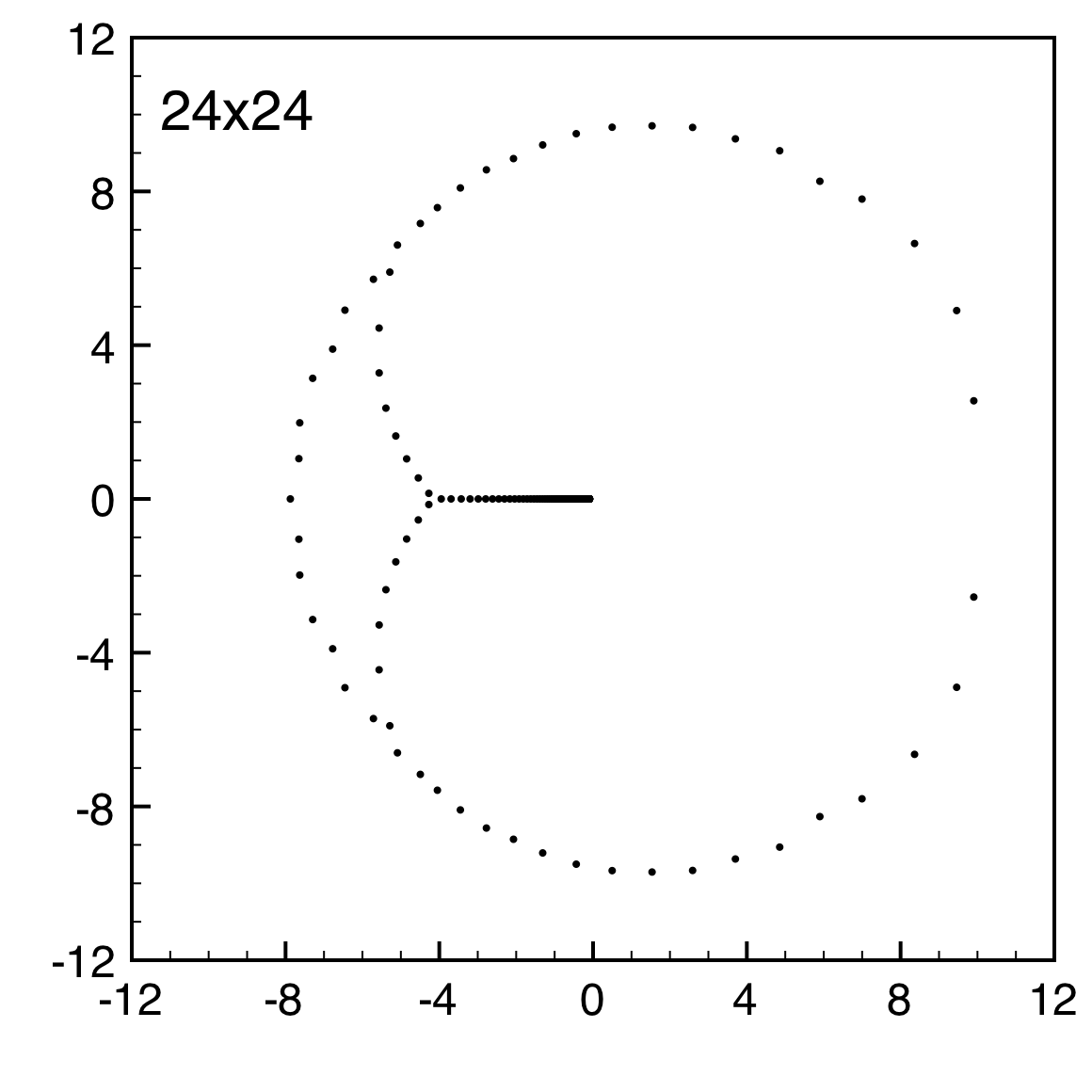}}
\put(200,0){\includegraphics[width=6cm]{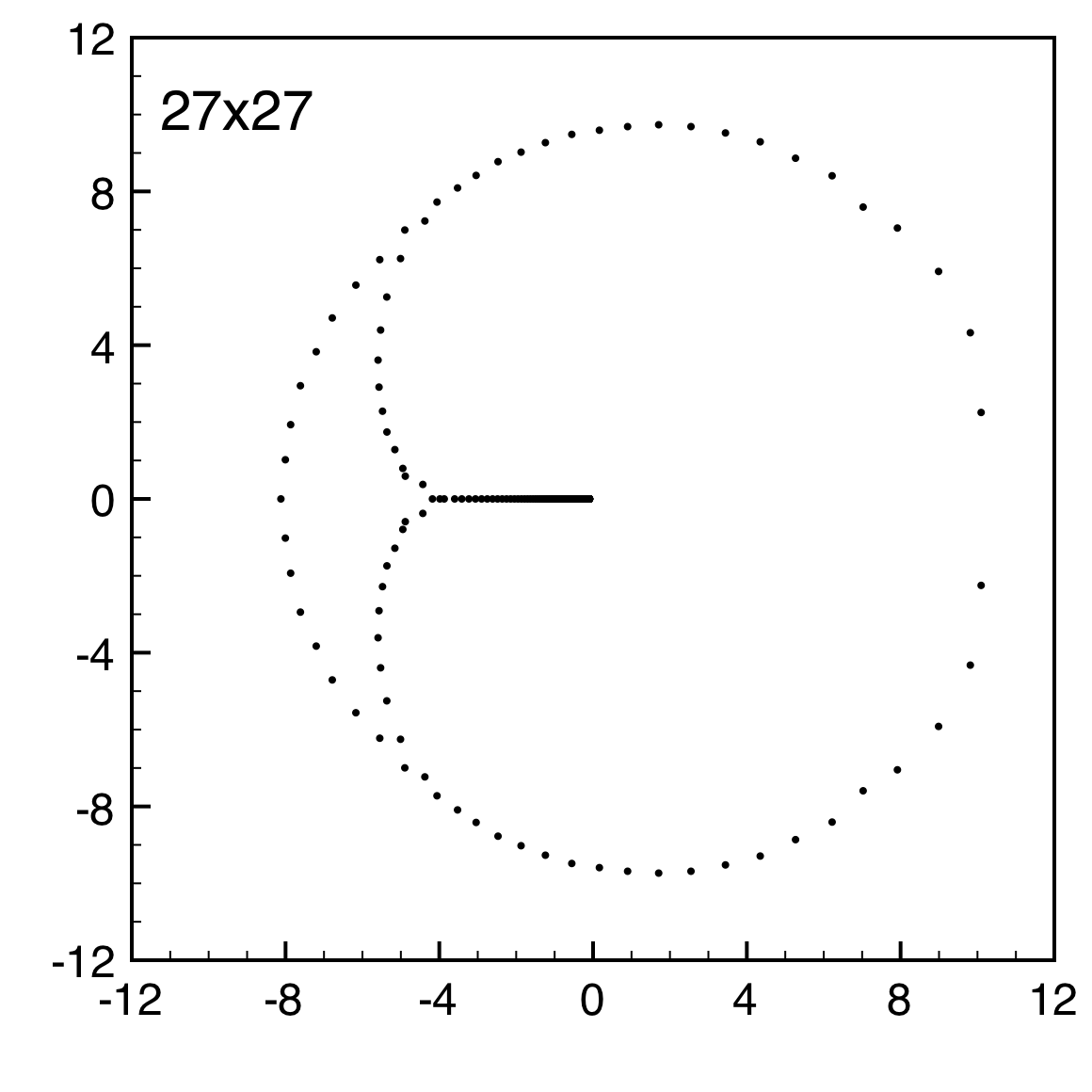}}
\end{picture}
\end{center}

\caption{\label{fig:hhpbzero} Plots of partition function zeros in the complex fugacity
  plane $z$  for  hard hexagon  with toroidal boundary
conditions of size $L\times L $ with $L=\,12,~15,~18,~21,~24,~27$. 
The value of $L\times L$
is given in the upper left hand corner of the plots.}
\end{figure}

It is numerically more difficult to compute partition function zeros
for toroidal boundary conditions and the maximum size we have been
able to study is $27\,\times \,27$. These results are plotted in figure~\ref{fig:hhpbzero}.
There is a necklace for $L\geq 12$ and there are zeros in the
necklace region for $15\,\times\, 15$ through $21 \,\times \, 21$. For
$24\, \times\, 24$ and $27\, \times \, 27$ there are no zeros in the 
necklace region.
 

\clearpage

\subsubsection{\label{sec:zeropbcomp} Comparison with the equimodular curve of $\kappa_{\pm}(z)$ \newline }

\begin{figure}[h!]

\begin{center}
 \begin{picture}(200,200)
\put(0,0){\includegraphics[width=7cm]{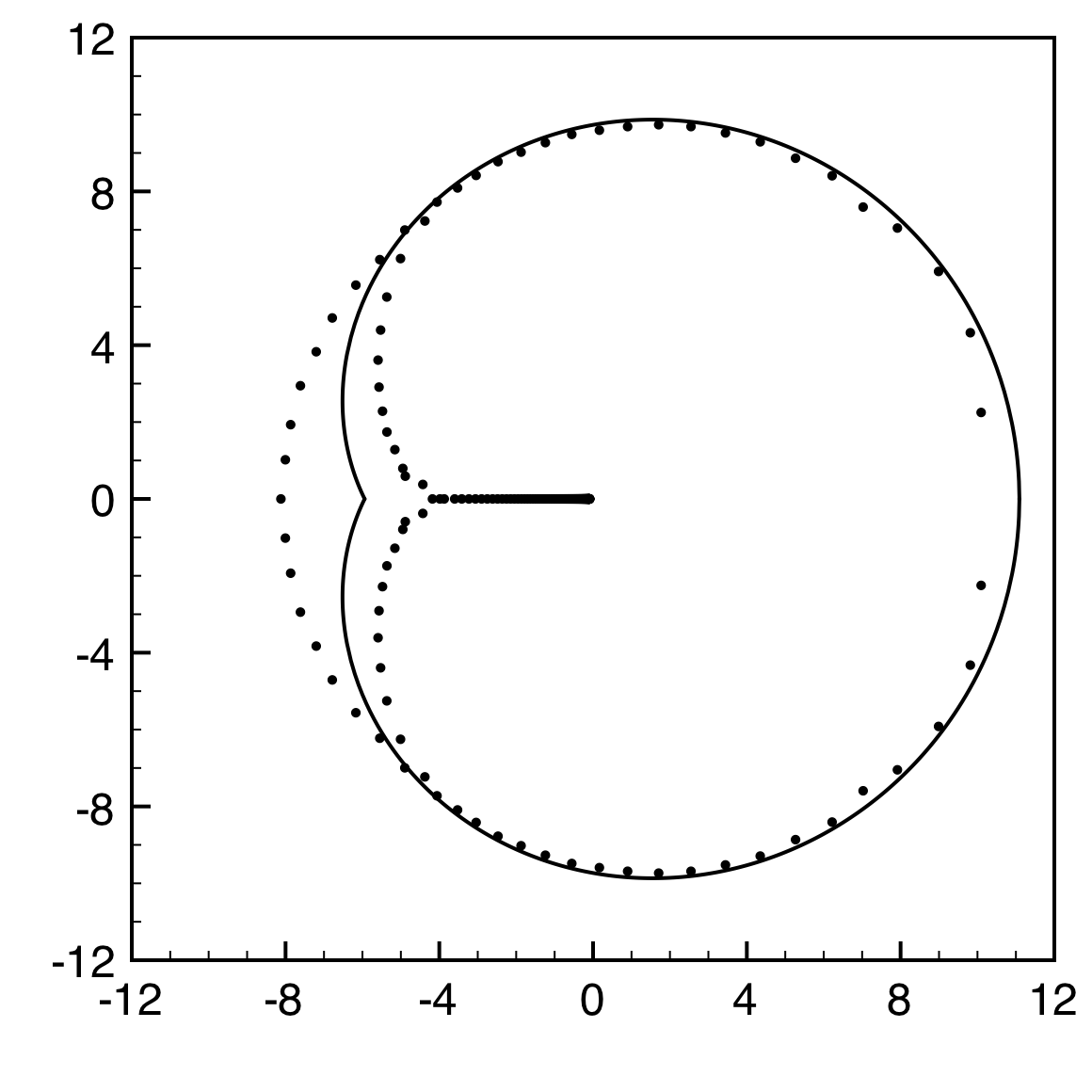}} 
\end{picture}
\end{center}

\caption{\label{fig:baxterw27} The equimodular curves for $|\kappa_{-}(z)|\,=\,\,|\kappa_{+}(z)|$
in the complex $z$ plane and the partition function zeros for
toroidal boundary conditions on the $27\times  27$ lattice } 

\end{figure}

In figure~\ref{fig:baxterw27} we compare the partition function zeros for toroidal
boundary conditions on the $27\, \times\, 27$ lattice with the equimodular curve 
of $\kappa_{\pm}(z).$ Outside of the necklace region the agreement is
much closer than it was for the cylindrical case for the $ 39\,\times\, 39$
lattice. It is appealing to attribute this agreement with the absence
of boundary effects.

\subsubsection{\label{sec:zeropbas} Dependence on the aspect ratio $L_v/L_h$ \newline }

\begin{figure}[h!]

\begin{center}
\hspace{0cm} \mbox{
\begin{picture}(400,525)
\put(0,350){\includegraphics[width=6cm]{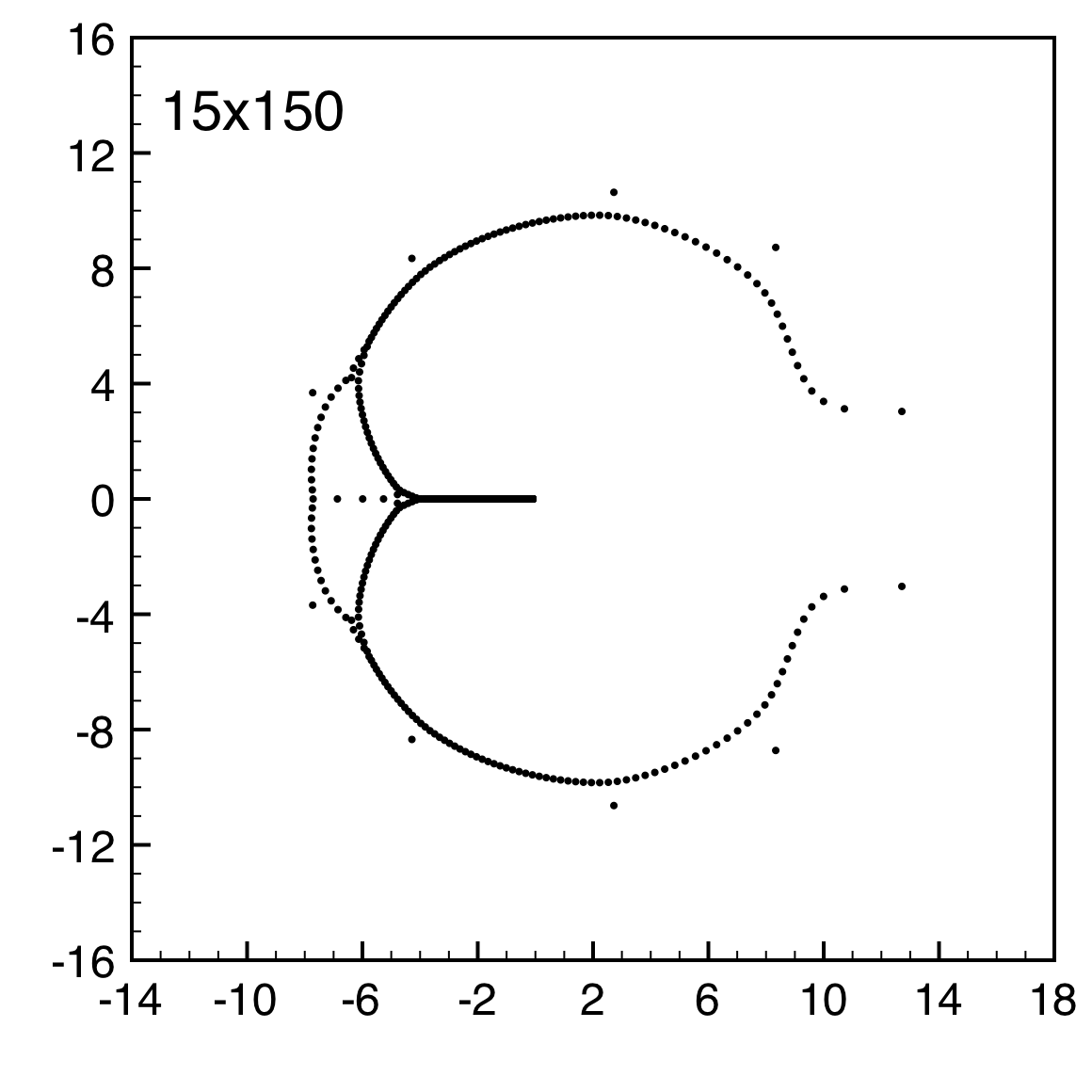}} 
\put(200,350){\includegraphics[width=6cm]{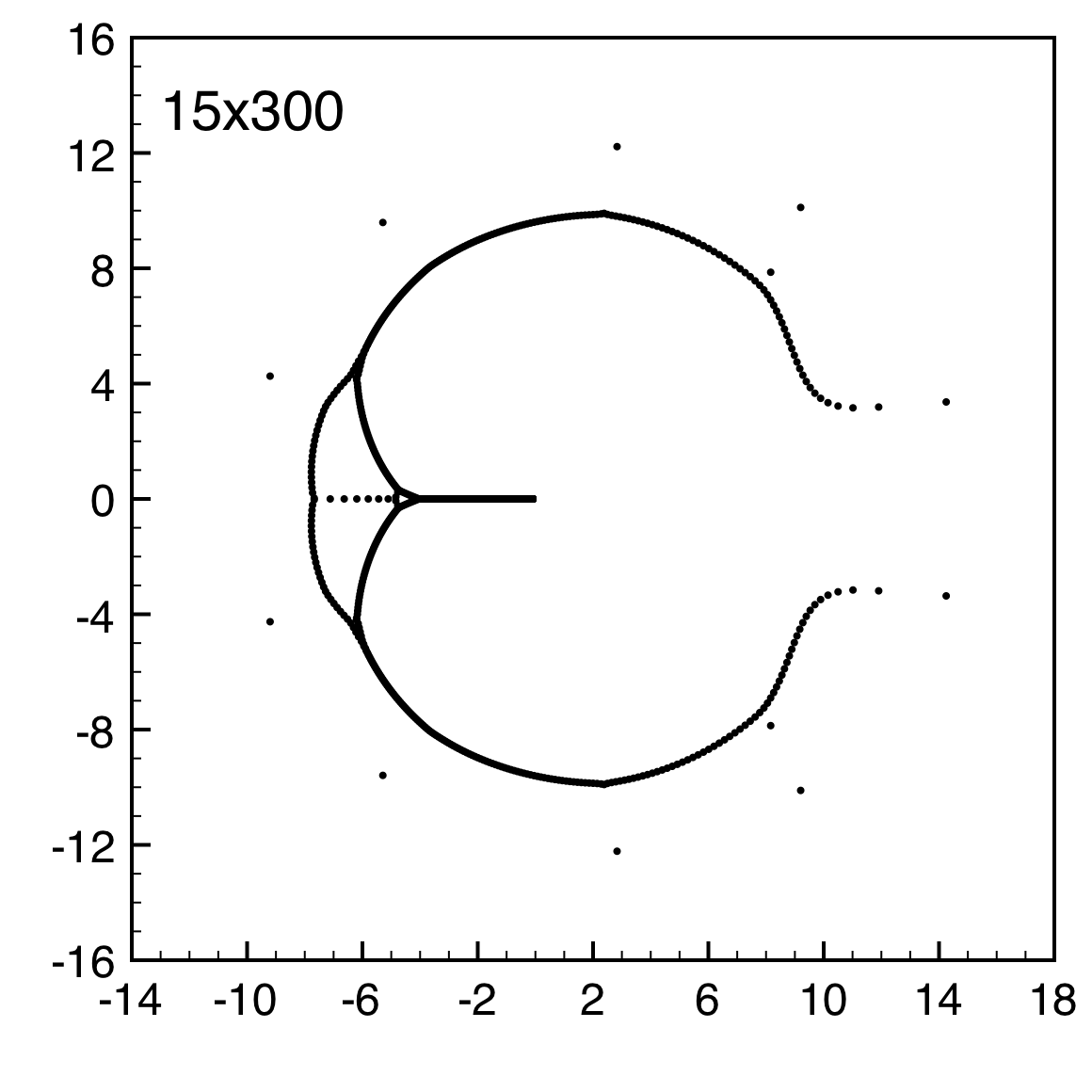}} 
\put(0,175){\includegraphics[width=6cm]{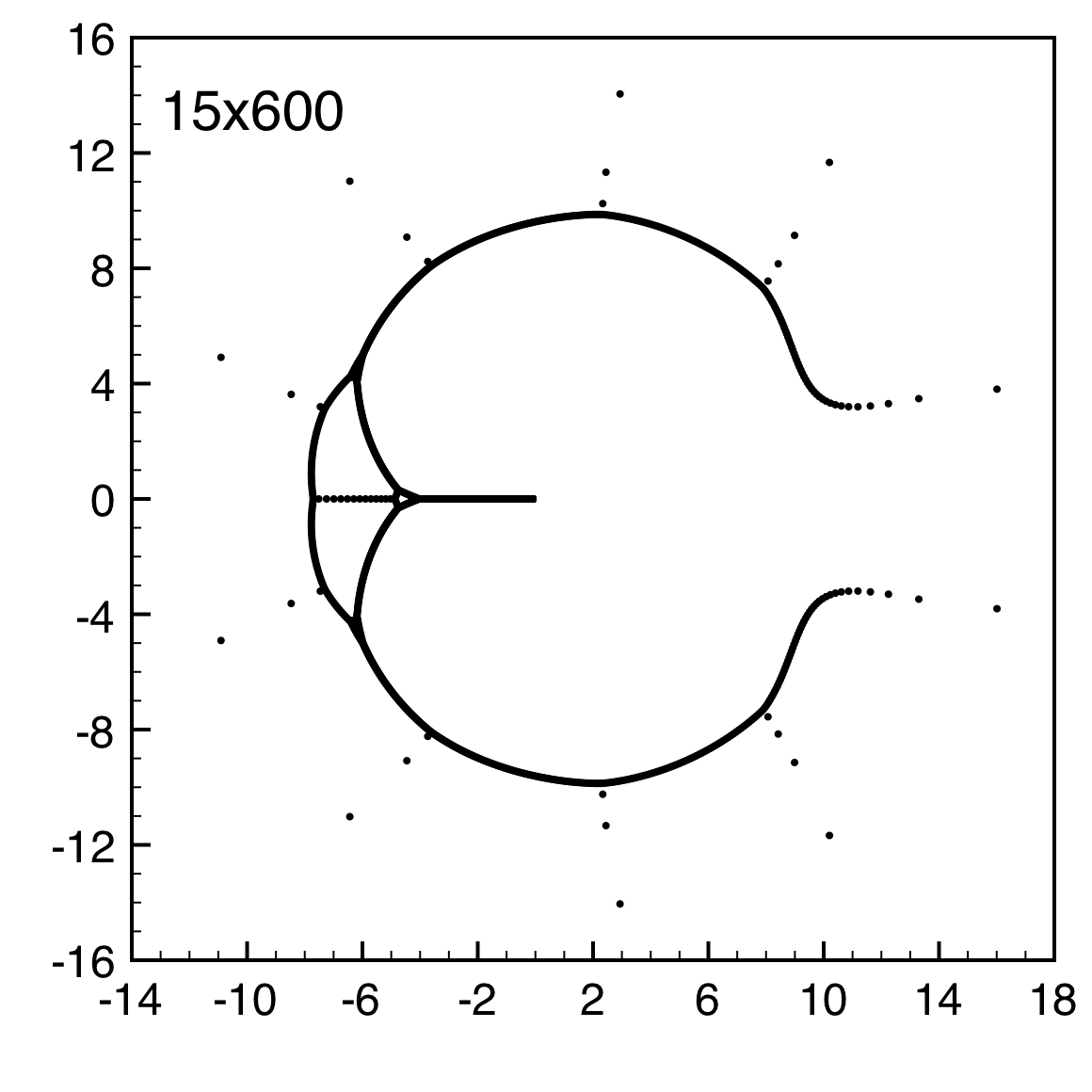}} 
\put(200,175){\includegraphics[width=6cm]{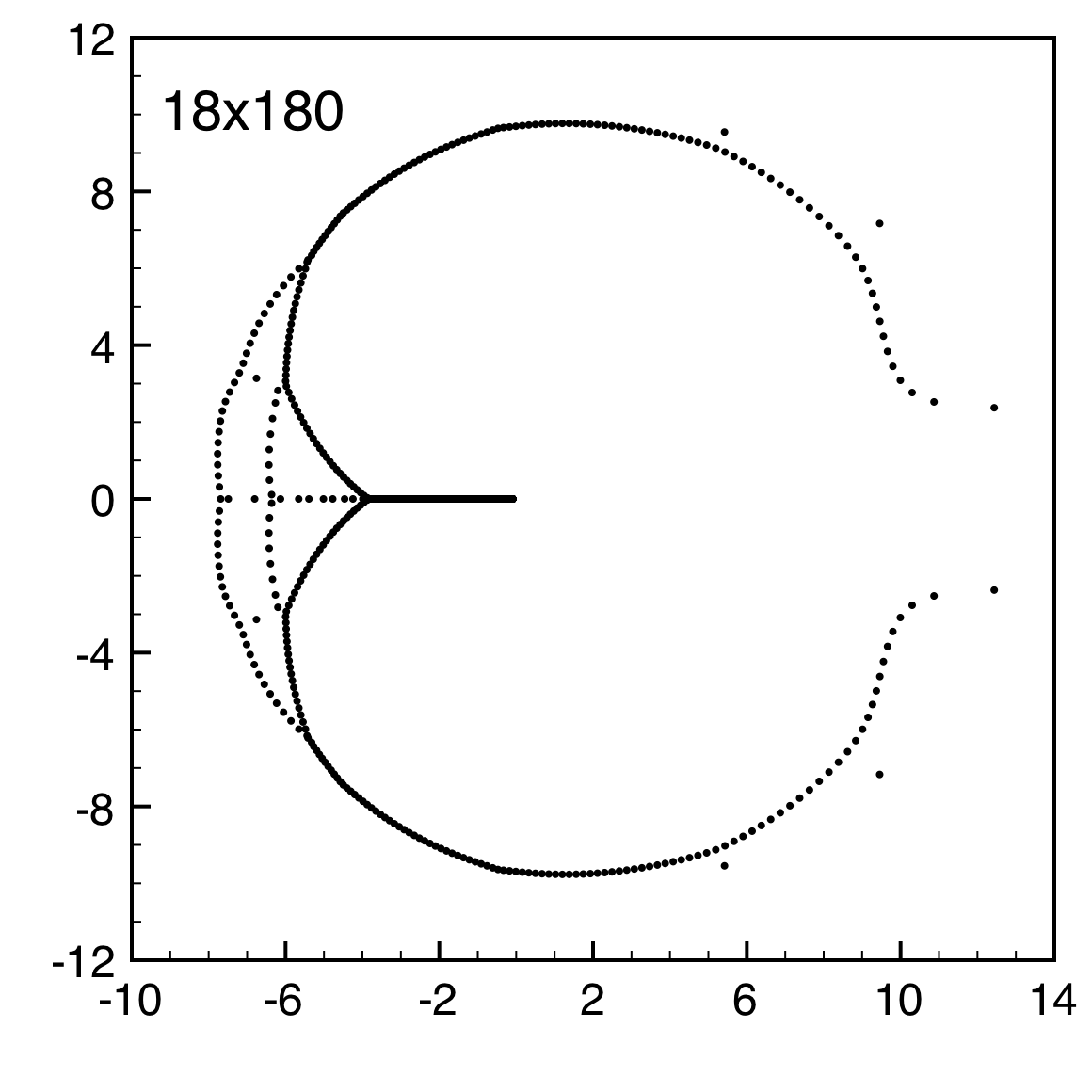}} 
\put(0,0){\includegraphics[width=6cm]{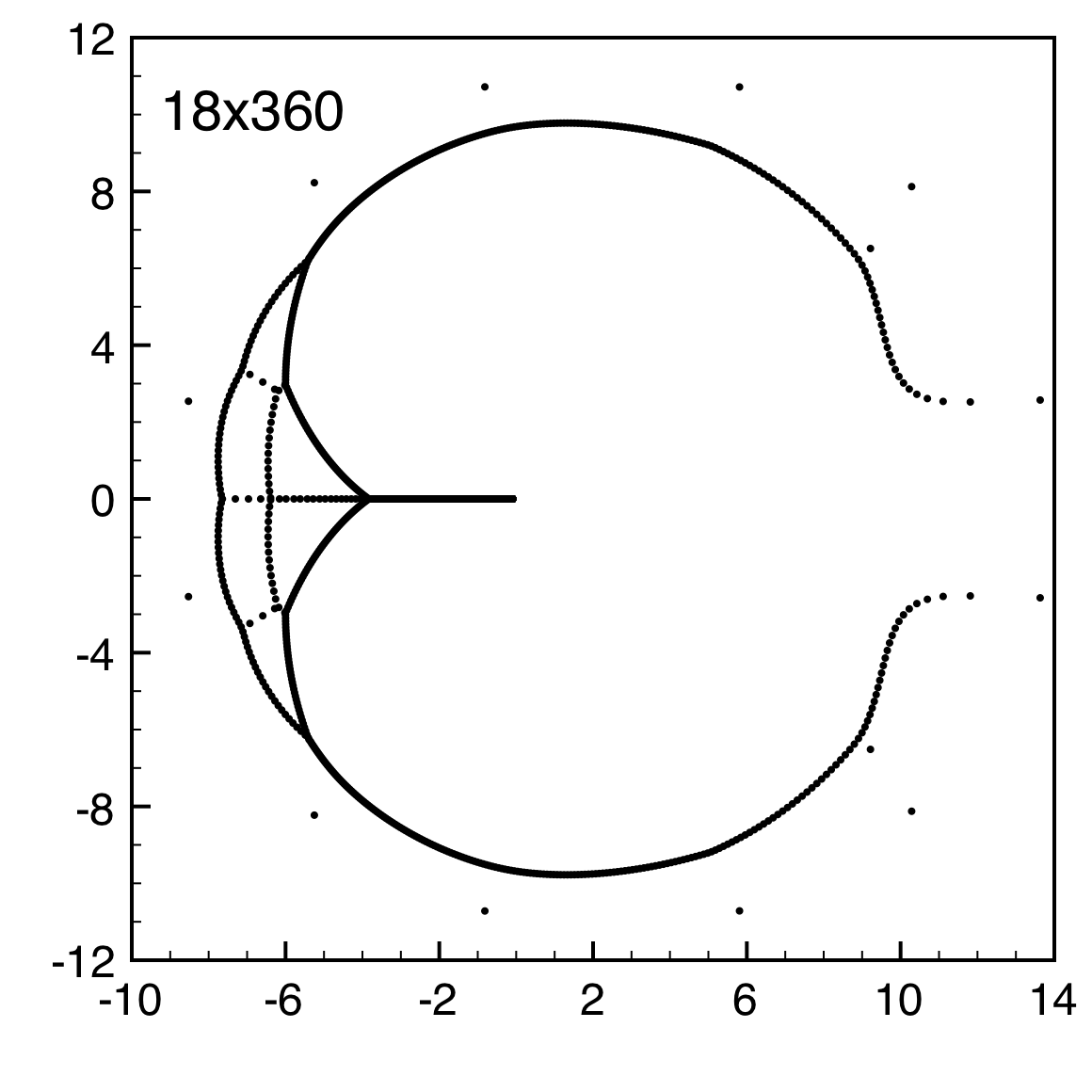}} 
\put(200,0){\includegraphics[width=6cm]{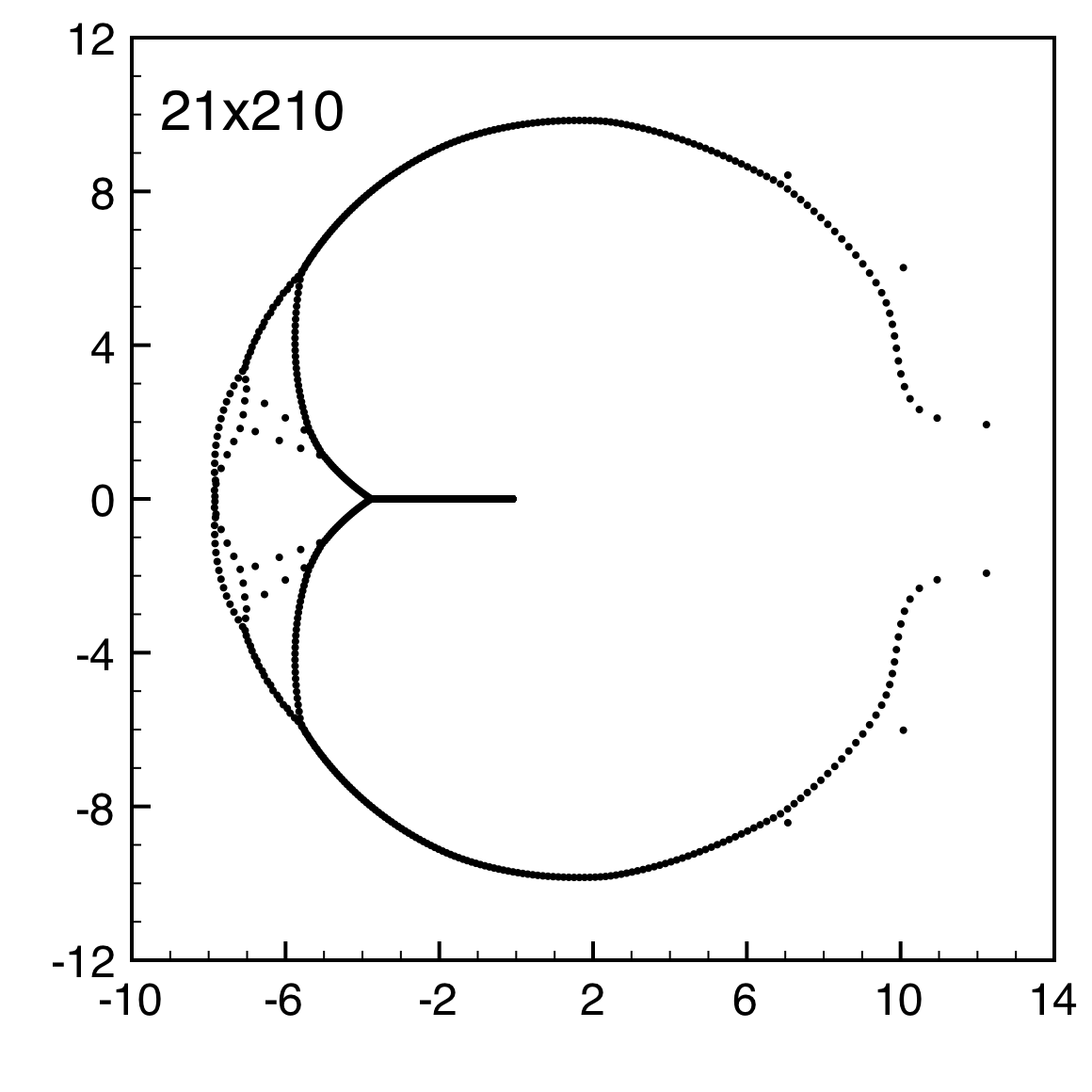}} 
\end{picture}
}
\end{center}
\
\caption{\label{fig:hhpbzerolong} Partition function zeros for toroidal boundary conditions on
  the lattices $15\,\times \, 150,~15\times 300,~15\times 600$, the lattices  
$18\times 180,~18\times 360$ and the lattice $21\times 210$. The
  number of points off of the main curve for fixed aspect ratio
  $L_v/L_h$ decreases with increasing $L_h$. }  

\end{figure}

We conclude our study of partition function zeros by examining the
dependence of the zeros on the aspect ratio $L_v/L_h$ of the
$L_h\,\times \,L_v$ lattices.  In figure~\ref{fig:hhpbzerolong} we plot the
partition function zeros for the toroidal lattices of  
various ratios $L_v/L_h$ as large as  40 for $L_h\,=\,15, \,18, \,21$.
We see that the number of zeros outside the main curve increases 
for fixed $L_h$ with increasing aspect ratio and for fixed aspect ratio
decreases with increasing $L_h$. It is furthermore obvious that even
for an aspect ratio of $40$ there are remarkably few zeros on the rays
seen in the transfer matrix equimodular curves of figure~\ref{fig:hhpblocus}.
From this we conclude, for
fixed $L_v/L_h\, <\,\infty$ with $L_h\,\rightarrow \,\,\infty$, that the
partition function zeros of the $L_h\, \times \, L_v$ on the toroidal 
lattice will not have any rays of zeros which extend to infinity.
 
 \clearpage
 
 \clearpage

\subsection{\label{sec:zerodens} Density of zeros for $z_y\leq z \leq z_d$}

On the negative $z$ axis we label as $z_j$ the position of the
$j^{th}$ zero where $z_1$ is the zero nearest to $z=0$ and $z_y$ is
the zero closest to the Y branching on the negative $z$ axis. Then 
calling $N_L$ the number of  zeros in the interval 
$z_y \,\leq \, z \, \leq \, 0 \,$ on a finite lattice of size 
$L\, \times \, L$ the density $D(z)$ in the 
thermodynamic limit is proportional to
\begin{eqnarray} 
\hspace{-0.9in}&&  \quad  \quad 
D(z)\,\,=\,\,\,\lim_{L\rightarrow \infty}\, \, D_L(z_j)
\quad \, \, \, \quad \hbox{where} \qquad  \,\,
 D_L(z_j)\,=\,\,\, \frac{1}{N_L \cdot \, (z_j-z_{j+1})}.
\end{eqnarray} 

This density of zeros will diverge at $z_d$ as $(1-z/z_d)^{-1/6}$, which is
obtained from the leading term in the expansion of 
$\rho_{-}(z)$. This  expansion is obtained, in \ref{app:rhoexp}, from
the algebraic equation (see eqn. (12.10) in~\cite{joyce}) as 
\begin{eqnarray} 
\label{rhomexp}
\hspace{-0.9in}&&  \quad  \quad  \quad  
\rho_{-}(z)\,\,\, =\, \, \,\,\,\,\,
t_d^{-1/6} \cdot \,\Sigma_0(t_d)\,\,\,\, +\Sigma_1(t_d)\,\,\,\,
+t_d^{2/3} \cdot  \, \Sigma_2(t_d)\,\,\,\, +t_d^{3/2} \cdot  \, \Sigma_3(t_d) 
\nonumber \\ 
\hspace{-0.9in}&&   \qquad   \qquad \qquad \quad \quad 
\, \,+t_d^{7/3} \cdot \, \Sigma_4(t_d) \, \,\,\, +t_d^{19/6} \cdot \, \Sigma_5(t_d), 
\end{eqnarray}
where $\,\, t_d\, =\,\,\, 5^{-3/2} \cdot \, (1-z/z_d)$, the fractional
powers are all defined positive for positive $t_d$
and where the $\, \Sigma_i(t_d)$ read
\begin{eqnarray}
\hspace{-0.95in}&& 
\Sigma_0 \,=\, \,-\frac{1}{\sqrt{5}} 
\, +\frac{1}{12}\left(5 +\frac{11}{\sqrt{5}}\right) t_d 
\,+ \frac{1}{144}\left(275 +\frac{639}{\sqrt{5}}\right) t_d^2  
\,+ \frac{1}{1296}\left(17765 +\frac{37312}{\sqrt{5}}\right) t_d^3 
\,+ \, \cdots   \nonumber  \\ 
\hspace{-0.95in}&&  
\Sigma_1\, =\, \,
\frac{1}{2}\left(1 + \frac{1}{\sqrt{5}}\right) 
+ \frac{1}{\sqrt{5}}\,t_d 
+ \frac{1}{2}\left(5 - \frac{1}{\sqrt{5}}\right)t_d^2 
- \frac{1}{2}\left(5 - \frac{83}{\sqrt{5}}\right)t_d^3 
\,\, + \,  \cdots  
\nonumber \\
\hspace{-0.95in}&&  
\Sigma_2\, =\,\, - \frac{2}{\sqrt{5}}\,
\, - \frac{2}{15}(25-4\sqrt{5})t_d 
\,+ \frac{4}{45}(125-108\sqrt{5})t_d^2 
\,- \frac{4}{405}(16775-4621\sqrt{5})t_d^3 +  \cdots  
\nonumber \\
\hspace{-0.95in}&&  
\Sigma_3\, = \, -\frac{3}{\sqrt{5}} 
\,- \frac{3}{4}\left(15 - \frac{7}{\sqrt{5}}\right) t_d 
\,+ \frac{3}{16}\left(175 - \frac{1189}{\sqrt{5}}\right) t_d^2 
\,- \frac{21}{16}\left(705 - \frac{646}{\sqrt{5}}\right) t_d^3 
\,\, +  \cdots  \nonumber \\ 
\hspace{-0.95in}&&  
\Sigma_4\, =\,\, -\frac{4}{\sqrt{5}} 
- \frac{2}{15} (175-13\sqrt{5}) t_d 
+ \frac{2}{45} (1625-2637\sqrt{5}) t_d^2 
- \frac{52}{405}(22100-3499\sqrt{5}) t_d^3
\, + \,  \cdots 
\nonumber \\
\hspace{-0.95in}&& 
\Sigma_5\, =\,\, -\frac{6}{\sqrt{5}} 
\, - \frac{1}{2}\left(95-\frac{31}{\sqrt{5}}\right)t_d 
\, + \frac{1}{24}\left(3875-\frac{34641}{\sqrt{5}}\right) t_d^2 
\, - \frac{31}{216}\left(55685-\frac{40892}{\sqrt{5}}\right) t_d^3
\, + \,  \cdots  
\nonumber \\
\hspace{-0.95in}&& 
\label{sigmaseries}
\end{eqnarray}

The term in $t_d^{2/3}$ was first obtained by Dhar~\cite{dhar} but the
full expansion has not been previously reported.
The form (\ref{rhomexp}) follows from the renormalization group 
expansion~\cite{cardy} of the singular part of the free energy at 
$z\,=\,z_d$ 
\begin{eqnarray} 
\label{rg}
\hspace{-0.9in}&& \qquad \quad  \quad  \quad \quad  
f_s\,\,\,\, =\,\,\,\, \, 
t_d^{2/y} \cdot \,\sum_{n=0}^{4}\, t_d^{-n(y'/y)} \cdot \,
\sum_{m=0}^{\infty}\, a_{n;m} \cdot \,t_d^m, 
\end{eqnarray}
where $y\,=\,12/5$ is the leading renormalization group exponent for the
Yang-Lee edge, and $y'\,=\,-2$, the exponent for the  contributing
irrelevant operator which breaks rotational invariance on the
triangular lattice, is determined from the term $t_d^{2/3}$ in
(\ref{rhomexp}). 
 
The density $\rho_{-}(z)$ has singularities only at $z_d$ and
$z_c$ in the plane cut on $\infty \leq z \leq z_d$ and $z_c\leq z \leq
\infty$. However, this does not require that the series
(\ref{sigmaseries}) for $\Sigma_j(t_d)$
will have $t_d$ evaluated at $z_c$ as their radii of convergence. We
have investigated this by computing the coefficients $c_j(n)$ of $z^n$
in the series for $\Sigma_j(t_d)$ using Maple up to $n=1200$.
 For $\Sigma_0(t_d)$ these
coefficients are all positive for $n>1$ and for $\Sigma_j(t_d)$ with
$j=2,3,4$ all coefficients are negative. However, for $\Sigma_5(t_d)$
the coefficients are negative for $0\leq n\leq 19$ and positive for
$n\geq 20$. For $\Sigma_1(t_d)$ the coefficients are positive for
$0\leq n \leq 554$ and negative for $n\geq 555$. Furthermore the
ratios $r_j=c_j(n)/c_j(n+1)$ seem to be 
converging to $2^{-3/2}=0.08944271\cdots$ which corresponds to $z=0$.

\begin{figure}[h!]

\begin{center}
\hspace{0cm} \mbox{
\begin{picture}(400,150)
\put(0,0){\includegraphics[width=6cm]{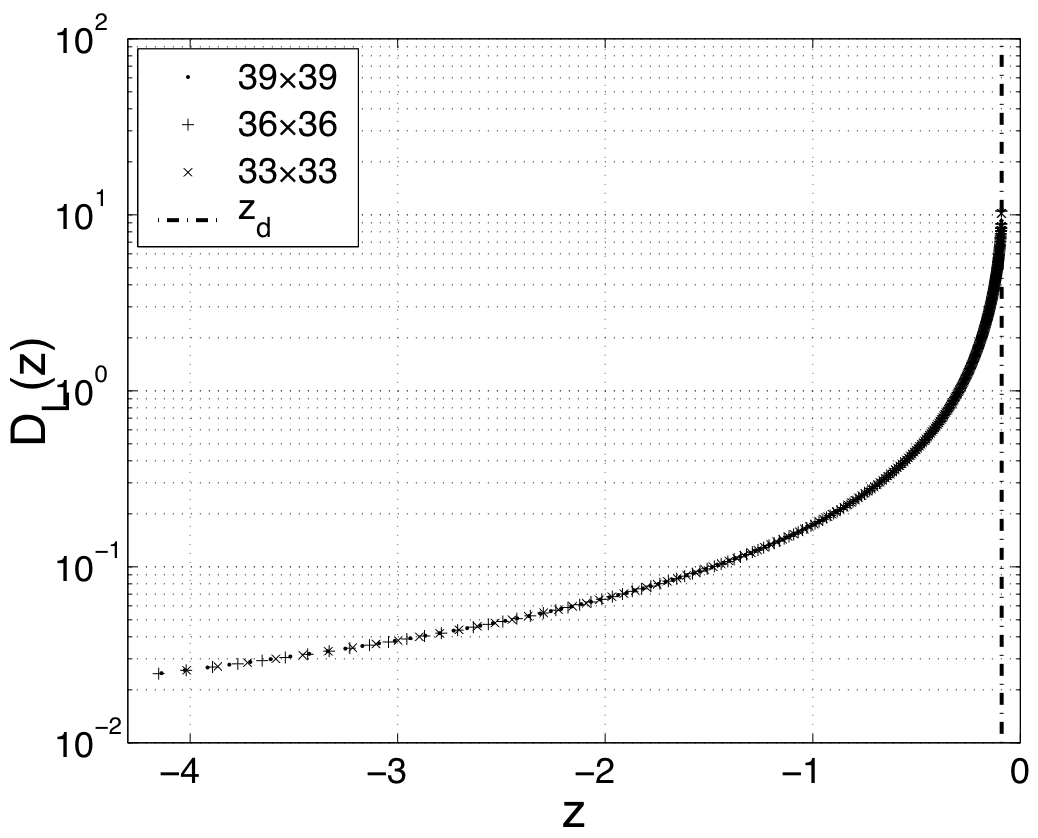}} 
\put(200,0){\includegraphics[width=6cm]{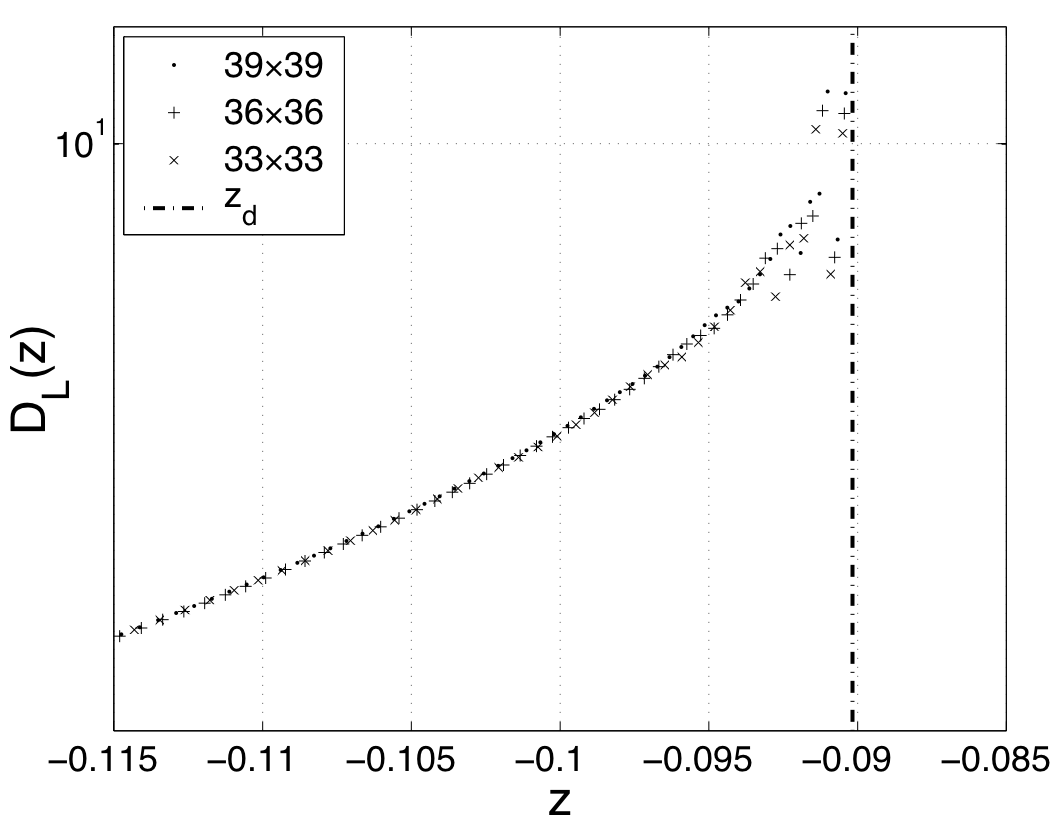}} 
\end{picture}
}
\end{center}

\caption{\label{fig:hclogall} Log plots of the density of zeros $D_L(z_j)$ on the negative $z$
  axis for $L\,\times\, L$ lattices with cylindrical boundary
  conditions. The figure on the right is an expanded scale near the
  singular point $z_d$. }  

\end{figure}

We investigate the density $\rho_{-}(z)$ further by plotting, 
in figure~\ref{fig:hclogall}, $ \, D_L(z_j)$ as  a
function of $z$  computed from the zeros of the $L\, \times\, L \, $ lattice
with cylindrical boundary conditions for $L\,=\,33,\, 36,\, 39$. The values of
$D_L(z_j)$ for all three lattices lie remarkably close to  the same curve 
except for the region $-0.093\,<\,z \,<\, z_d$, where some scatter is observed
which is caused by the finite size of the lattice. 

\begin{figure}[h!]

\begin{center}
\hspace{0cm} \mbox{
\begin{picture}(400,150)
\put(0,0){\includegraphics[width=6cm]{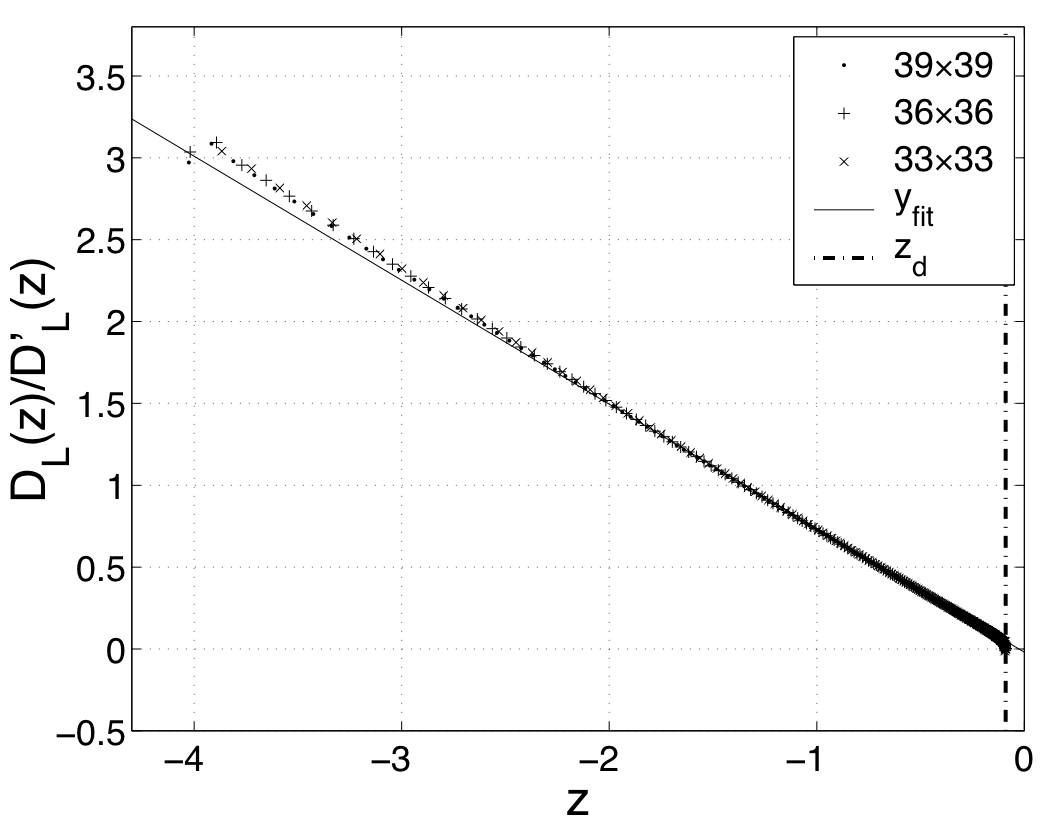}} 
\put(200,0){\includegraphics[width=6cm]{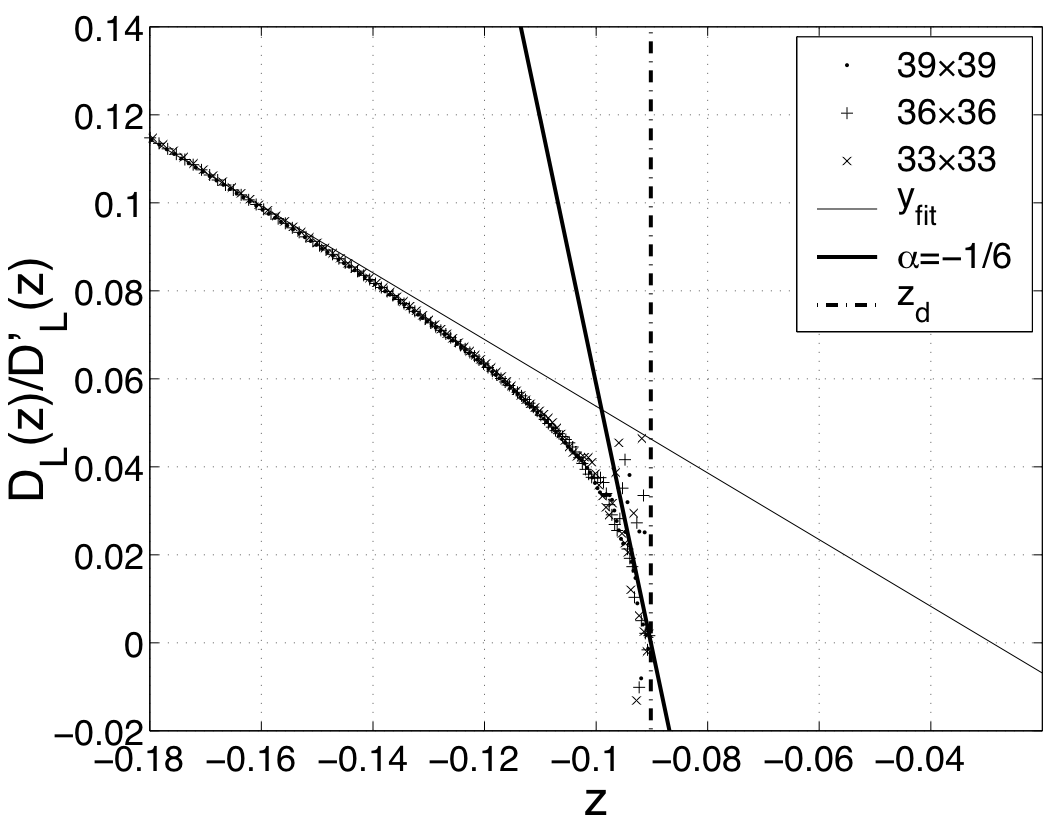}} 
\end{picture}
}
\end{center}

\caption{\label{fig:hcalldiff} Plots of $D_L(z_j)/D'_L(z_j)$ on the negative $z$
  axis for $L\times L$ lattices with cylindrical boundary
  conditions. For the plot on the left it is impressive that for the range
  $-4.0\,\leq z \,\leq\, -0.14$ the data is extremely well fitted by the power
  law form (\ref{zerofit}) with an exponent $-1.32$ (which
  corresponds to a slope of $-0.76$) and an intercept $z_f\,=\,-0.029$. 
The plot on the right is an 
  expanded scale near $z_d$
  and the line passing through $z=z_d$ with  slope of $-6$ is
  plotted for comparison which corresponds to the true exponent
  $=\,-1/6$ which only is observed in a very narrow range near $z_d$ of
 $ -0.095\,\leq \,z \,\leq z_d\,=\,-0.0901\cdots$.}  

\end{figure}

To estimate the divergence of $D(z)$, at $z\,=\,z_d$, we write for $z$ 
near $z_d$
\begin{eqnarray} 
\label{zerofit}
\hspace{-0.9in}&& \quad \quad  \quad 
D(z)\,\,\sim \,\, A \cdot \, (z_d-z)^{\alpha} 
\quad \,\, \,\, \hbox{and thus} \qquad \,\,\,\, 
 {{D(z) } \over {D'(z) }} \,\,\sim \,\, 
{{z_d-z } \over {\alpha }},  
\end{eqnarray}
where $D'(z)$ is the derivative of $D(z)$.
In figure~\ref{fig:hcalldiff} we plot $ D_L(z_j)/D'_L(z_j)$, where we define
\begin{eqnarray} 
\hspace{-0.9in}&& \qquad \quad  \quad  \quad 
D'_L(z_j)\,\,=\,\,\,\,\,
\frac{D_L(z_{j+1}) -D_L(z_j)}{z_{j+1} -z_j}. 
\end{eqnarray}
For $z$, away from $z_d$, the plot is very well fitted by the $\, z$-line
\begin{eqnarray} 
\hspace{-0.9in}&& \qquad \quad  \quad \quad 
 {{D_L(z) } \over {D_L'(z) }} \,\,\, \sim \,\,\,\, \, {{z_f -z } \over {\alpha }},  
\end{eqnarray}
with $ \, \alpha_f\,=\,\,-1.32 \,\, \cdots \,$ (or $\alpha_f^{-1}\,=\,\,-0.76 \, \cdots$), 
and $\, z_f\,=\,\,-0.029 \, \cdots \,$
However, for $-0.14 \,\leq \, z\, \leq \,z_d$ this fit is no longer valid. 
We make in figure~\ref{fig:hcalldiff}  a comparison with the true value of  
$\alpha\,=\,\, -1/6$, which is obtained from $\kappa_{-}(z)$.  This 
figure vividly illustrates the very limited range of validity  for the use
of  perturbed conformal field theory and scaling arguments to describe 
systems away from critical points.  The same phenomenon has
  been seen in \cite[eqn. (4.8) and Fig. 4]{jesper} for Hamiltonian chains.  

\section{\label{sec:disc} Discussions}

The results of the numerical studies presented above allow us to
discuss in some detail the relation of the functions $\kappa_{\pm}(z)$, 
which completely describe the
hard hexagon partition function per site on the positive $z$ axis, 
with the partition function per site in the full
complex $z$ plane. In particular  the
approach to the thermodynamic limit, the existence of the necklace,
the relation to the renormalization group  and 
questions of analyticity will be addressed.

\subsection{\label{sec:disctdl} The thermodynamic limit}

When the fugacity $z$ is real and positive the free energy and the
partition function per site is independent of the aspect ratio
$L_v/L_h$ of the $L_h\, \times \,L_v$ lattice as 
$L_h,\, L_v \,\rightarrow \,\infty$, and will be the same for both cylindrical 
and toroidal boundary conditions. This is a necessary condition for 
thermodynamics to be valid.

However, the example of hard squares at $z=-1$, where it is 
found~\cite{fse}-\cite{baxneg} that the partition function $Z_{L_v,L_h}(-1)$
depends on number theoretic properties of $L_v$ and $L_h$, demonstrates
that there may be places in the complex $z$ plane where a
thermodynamic limit independent of $L_v/L_h$ does not exist. We have
investigated in sections~\ref{sec:ev} and \ref{sec:zero} the extent to which our data supports the
conclusion that for complex $z$ there is a thermodynamic limit
independent of the aspect ratio $L_v/L_h$ as stated in (\ref{limitint}). 
If this independence holds then the limiting locus of partition function 
zeros will lie on the
limiting locus of transfer matrix equimodular curves. However, 
the converse does not need to be true and there is no 
guarantee that zeros will lie on all limiting loci of transfer matrix
eigenvalues. In particular we have argued  in sections~\ref{sec:evpb} and \ref{sec:zeropbas} that
for toroidal boundary conditions there will be no zeros in the rays 
which go to infinity.  
 
\subsection{\label{sec:discneck} Existence of the necklace}

All the data both for partition function zeros and transfer matrix
eigenvalues contain a necklace in the left half plane. Such a necklace
is incompatible with a partition function which only includes the
functions $\kappa_{\pm}(z)$. 

For cylindrical boundary conditions it is clear from figures~\ref{fig:hhcflocus} and \ref{fig:comp1} that the
limiting locus of transfer matrix equimodular curves in the necklace 
region has not yet been obtained. Furthermore in figures~\ref{fig:hhcflocus} and \ref{fig:comp1} there are
equimodular curves inside the necklace region beginning at $L_h\,=\, 18$ whereas
in figure~\ref{fig:hhzero} partition function zeros only appear  clearly
 inside the  necklace region for lattices $30\,\times\, 30$ and greater.

For toroidal boundary conditions the dominance of the $P\,=\,0$ sector
in the limit $L_h\,\rightarrow\, \infty$, discussed in section~\ref{sec:evdom},
implies that in the thermodynamic limit the necklace will be the same
as for cylindrical boundary conditions.

The simplest mechanism which will account for this behavior is for
there to be one (or  more) extra eigenvalue(s) of the transfer 
matrix which becomes
dominant in the necklace region. Further analytic computation is
needed to verify this mechanism. However, the  present data does not
rule out the possibility that for sufficiently large systems the
necklace region could be filled with zeros.

\subsection{\label{sec:discrg} Relation to the renormalization group}

The form of the singularity of the density $\rho_{-}(z)$ at $z\,=\,z_d$ 
given in section~\ref{sec:zerodens} in (\ref{rg})
is not the most general form, allowed by the renormalization group.
The most general form allows the singular part of the free energy
to have $y'\,=\,-1$, which would give a term 
in (\ref{rhomexp}) with exponent $t_d^{1/4}$ (which is, in fact, not present). 
This may be explained by the following renormalization group 
argument given by Cardy~\cite{private}. The
integer corrections given by $ny'$ are conformal descendants of the
identity operator. The total scaling dimension of these operators is
$N +{\bar N}$. Their conformal spin is $N -{\bar N}$, where $N$ and
$\bar N$ are nonnegative integers, and the corresponding exponent $y'$ is 
$2-N -{\bar N}$. However, the six fold lattice symmetry of the hard
hexagon model allows only operators with $N-{\bar N} \, \equiv\, 0$ (mod 6).
Therefore the dimensions $y'$ cannot be odd which is what is observed
in (\ref{rhomexp}). The same conclusion will apply also to hard
squares but not for hard triangles. 

\subsection{\label{sec:discana} Analyticity of the partition function}

The final property to be discussed is the relation of the analyticity
of the free energy obtained by analytically continuing the free energy
from the positive $z$ axis into the complex $z$ plane.

For hard hexagons the functions $\kappa_{\pm}(z)$ have singularities
only at $z\,=\,z_d, \, z_c, \, \infty$, whereas the partition 
function per site in the
complex plane fails to be analytic at those boundaries which are the
thermodynamic limit of the equimodular curves. It is obvious for hard
hexagons that these boundaries have nothing to do with the analyticity
of $\kappa_{\pm}(z)$. However it is unknown if in general the partition
function per site on the real axis can be continued 
analytically beyond the region where it corresponds to the 
maximum modulus of the transfer matrix eigenvalue.

The other locus where the hard hexagon model has partition function
zeros is on the negative real axis for $z \, \leq \,  z_d$. These zeros
correspond to the
complex conjugate solutions for $\kappa_{-}(z)$ alone and have no
connection with $\kappa_{+}(z)$.  The leading singular behavior of 
hard hexagons at $z_d$ is believed to be a property shared 
by all systems with purely repulsive (positive) potentials and is a universal
repulsive-core singularity  \cite{fisher1,fisher2}. Therefore 
 it is of considerable interest to determine whether the ability 
to continue through this locus of zeros, which is the case for the
integrable system of hard hexagons, will hold for all the non-integrable
models in the same universality class.  

\section{\label{sec:conc} Conclusion}

The hard hexagon model solved by Baxter~\cite{baxterhh,baxbook} 
not only satisfies the Yang-Baxter equation but also as 
shown by Joyce~\cite{joyce} and Tracy \etal~\cite{Tracy1,Tracy}
has a remarkable structure in terms of
 algebraic modular functions and their associated 
 Hauptmoduls. It is thus 
a good candidate for a global analysis in the whole
complex plane  of the partition function per site 
which is complementary to a local analysis based 
on series expansions and perturbation theory.

In this paper we have made a precision finite size study for the hard
hexagon model  of the zeros of the partition function and of the
equimodular curves of the transfer matrix in the complex fugacity
plane $z$. This study reveals that the partition function per site has 
more structure for complex $z$ than has been seen in previous studies
on much smaller systems~\cite{woodzero}-\cite{wood3}. 
In particular our results demonstrate that the conjecture on zeros 
of the hard hexagon partition function made in~\cite{wood3} is
incorrect, and 
corresponds to a too simple high density equimodular condition of
the Hauptmodul being real. 
This condition has to be replaced by more involved
equimodular conditions involving both the low and high density
partition functions. Furthermore we have found 
that the results of~\cite{baxterhh,baxbook} 
on the positive $z$ axis are not sufficient to determine all of the 
analytic structure of the partition function per site 
in the complex $z$ plane. 

The full 
significance of our results is to be seen in the comparison with hard
squares and with the Ising model in a magnetic field which do not
satisfy a Yang-Baxter equation and will not have the global properties
of modular functions.
 In particular we note  that in section~\ref{sec:evmul} it was found that on the 
negative $z$ axis the zeros of the resultant of the characteristic equation 
of the transfer matrix have the remarkable property that their multiplicity 
is two. This is in distinct contrast with hard squares where the
multiplicity of the roots of the resultant is one. 

Hard squares and hexagons are the limiting case of Ising models in a
magnetic field when the field becomes infinite. The
results of this paper have extensions to Ising models in a 
finite magnetic field which will be presented elsewhere.

\vskip .2cm 

\vskip .3cm

{\bf  Acknowledgments} 

We are pleased to acknowledge fruitful discussions 
with J.L. Cardy and A.J. Guttmann. One of us (JJ) is pleased to thank
the Institut Universitaire de France and Agence Nationale de la
Recherche under grant ANR-10-BLAN-0401 and the Simons Center for
Geometry and Physics for their hospitality. One of us (IJ) was
supported by an award under the Merit Allocation Scheme of the NCI
National facility at the ANU, where the bulk of the large scale
numerical computations were performed, and by funding under the
Australian Research Council's Discovery Projects scheme by the grant
DP120101593. We also made extensive use of the High Performance
Computing services offered by ITS Research Services at the University
of Melbourne.

\vskip .2cm

\vskip .3cm

\appendix

\section{\label{app:sing} The singularities of  $\kappa_{\pm}(z)$}

The partition functions per site $\kappa_{\pm}(z)$ are singular at
$z_c, \,z_d$ and $\infty$.
At $z\,=\,\,z_{c}$, and $ \, z\,=\,\,z_{d}$, the values of the three 
$\Omega_i$ read respectively
\begin{eqnarray} 
\label{omzc}
\hspace{-0.9in}&& \quad  \quad 
\Omega_1(z_{c})\,=\,\, 0, \quad \quad \,\,
\Omega_2(z_{c})\,=\,\,(5^{5/2}z_{c})^2,
\quad \quad \, \,
\Omega_3(z_c)\,=\,\,-(5^{5/2}z_c)^3,  
\end{eqnarray}
\begin{eqnarray}
\label{omzd}
 \hspace{-0.9in}&&  \quad  \quad 
\Omega_1(z_{d})\,= \,\, 0, \quad \quad\,\, 
 \Omega_2(z_{d})\,=\,\,(5^{5/2}z_{d})^2, \quad \quad\,\, 
 \Omega_3(z_d)\,=\,\,(5^{5/2}z_d)^3. 
\end{eqnarray}

\subsection{\label{app:singhigh} High density}

As $z \, \rightarrow \, \infty$ the physical $\kappa_+(z)$, which 
satisfies the algebraic equation (\ref{alghi}), diverges. There is 
only one such real solution and by direct expansion of 
(\ref{alghi}) we find that
\begin{eqnarray}
\label{kappalz}
\hspace{-0.9in}&&  \quad   
\kappa_{+}(z) \,\, = \,\,\,\, z^{1/3}\,\, +\frac{1}{3}\,z^{-2/3}
+\frac{5}{9}\,z^{-5/3}\,\, +{\frac {158}{81}}\,z^{-8/3}\,\,
+{\frac {2348}{243}}\,z^{-11/3}\, \,\, + \, \cdots 
\end{eqnarray}
which agrees with eqn. (7.14) in~\cite{joyce}. It follows from
(\ref{kappalz}) that $\kappa_{+}(z)$ has a branch cut on the segment
$-\infty<\,z\,\leq \,z_d$ and that on this  segment the phase is 
\begin{eqnarray} 
\label{kpphase}
\hspace{-0.9in}&& \qquad \quad  \quad  \quad 
e^{\pm\pi i/3} \quad \quad  \quad  {\rm for}
 \quad  \quad \quad   {\rm Im}z \,\,=\,\,\, 
\pm \epsilon\, \, \rightarrow \, \,0. 
\end{eqnarray}

When $\, z_c\,< \,z\,<\,\infty$ there is  one real positive, one real negative, 
and one complex conjugate pair of solutions to the fourth
order equation (\ref{alghi}) for $\kappa_{+}^6$. The negative solution is 
larger in magnitude than the positive solution, and, thus, cannot correspond 
to any eigenvalue of the transfer matrix. However, the magnitude of the
complex conjugate pair of solutions is less than the value of the real
positive root. At $z\,=\,z_c$ the real positive root collides
with the complex conjugate pair. 

When $z\,=\,z_c$, introducing the rescaled variable 
\begin{eqnarray} 
\label{wdefhi}
\hspace{-0.9in}&& \qquad \quad  \quad 
w_{c+}\,\,\,=\,\,\, \, \Omega_3(z_c) \cdot \, {{\kappa_{+}^6(z_c)} \over {z_c^6}}
\,\,\,=\,\,\, \, -(5^{5/2}/z_c)^3 \cdot \,\kappa_{+}^6(z_c), 
\end{eqnarray}
we find that (\ref{alghi}) reads $\, (w_{c+} +3^{9})^3 \, = \, \, 0$.
Thus, using (\ref{wdefhi}) and the fact that $\kappa_{+}(z_c)$ 
must be positive, we obtain
\begin{eqnarray}
\label{kpc} 
\hspace{-0.9in}&& \qquad \quad \quad  \quad 
\kappa_{+}(z_c)\,\,=\,\,\, \, 
(3^3\cdot 5^{-5/2}\,z_c)^{1/2} \, \,=\,\,\, 2.3144003 \,\,  \cdots
\end{eqnarray}
which is (7.17) of~\cite{joyce}.   

For $z\,=\,z_d$, introducing the rescaled variable 
$\, w_{d+}\,\,=\,\,\, \Omega_3(z_d) \cdot \, \kappa_{+}^6(z_d)/z_d^6$
$\,=\,\,  (5^{5/2}/z_d)^3 \cdot \,\kappa_{+}^6(z_d)$,
we find that (\ref{alghi}) also reads $\, (w_{d+} +3^{9})^3 \, = \, \, 0$.
Using (\ref{kpphase}), one gets  
$\, \kappa_{+}(z_d)^6\, =\, \, 3^9/5^8 \, (1525 \, -682\,\, 5^{1/2})$
or $\,  \kappa_{+}(z_d)\, \,=\, \,\, e^{\pm\pi i/3}\,\, 0.208689\, \, \cdots$. 

\subsection{\label{app:singlow} Low density}

When $z\,=\,z_c$ equation (\ref{alglo}) reduces using (\ref{omzc}) 
to the eleventh order equation
\begin{eqnarray}
\label{algloc}
\hspace{-0.9in}&& \qquad \quad \quad  \quad 
f_{-}(z_c,w_{c-})\,\, =\, \, \,\, \sum_{k=0}^{11} \, {\tilde C}^{-}_k \cdot \, w_{c-}^k
\,\,\,= \,\, \, 0
\end{eqnarray}
with  $\, w_{c-}\,=\,\, 5^{5/2}\, \kappa^2_{-}(z_c)/z_c$  
and
\begin{eqnarray}
&&{\tilde C}^{-}_0\,=\,\, -2^{32}\cdot 3^{27},\nonumber\\
&&{\tilde C}^{-}_1\,=\,\, 0,\nonumber \\
&&{\tilde C}^{-}_2\,=\,\, 2^{26}\cdot 3^{23}\cdot 31,\nonumber\\ 
&&{\tilde C}^{-}_3\,=\,\,  -2^{26}\cdot 3^{19}\cdot 47, 
\nonumber\\
&&{\tilde C}^{-}_4\,=\,\,  -2^{17}\cdot 3^{18}\cdot 5701,\nonumber\\  
&&{\tilde C}^{-}_5\,=\,\,  2^{16}\cdot 3^{14}\cdot 7^2 \cdot 19 \cdot 37, 
\nonumber\\  
&&{\tilde C}^{-}_6\,=\,\, -2^{10}\cdot 3^{10}\cdot 7\cdot 273001, 
\nonumber\\
&&{\tilde C}^{-}_7\,=\,\, 2^9 \cdot 3^{10} \cdot 11 \cdot 13 \cdot 139, 
\nonumber\\
&&{\tilde C}^{-}_8\,=\,\, -3^5\cdot7\cdot 1028327, \nonumber\\
&&{\tilde C}^{-}_9\, =\,37\cdot 79087, \nonumber\\
&&{\tilde C}^{-}_{10}\, =\,\,  -19\cdot 139, \nonumber\\ 
&&{\tilde C}^{-}_{11}\, =\,\,  1, 
\end{eqnarray}
which factorizes as
\begin{eqnarray} 
\label{factorlo}
\hspace{-0.9in}&& \quad  \quad 
f_{-}(z_c, \, w_{c-})\, \,  =\,\,\,  
(w_{c-}+2^4)^2 \cdot \,(w_{c-}-3^3)^3 \cdot \,  (w_{c-}\, -2^4 \cdot 3^3)^6
\,\, =\,\,\,  0.
\end{eqnarray}
From the second factor in (\ref{factorlo}) we obtain the solution
\begin{eqnarray} 
\label{equal}
\hspace{-0.9in}&& \qquad \quad  \quad 
\kappa_{-}(z_c)\,\,\,  =\,\, \, \, \,  (3^3\cdot 5^{-5/2}\, z_c)^{1/2} 
\,\,\,  =\,\, \,\,   \kappa_{+}(z_c), 
\end{eqnarray}
as required by continuity. At $ z\,=\,z_c$ the solution for 
$\kappa_{-}(z_c)$ is three
fold degenerate which also agrees with the degeneracy of
$\kappa_{+}(z_c)$.

For $0 \, <\,z\,< \, z_c\,$ there is one real positive, one real negative 
and five complex conjugate solutions of the 12th order equation
(\ref{alglo}). Three of the complex conjugate pairs have a modulus
less than the real positive solution. At $z=\, z_c$ a collision of the
real positive root with one of the complex conjugate pairs occurs.

When $z= \, z_d$ an analogous reduction can be made by use 
of (\ref{omzd}) and of the rescaling 
$\, w_{d-}\,\, =\,\,\, 5^{5/2}\, \kappa_{-}^2(z_d)/z_d$. 
We find, in analogy to (\ref{factorlo}), the factorization
\begin{eqnarray}
\label{factorlod} 
\hspace{-0.9in}&&  \quad 
f_{-}(z_d,w_{d-})
\,\,  =\,\,\, \, 
 -(w_{d-}-2^4)^2 \cdot(w_{d-}\,+3^3)^3 \cdot (w_{d-}\, +2^4\cdot 3^3)^6
\,\,\,  =\,\,\,\,  0.
\end{eqnarray}

 At $z\,=\,z_d$ the last factor in (\ref{factorlod}) vanishes 
and we find
 \begin{eqnarray}
\label{834}
 \kappa_{-}(z_d)\,\, &&=\, \,\, (-2^4\cdot 3^3 \cdot 5^{-5/2}\, z_d)^{1/2}
 \, \,=\,\,\,  (2^4\cdot 3^3 \cdot 5^{-5/2}/z_c)^{1/2}
 \nonumber\\&& =\, \,\,\,   
4\,|\kappa_{+}(z_d)| \, \,=\,\,  \,\, 0.83475738\,\, \cdots .
\end{eqnarray}

 For $z_d\, < \, z\,  <\, 0$ there are three real positive, three 
 real negative, and three complex conjugate solutions of the 
  polynomial relation (\ref{alglo}) of degree twelve in $\kappa_{-}^2$,
 and all three of the complex conjugate solutions have a modulus smaller
 than the largest positive real solution. The largest positive real
 solution is the dominant eigenvalue, until $z= \, z_d$, when a collision
 with the next real largest solution and two complex conjugate pairs
 occurs.

\section{\label{app:rhoexp} Expansion of $\rho_{-}(z)$ at $z_c$ and $z_d$}
 
The low density function $\rho_{-}(z)$ satisfies the polynomial equation 
of degree twelve in $\rho_{-}$ and degree four in $\, z$
 (see eqn. (12.10) in~\cite{joyce})
\begin{eqnarray}
\hspace{-0.9in}&&  \quad 
\rho_{-}^{11} \cdot \, (\rho_{-}-1) \cdot \, z^4 
\,\,\, \, -[\rho_{-}^5\, z^3\, -(\rho_{-}-1)^5 \, z] \cdot \, p_7
\nonumber\\
\hspace{-0.9in}&&  \quad \quad \quad \quad 
\,\, +\rho_{-}^2\,\cdot \, (\rho_{-}-1)^2 \cdot \, p_8 \cdot \, z^2
\,\,\, \,+\rho_{-}\,\cdot \,  (\rho_{-}-1)^{11}
\, \, \,=\, \,\, \,  0  
\end{eqnarray}
where
\begin{eqnarray}
\hspace{-0.9in}&&  \quad  \, \, 
p_7  \,  \, = \, \,  \,
 22\rho_{-}^7\,-77\,\rho_{-}^6\,+165\rho_{-}^5\,-220\rho_{-}^4
\, +165\rho_{-}^3\,-66\rho_{-}^2\,+13\rho_{-}\,-1, 
\nonumber \\
\hspace{-0.9in}&&  \quad \, \, 
p_8  \,  \, = \, \,  \,
119\rho_{-}^8\,-476\rho_{-}^7\, +689\rho_{-}^6\,-401\rho_{-}^5\,
-6\rho_{-}^4\,+125\rho_{-}^3\,-63\rho_{-}^2\, +13\rho_{-} \, -1. 
\end{eqnarray}
This equation has the remarkable property that at 
$z\, =\,z_c,~z_d$ it reduces to a fifth order equation
\begin{eqnarray}
\label{rhomc1}
\hspace{-0.9in}&& \quad \quad 
-\frac{275 +123\sqrt{5}}{8000} \cdot \, (10\rho_{-}\,
 -5\, +\sqrt{5})^5
 \,  \, = \, \,  \, 0 
\quad \quad \, \quad  {\rm for}\, \,     \quad
 z\, =\,z_c, 
\\
\hspace{-0.9in}&&   \quad \quad 
-\frac{275-123\sqrt{5}}{8000} \cdot \,  (10\rho_{-}\, 
-5\,-\sqrt{5})^5\,\,=\,\,\, 0
 \quad \quad \, \quad  {\rm for} \, \,   \quad
 z\,=\,\,z_d. 
\end{eqnarray}

There are four distinct Puiseux
expansions of $\rho_{-}$ about $z_c$ which are real for $z\,<\,z_c$. 
The  leading exponents of these expansions are $-1,\,-1/6,\,0,\,0$.
The physical solution must be finite at $z\,=\,z_c$ and we see from 
(\ref{rhomc1}) that the two solutions which are constant
at $z\,=\,z_c$ have the value $\rho_{-}(z_c)\,=\,\,(1-5^{-1/2})/2$. To decide
which of these two Puiseux expansions is the correct physical solution
we need the independent condition that the leading nonanalytic term
has exponent $2/3$. The result~\cite[12.15]{joyce} follows from this
additional condition.

At $z=z_d$ there are also  four Puiseux expansions of $\rho_{-}(z)$
which are real for $z_d\,< \,z$. The leading exponents are, again,
$-1,\,-1/6,\,0,\,0$. Now, unlike $\rho_{-}(z_c)$, the density is not
constant at $z= \, z_d$, but diverges with exponent $-1/6$. Furthermore,
in the cluster expansion of $\rho_{-}(z)$ about $z\,=\,0$, it follows,
 from a theorem of Groeneveld~\cite{gr}, that because the sign of the
coefficient of $z^n$ is $(-1)^{n-1}$, the density must be negative in
the segment $z_d \, < \, z\,< \,0$. The leading term of the Puiseux 
expansion with exponent $-1$ is positive and is, thus, excluded. There 
are six conjugate solutions with exponent $-1/6$. The member of this
class which has the correct negative behavior 
$z \,\, \rightarrow\, \, z_d+ \,$ 
is the result given in (\ref{rhomexp}).    

\section{\label{app:haupt} The Hauptmodul equations and the $\kappa_{\pm}$ equimodular curves}

The equations (\ref{alghi}) and (\ref{alglo}) for $\kappa_{\pm}$
may be usefully re-expressed in terms of the Hauptmodul $\, H$ 
\begin{eqnarray} 
\label{hdef}
\hspace{-0.9in}&& \qquad \qquad \quad  \quad 
H\,\,\,=\,\,\,\, 1728 \, z \cdot \, 
\frac{ \Omega_1^5(z)}{\Omega_3^2(z)}, 
\end{eqnarray}
by making the rescaling
\begin{eqnarray} 
\label{wdef}
\hspace{-0.9in}&& \qquad \qquad  \quad  \quad 
W_{\pm}\,\,=\,\,\, \,\,
\Omega_3(z) \cdot \, 
\Bigl( {{\kappa_{\pm} } \over {z}} \Bigr)^6.
\end{eqnarray}

For high density it is straight forward to use (\ref{hdef}) and
(\ref{wdef}) in (\ref{alghi}) to obtain 
\begin{eqnarray}
\label{polhigh}
\hspace{-0.9in}&&\quad \quad \, \, \, 
 P_{+}(W_{+}, \, H) \, \,  = \, \, \,  
{H}^{2} \cdot \, W_{+}^{4} \,  \, 
+ \, 2^7\cdot \, 3^6 \, \cdot \,  (27\,H-32) \cdot \,  W_{+}^{3}
 \nonumber \\ 
\hspace{-0.9in}&& \qquad \quad  \,  \,  \, 
+\, 2^7\cdot \, 3^{16} \, \cdot \,  (45\,H-32) \cdot \, W_{+}^{2}
\, \,  -2^{12} \cdot \, 3^{25} \, W_{+} \,  \, 
- 2^{12} \cdot \, 3^{33}\, \,  =\, \, \,   0.
\end{eqnarray}
The algebraic curve $\, P_{+}(W_{+}, \, H)\,\,    =\,\, 0$ 
is the union of two genus zero curves. 

\vskip .1cm 

For low density the polynomial relation (\ref{alglo}) on $\kappa_{-}$
in the $\, z$ variable can be written 
in terms of the Hauptmodul (\ref{hdef}), and of the rescaled 
variable $W_{-}$ (\ref{wdef}), as follows
\begin{eqnarray}
\label{pollow}
\hspace{-0.9in}&& \quad 
P_{-}(W_{-},H) \, \,   = \, \, \,  \,    
 H^6 \cdot \, W_{-}^{12} \, \,\,\,   \,  \, 
+2^{12} \cdot \,  3^{7} \, \cdot \, P_{11} \cdot \,  W_{-}^{11} \, \,  \, \, 
\,\, +\,2^{19} \cdot \, 3^{13}  \, \cdot \, P_{10} \cdot \,  W_{-}^{10}
\nonumber \\ 
\hspace{-0.9in}&& \qquad \quad 
-2^{32} \cdot \, 3^{18} \cdot \, P_{9} \cdot \,  W_{-}^9  \, \,
-2^{36} \cdot \, 3^{29} \cdot \,  P_{8} \cdot \,  W_{-}^8  \,\,\,  \,  \,  
+2^{52} \cdot \, 3^{38} \cdot \, P_{7} \cdot \,  W_{-}^7 \,\,  \,  \,  
\nonumber \\
\hspace{-0.9in}&& \qquad \quad 
+2^{62} \cdot \, 3^{46} \cdot \, P_{6} \cdot \,  W_{-}^6  \,  \,
-2^{77} \cdot \, 3^{56} \cdot \, P_{5} \cdot \,  W_{-}^5 \, \, \, \, \, 
-2^{85} \cdot \, 3^{65} \cdot \, P_{4} \cdot \,  W_{-}^4
\nonumber \\
\hspace{-0.9in}&& \qquad \quad 
+2^{100} \cdot \, 3^{73} \cdot \, P_{3}   \cdot \,  W_{-}^3 \,\,  \, \, 
 -2^{110} \cdot \, 3^{83} \cdot \, P_{2} \cdot \,  W_{-}^2 \,\,  \, \, 
 +47 \cdot \, 2^{126} \cdot \, 3^{92} \cdot \, W_{-} 
 \nonumber\\
\hspace{-0.9in}&&\qquad \quad 
-2^{132} \cdot \, 3^{99}  \,\, = \,\, \, 0, 
\end{eqnarray}
where the polynomials $\, P_n$ read:
\begin{eqnarray}
\label{Ppollow}
\hspace{-0.9in}&&
P_{11}  \, = \, \, \, 85423588659\, H^5\, \,
-1273194070087\, H^4 \,\, +5683675368960\, H^3 \, \,
\nonumber \\
\hspace{-0.9in}&& \qquad 
-3624245 \cdot \,  2^{12} \cdot \, 3^{6} \, H^2 \,\,
 +901 \cdot \,  2^{19} \cdot \, 3^{9}\, H \,
 \, -2^{24} \cdot \, 3^{11}, 
\nonumber \\
\hspace{-0.9in}&&
P_{10} \, = \, \, \,
2098366262345322754767\, H^5 \, 
-4991131592299977169590\, H^4 \,
\nonumber \\
\hspace{-0.9in}&& \qquad 
 +3893219286516719759223\, H^3\, 
-1056221406812154079936\, H^2 \,\,
\nonumber \\
\hspace{-0.9in}&& \qquad 
 +56427952366139092992\, H \, \,
-483780265 \cdot \, 2^{17} \cdot \, 3^{5},
\nonumber \\
\hspace{-0.9in}&&
P_{9}  \, = \, \, \,
15382723254412673871318753\, H^4\, \,
+26277083153777345473689849\, H^3 \,
\nonumber \\
\hspace{-0.9in}&& \qquad 
+4098422120568047655974595\, H^2\, \,
+37921229707060286737587\, H \,\,
\nonumber \\
\hspace{-0.9in}&& \qquad 
+1560354561975860656, 
\nonumber \\
\hspace{-0.9in}&&
P_{8}  \, = \, \,  \,
1020939125266735071750904401\, H^4\,
\, -1161800973997140083525143956\, H^3 \,
\nonumber \\
\hspace{-0.9in}&& \qquad 
 +214393801490313112726470774\, H^2
-2006070488338798415238516\, H \,
\nonumber \\
\hspace{-0.9in}&& \qquad 
 +59190955246329648961, 
\nonumber \\
\hspace{-0.9in}&&
P_{7}  \, = \, \, \,
508697400997842959916351\,H^3\,\,
 -554351605658908065490725\,H^2\,
\nonumber \\
\hspace{-0.9in}&& \qquad 
 -35192800976394203832051\,H \,
 -2775596721861024679,
\nonumber \\
\hspace{-0.9in}&&
P_{6}  \, = \, \, \,
1245962466251450908065\,H^3\,\,
-15255449815782496728645\,H^2\,\,
\nonumber \\
\hspace{-0.9in}&&\qquad 
+8457596543456744207175\,H\,
 -13332664262978720611, 
\nonumber \\
\hspace{-0.9in}&&
P_{5}  \, = \, \, \,
114630292396020573\, H^2\,\,
 -366034684810378734\, H \,\,
 +92792159042784817, 
\nonumber \\
\hspace{-0.9in}&&
P_{4}  \, = \, \, \,
938107512437391\, H^2 
\, \,-1026461977730478\, H\,
+933965999427127, 
\nonumber \\
\hspace{-0.9in}&&
P_{3}  \, = \, \, \,
121395557277\, H \, -59327302513, 
\nonumber \\
\hspace{-0.9in}&&
P_{2}  \, = \, \, \, 
11532609\, H \, -1281659. 
\end{eqnarray}

Do note that the algebraic curve $\, P_{-}(W_{-}, \, H)\,\,    =\,\, 0$ 
is actually a genus zero curve. The algebraic curve (\ref{pollow})
is the sum of 43 monomials of degree six in $\, H$ and 
degree 12 in $\, W_{-}$, 
as compared to a sum of 157 monomials of degree 22 in$\, z$
 and degree 24 in $\kappa_{-}$ for (\ref{alglo}). At first sight, 
the polynomial 
relation (\ref{pollow}), in the Hauptmodul
and the rescaled variable $\, W_{-}$, looks
quite different from (\ref{alglo}). In fact, the two polynomial relations
(\ref{pollow}) and (\ref{alglo}) are in agreement, as can be seen 
on the quite remarkable identity 
\begin{eqnarray} 
\label{lhs}
\hspace{-0.95in}&& \,   
z^{66} \cdot \, P_{-}(W_{-}, \, H)\,\,    =\,\,   \,   
12^{18} \cdot \,  f_{-}(z, \kappa_{-}) \cdot \, 
f_{-}(z, e^{2\pi i/3}\,  \kappa_{-}) \cdot \, 
 f_{-}(z, e^{-2\pi i/3}\,  \kappa_{-}),
\end{eqnarray}
where the l.h.s. of (\ref{lhs})  is actually
a polynomial expression in terms of $\,\kappa_{-}$ and $\, z$. 

\vskip .1cm 

\subsection{\label{app:H+-} $ \quad \kappa_{+}$ versus $\kappa_{-}$ }

The functions $\kappa_{-}$ and $\kappa_{-}$ are not related by
analytic continuation. However, because both $W_{+}$ and $W_{-}$ are
algebraic functions of the same Hauptmodul we can eliminate $H$
between (\ref{polhigh}) and (\ref{pollow}) to obtain the following
algebraic relation between $W_{+}$ and $W_{-}$
\begin{eqnarray} 
\label{versus}
\hspace{-0.9in}&&  
{W_{-}}^{4} \, {W_{+}}^{6}  \, \, \,
+32\,{W_{-}}^{3} \, W_{+}^{5} \cdot \, 
(1509\,W_{-}-512\,W_{+})
\nonumber \\ 
\hspace{-0.9in}&&   
 -2 \,\, W_{-}^{2}\, W_{+}^{3}  \cdot \,
(W_{-}^{3} \, -411832512\,W_{-}^{2} \,W_{+}
+937623552\,W_{-} W_{+}^{2} \, -50331648\, W_{+}^{3}) \, 
\nonumber \\ 
\hspace{-0.9in}&& 
-32\,W_{-} \,W_{+}^{2}  \cdot \, (34791\,W_{-}^{4}
-182579836224\,W_{-}^{3} W_{+} \, -1128985165824\, W_{-}^{2} \, W_{+}^{2}
\nonumber \\ 
\hspace{-0.9in}&& \qquad \qquad 
-549067948032\,W_{-} \, W_{+}^{3} +8589934592\, W_{+}^{4})\, 
\nonumber \\ 
\hspace{-0.9in}&& 
 + \, (W_{-}^{4} -84091500544\,W_{-}^{3} \, W_{+} -1482164797440\,W_{-}^{2}\, W_{+}^{2}
-8145942347776 \,W_{-} \, W_{+}^{3}
\nonumber \\ 
\hspace{-0.9in}&& \qquad   \, +68719476736\,W_{+}^{4}) \cdot \, 
(W_{-}^{2}-172928\,W_{-} \, W_{+} +4096\,W_{+}^{2})
 \, \, \, = \, \, \, \, 0.
\end{eqnarray}
This remarkable algebraic relation for hard hexagons follows from the
modular properties of $\kappa_{+}$ and $\kappa_{-}$ and is not expected
to exist for a generic system.

One verifies easily that eliminating $\, W_{+}$ between (\ref{polhigh})
and (\ref{versus}) one recovers (\ref{pollow}),
and that eliminating $\, W_{-}$ between (\ref{pollow})
and (\ref{versus}) one recovers (\ref{polhigh}). 

The polynomial relation (\ref{versus}) of degree six in
$\, W_{+}$ and $\, W_{-}$, is actually also a  genus zero
algebraic curve. 

\vskip .1cm 

The situations where $\, \kappa_{-} \, = \, \, \kappa_{+}$ (see (\ref{equal}))
correspond to $\, W_{-} \, = \, \, W_{+}$ in (\ref{versus}). It yields 
the values  $\, 0, \, -3^9, \, -57707, \, 22743 \, \pm 30268 \, i$, 
corresponding to $\,  W_{-} \, = \, \, W_{+} \, = \, -3^9$.
Note that $\, \kappa_{-} \, = \, .83475738 \, \cdots$
in (\ref{834}) corresponds to the integer value 
$\, W_{-} \, = \, \, -2^{12} \, 3^9$. 

\vskip .1cm 

\subsection{\label{app:Hcurve} The $\kappa_{\pm}$ equimodular curves}

The $\kappa_{\pm}$ equimodular condition reads
 $\, |W_{+}|\,= \, |W_{-}|$ in terms of $\, W_{\pm}$.  Setting the ratio
\begin{eqnarray} 
\hspace{-0.9in}&& \qquad \qquad \qquad \quad  \quad 
r\,=\,\, {{W_{+}  } \over {W_{-} }}, 
\end{eqnarray}
we can obtain a polynomial relation 
between this ratio $r$ and the Hauptmodul $\, H$,
eliminating $\, W_{-}$ between $\, P_{-}( W_{-},\,H) \, = \, \, 0$
and  $\, P_{+}( r \cdot \, W_{-},\,H) \, = \, \, 0 $, by  
performing a resultant. This resultant calculation yields a polynomial 
condition $\, P(r,\,H) \, \, = \, \, \, 0$,
where the polynomial, of degree  36 in $r$ and degree 18 in $\, H$, 
is the sum of 577 monomials.  When $H\,=\,\, 0$ this polynomial 
reduces to 
\begin{eqnarray}
\label{PatHzero}
\hspace{-0.9in}P(r,\,0) \, \, = \, \, \, 2^{108} \cdot \, r^3 \cdot \, 
(4096 \, r\, + 19683)^6 \cdot \, (4096 \, r\, - 1)^{12} \cdot \, (r \, -1)^6,
\end{eqnarray}
and when $H \, = \, 1$, it reduces to
\begin{eqnarray}
\hspace{-0.9in}&&\quad P(r,1) \, \, = \, \, \,
(330225942528\, r^3 \, +216854102016 \, r^2 \, +72695294208 \, r \, +1)^3 
\nonumber \\
\hspace{-0.9in}&& \quad \quad \, \, \, 
\times (16777216\, r^3 \, 
-297467904\, r^2 \, +2692418304\, r \, -1)^6
\cdot \, (256\, r \, +27)^9. 
\end{eqnarray}

The equimodularity condition
$\, |\kappa_{+}| \, = \, \,|\kappa_{-}|$ corresponds to an algebraic curve 
in the $(x, \, y)$ complex plane ($z \, = \, \, x \, + i \, y$). This curve 
can be obtained by writing the Hauptmodul as a function of $\, x$ and $\, y$,
namely $\, H \, = \, \, X(x, \, y) \, + \,  i \,Y(x, \, y)$, where 
$\,X(x, \, y)$ and $\, Y(x, \, y)$ are quite large rational expressions 
of $\, x$ and $\, y$, and then parametrising the equimodularity condition 
$\, |r| \, = \, \, 1$  as
 $\,r \, = \, (1-t^2)/(1+t^2) \, + \,\, 2\, i \, t/(1+t^2)$,
where $\, t$ is a real variable. This amounts to writing
\begin{eqnarray}
\hspace{-0.95in}&&P\left({{1-t^2} \over {1+t^2}} \,
+ \,  i \cdot \, {{2\, t} \over {1+t^2}},  \,  \, X(x, \, y) \, + \,  
i \,Y(x, \,y)\right) 
\, \, = \, \, \, {\cal P}(x, \, y, \, t) \, 
+ \, \, \, \, i \cdot \, {\cal Q}(x, \, y, \, t)  
\, \, = \, \, \, 0.  
\nonumber 
\end{eqnarray}
where  $\, {\cal P}(x, \, y, \, t)$ and $\, {\cal Q}(x, \, y, \, t)$ 
are quite large rational expressions of the real  variables 
$\, x$, $\, y$ and $\, t$. Let us denote  $\, {\cal N}_1(x, \, y, \, t)$
the numerator of  $\, {\cal P}(x, \, y, \, t)$
and  $\, {\cal N}_2(x, \, y, \, t)$ the numerator of
  $\, {\cal Q}(x, \, y, \, t)$.
Eliminating $\, t$ between  $\, {\cal N}_1(x, \, y, \, t) \, = \, \, 0$
and $\, {\cal N}_2(x, \, y, \, t) \, = \, \, 0$, performing a resultant,
one will get finally a quite large polynomial condition
 $\,   {\cal P}(x, \, y) \, = \, \, 0$, corresponding to the 
algebraic equation of the equimodularity condition 
$\, |\kappa_{+}| \, = \, \,|\kappa_{-}|$.

\subsubsection{\label{app:Hsym} Icosahedral symmetry of the equimodular curve\newline}

The  polynomial condition is too large to be given explicitly here.
It is, however, worth noting that, since the equimodular curve is 
deduced from polynomial expressions that depend only on the Hauptmodul
$\, H$, the equimodular curve has the quite non-trivial
property that it is compatible with the icosahedral symmetry 
of the hard hexagon model~\cite{joyce2}. This icosahedral symmetry
corresponds to the following symmetry of the Hauptmodul
 (\ref{hdef}). Let us introduce the complex variable $\,\zeta$
defined by $\, z \, = \, \zeta^5$,  
the fifth root of unity $\, \omega$ and the  golden number 
$\, \tau$
\begin{eqnarray}
\hspace{-0.9in}&& \quad \quad \quad 
\omega \, \,  = \, \, \,  1/4\,\sqrt {5} \, -1/4 \,
 +1/4\,i\sqrt {2}\sqrt {5 \, +\sqrt {5}},
\quad \quad \quad  \tau   \, \,  = \, \, \,  
 {{1 \, +\sqrt {5}} \over {2}}.
\end{eqnarray}

Let us consider the order-five transformation $\, h_5$
\begin{eqnarray}
\hspace{-0.9in}&& \quad \quad \quad \quad \quad 
\zeta \, \,\quad   \longrightarrow \, \,  \,\quad  h_5(\zeta) \, \, = \, \, \, 
\tau  \cdot {{ \omega \, +(1 -\tau) \, \zeta } \over { \omega \, + \tau \,\zeta }}. 
\end{eqnarray}

It is a non-trivial but straightforward calculation to see 
that the Hauptmodul $\, H$, 
seen as a function of the complex variable $\,\zeta$, 
 is actually invariant by this order-five transformation $\, h_5$,
 by the involution $\, \zeta \, \,  \rightarrow \, \,  \, -1/\zeta$
as well as the order-five transformation 
 $\, \zeta \, \,  \rightarrow \, \,  \, \omega \cdot \, \zeta$:
\begin{eqnarray}
\hspace{-0.9in}&& \quad \quad \quad   \quad  \quad 
H(\zeta)  \, \, = \, \, \, H(h_5(\zeta))  \, \, = \, \, \,
 H\Bigl( {{-1} \over {\zeta}} \Bigr) 
 \, \, = \, \, \,H(\omega \cdot \, \zeta).
\end{eqnarray}

\subsubsection{\label{app:Hsel} A selected point of the equimodular curve \newline}

The algebraic equimodular curve $\,   {\cal P}(x, \, y) \, = \, \, 0$
intersects the real axis $\, y \, = \, \, 0$
at the critical value $z_c$ 
(i.e. $\, H \, = \, \, 0$, see (\ref{PatHzero}))
and at an algebraic value $\, z \, = \, \, -5.94254104 \cdots$ 
corresponding to the algebraic value of the Hauptmodul 
$\, H \, = \, \, 1.2699347  \cdots$, a root of 
the polynomial $\, P_{12}(H)$
 of degree twelve in $\, H$:
\begin{eqnarray}
\label{PolH}
\hspace{-0.9in}&& P_{12}(H) \, \, = \, \, \,  \, 
420659520093064357478960957541^2 \cdot \, H^{12}
\nonumber \\
\hspace{-0.9in}&& \qquad 
-3035676163450716673183784433435873765727935868148497169025
 \cdot \, 2^9 \cdot \, H^{11}
\nonumber \\
\hspace{-0.9in}&& \qquad 
+180032218185835528405034756761309783218694171171683152985
 \cdot \, 2^{16} \cdot \, H^{10}
\nonumber 
\end{eqnarray}
\begin{eqnarray}
\hspace{-0.9in}&& \qquad 
-963917598568487789731961832547602704647778692096330233
 \cdot \, 2^{25} \cdot \,H^9
\nonumber \\
\hspace{-0.9in}&& \qquad 
+4687917985071549790872555988500591318811098924601809
 \cdot \, 2^{33} \cdot \,H^8
\nonumber \\
\hspace{-0.9in}&& \qquad 
-11794524087347323954434252908699683281468087505905 
\cdot \, 2^{41} \cdot \,H^7
\nonumber 
\end{eqnarray}
\begin{eqnarray}
\hspace{-0.9in}&& \qquad 
+34111250660390601705930372758400977149413250857 
\cdot \, 2^{48} \cdot \,H^6
\nonumber \\
\hspace{-0.9in}&& \qquad 
-16562829715197286592872531393597405351924479
 \cdot \, 2^{57} \cdot \,H^5
\nonumber 
\end{eqnarray}
\begin{eqnarray}
\hspace{-0.9in}&& \qquad 
+10432786705236893496285793791292996147041 \cdot \, 2^{65} \cdot \,H^4
\nonumber \\
\hspace{-0.9in}&& \qquad 
-4452980987936971936196603653288348935 \cdot \,2^{73} \cdot \,H^3
\nonumber \\
\hspace{-0.9in}&& \qquad 
+2184609189525225289847951233328377 \cdot \,2^{80} \cdot \,H^2
\nonumber 
\end{eqnarray}
\begin{eqnarray}
\hspace{-0.9in}&& \qquad 
-14687865423363371951559480967168 \cdot \,176160768^3 \cdot \,H
\nonumber \\
\hspace{-0.9in}&& \qquad 
+956497920^6.
\end{eqnarray}

This algebraic point corresponds to the following algebraic values 
of $\, W_{\pm}$,  in (\ref{pollow}) and (\ref{polhigh}),
$\, W_{+} \, = \, \,-5404.2605\, \cdots  $ 
and $\, W_{-}\, = \, \,2118.9287 \, \cdots  \, + \,971.5363 \, \cdots \, i  $,
the ratio $\, r\, = \, \, W_{+}/W_{-}$ being, as it should, a 
complex number of unit modulus, namely
$\, -0.3920 \cdots \, + \, i \, .9199 \cdots$ 
This algebraic point is characterized by the fact that $\, W_{+}$
or $\, \kappa_{+}^6$ (but not $\, \kappa_{-}$) is a real number:
  $\, \kappa_{+}^6 \, = \, \,26.6786 \, \cdots $ but
 $\, \kappa_{-} \, = \, \,0.864  \, \cdots -1.497 \, \cdots \, i$.

\clearpage

\section{\label{app:card} Cardioid fitting of partition function zeros}

\begin{figure}[h!]

\begin{center}
\hspace{0cm} \mbox{
\begin{picture}(350,120)
\put(0,0){\includegraphics[width=4cm]{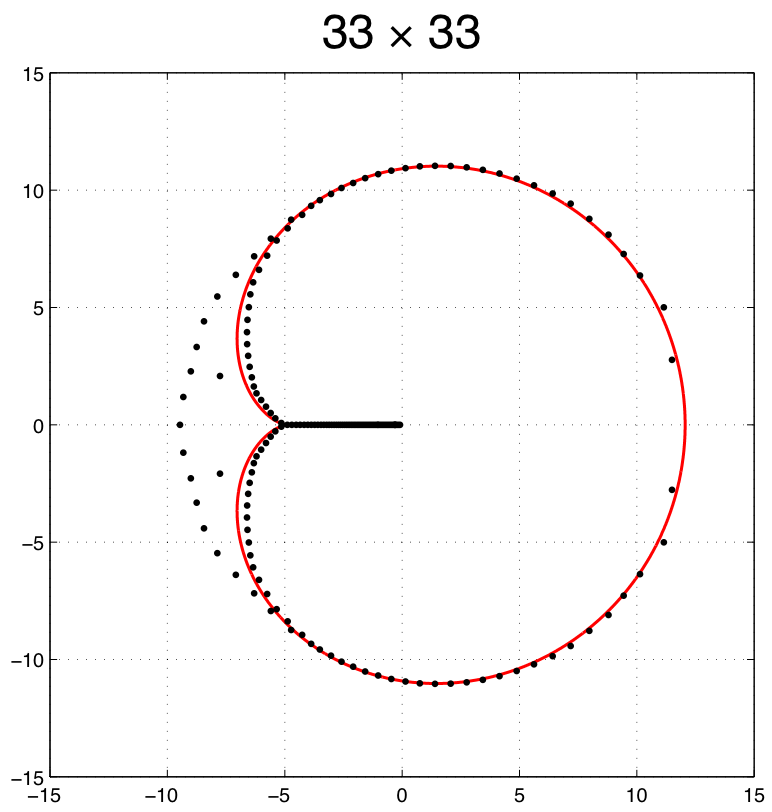}}
\put(120,0){\includegraphics[width=4cm]{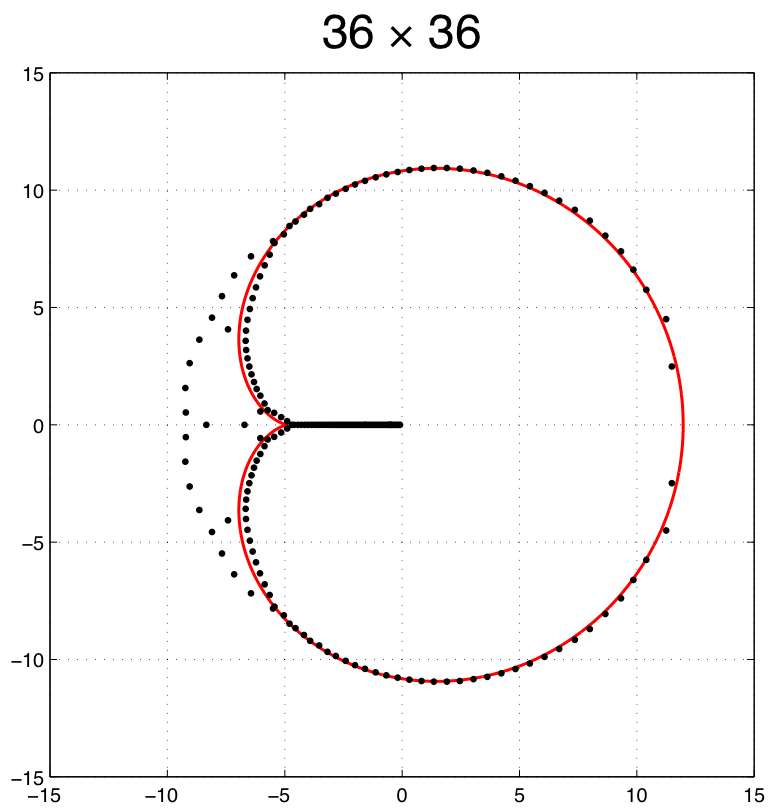}}
\put(240,0){\includegraphics[width=4cm]{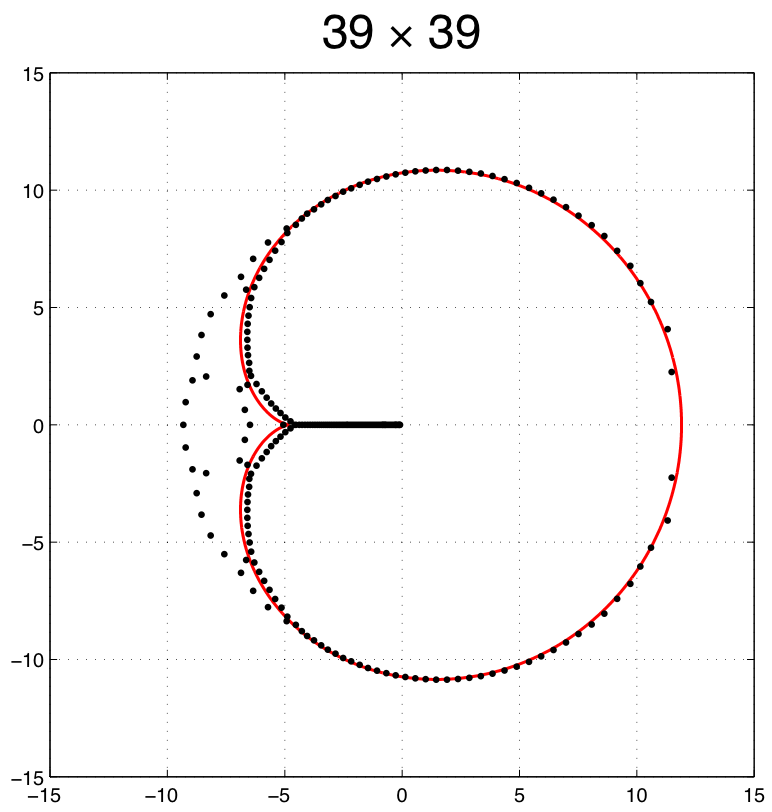}}

\end{picture}
}
\end{center}

\caption{\label{fig:cardfit} Fitting of the partition function zeros for cylindrical
  boundary conditions to the cardioid of (\ref{cardioid}) for the
  $33\times 33,~36\times 36$  and $39\times 39$ lattice.} 
\end{figure}

\begin{figure}[ht]

\begin{center}
\hspace{0cm} \mbox{
\begin{picture}(300,100)
\put(0,0){\includegraphics[width=5cm]{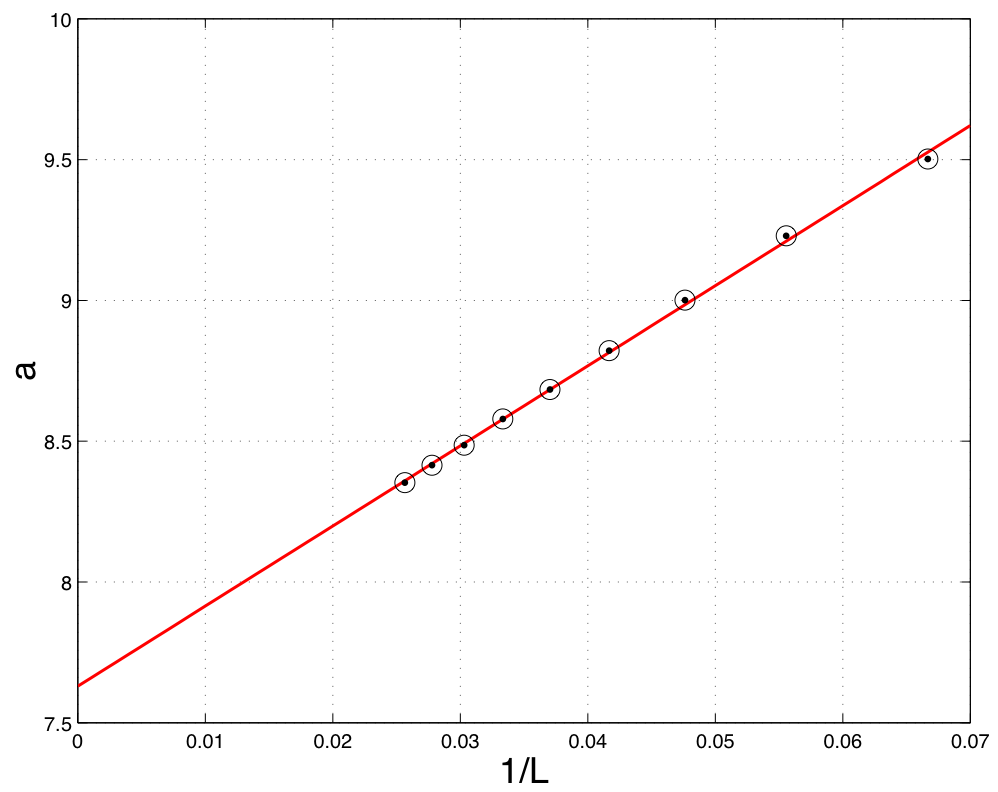}}
\put(150,0){\includegraphics[width=5cm]{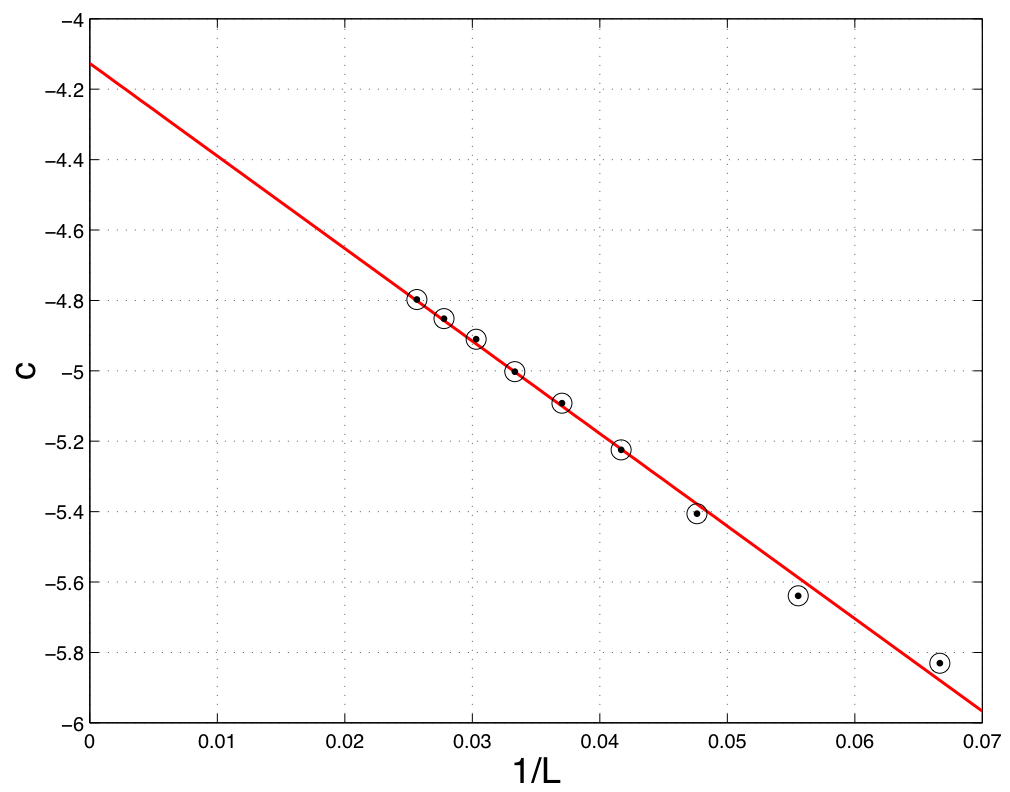}}

\end{picture}
}
\end{center}

\caption{\label{fig:cardlimit} The fitting parameters $a$ and $c$ for cylindrical
  boundary conditions of the cardioid (\ref{cardioid}) versus $L$ for
  the partition zeros of the $L\times L$ lattice.}
\end{figure}

Examples of the excellent fit by the cardioid curve (\ref{cardioid}) of 
the inner boundary of the partition function zero on cylindrical lattices 
referred to in section~\ref{sec:zerocbend} for the 
cases $33 \, \times \,33,~36\times 36$ and $39\,\times\, 39$ are plotted in
figure~\ref{fig:cardfit}. In figure~\ref{fig:cardlimit}  we plot the values of $a(L)$ and $c(L)$ the 
best fitted cardioid of (\ref{cardioid}) versus $1/L$ and observe that
they are remarkable well fitted by a straight line which extrapolates
as $L\,\rightarrow \,\infty$ to
\begin{eqnarray} 
\hspace{-0.9in}&& \qquad \quad  \quad 
a\,=\,\,7.6302 \,\cdots \quad \quad  c\,=\,\,-4.1268 \,\cdots 
\end{eqnarray}



\clearpage

\section{ \label{app:TM} Transfer-matrix algorithms }

\begin{figure}[htbp] 
   \centering
    \includegraphics[width=8cm]{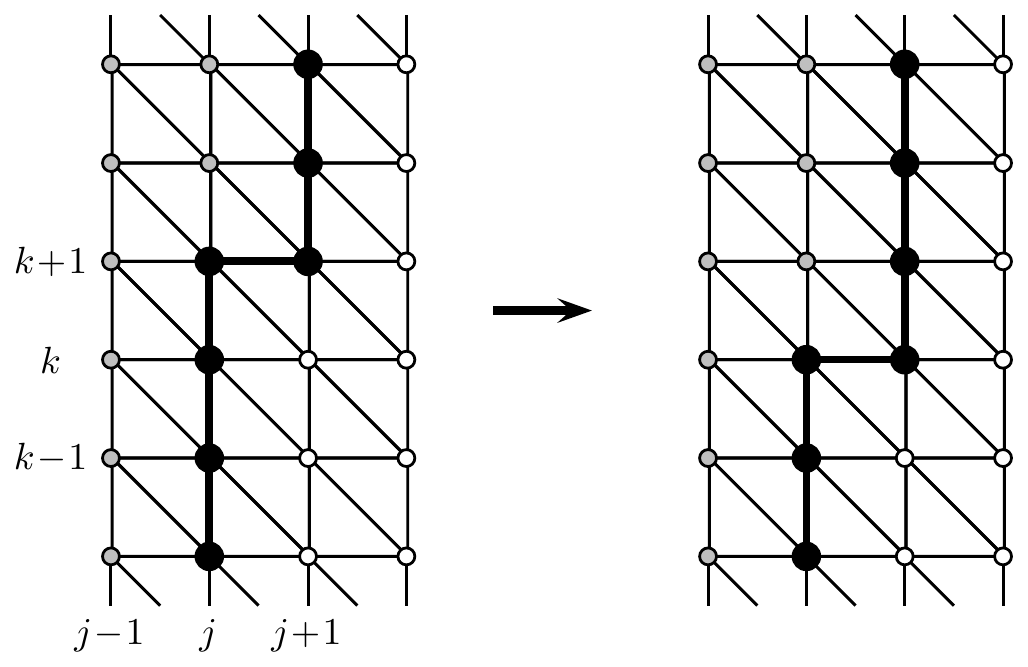} 
   \caption{ \label{fig:TTMGen} The movement of the transfer-matrix cut-line in the general case.}
  
\end{figure}

To calculate the partition functions $Z_{L_v,L_h}(z)$ we use what are known 
as `transfer matrix' techniques. These work by moving a cut-line through the lattice
and constructing a partial partition function for each possible configuration/state
of sites on the cut-line.  The most efficient way of calculating the partition function
is to build up the lattice site-by-site as illustrated in figure~\ref{fig:TTMGen}. Sites on the
cut-line are shown as large solid circles. Each of these sites can be either empty
or occupied. Because of the hard particle constraint, nearest neighbors cannot be
occupied simultaneously.  So any configuration can be mapped to a binary
integer in an obvious fashion with empty sites mapped to 0 and occupied sites to 1.  
The distinct configurations along the cut-line are thus circular $n$-bit strings with no repeated
$1$'s. Their number is given by the
Lucas numbers $L(n)$ (sequence A000204 in the OEIS \cite{oeis}), which have
the simple recurrence $L(n)=L(n-1)+L(n-2)$. In fact, $L(n)=(1+\sqrt{5})/2)^n+(1-\sqrt{5})/2)^n$,
so the number of allowed configurations has the growth constant $(1+\sqrt{5})/2=1.618\ldots$, and
are therefore exponentially rare among the integers. Hence as is standard in such a case
we use hash tables as our basic  data structure  to store and access the sparse array
representing the state space of the model.

For each configuration we maintain a partial partition function 
(sum over all states) for the lattice sites already visited with each occupied site given
weight $z$ and each empty site given weight 1. The shaded circles  in figure~\ref{fig:TTMGen} 
represent  sites already fully accounted for, that is, all possible occupancies have been summed over. 
The open circles are sites that are yet to be visited and hence are not yet accounted for.
The black circles are not yet fully accounted for since their possible `interactions' with the open sites
have not yet been included.  The movement of the cut-line in  figure~\ref{fig:TTMGen} 
consists of a move from the site at position $(j,k\!+\!1)$ to the `new' site at position $(j\!+\!1,k)$ with `interactions'
with the neighbor sites at $(j,k)$ and $(j+1,k+1)$. Notice that the update of the
partial partition functions does not depend on the state of any other sites on the cut-line. 
Formally we can view the update as a matrix multiplication  $\bf w= Tv$, mapping the vector
of partition functions $\bf v$ prior to the move to the vector of partition functions $\bf w$ after the move.
The great advantage of the site-by-site updating is that we need
not store the actual transfer matrix $\bf T$; it is given implicitly by a set of simple
updating rules depending only on the `local' configuration (states) of the sites
on the cut-line which are nearest neighbors to the new site.

We shall refer to the configuration of sites prior to the move as a `source' and denote its integer
representation with an $S$ while a configuration after the move is referred to as a `target' and
denoted with a $T$.  In an update we simply have to determine the allowed configuration of the 
`new' site, that is, whether the new site is occupied or empty. The `hard' constraint makes this very simple 
since the new site can be occupied only when  all neighbor sites in the source configuration 
are empty. Since each site can be either empty (0) or occupied or (1) the four sites around the face 
have 16 possible configurations but only 6 are allowed because of the hard constraint. 
The 3 sites along the left and top form the `local' source configuration  and there are 
5 (out of 8) allowed configurations. The local target configuration is given by the states of the 
3 sites along the bottom and right of the face (2 sites occur in both source and target). 
By writing down the 6 allowed local configurations one can easily deduce  the following simple updating rules

\begin{eqnarray}
Z(T_{000})&=&Z(S_{000})+Z(S_{010}), \nonumber \\
Z(T_{100})&=&Z(S_{100}),  \nonumber \\
Z(T_{010})&=&z\cdot Z(S_{000}), \label{updating} \\
Z(T_{001})&=&Z(S_{001}), \nonumber \\
Z(T_{101})&=&Z(S_{101}), \nonumber 
\end{eqnarray}
where the subscript triplets represent the states of the  `local' source sites at positions $(j,k)$, $(j,k+1)$,  $(j+1,k+1)$
and target sites at positions $(j,k)$, $(j+1,k)$,  $(j+1,k+1)$, respectively.

The transfer matrix algorithm described above takes care of the summation over sites in the
interior of the lattice. Special rules apply at the top and bottom of a column. When adding a site
at the top of a new column we include interactions between the site 
left of the new site and the site in the previous column on the bottom row (this interaction along 
a diagonal edge implements part of the periodic boundary condition in the $L_h$ direction). 
Finally after we have completed a new column  the site `left over' in the previous column is 
superfluous to requirements and we can `contract' the state space by summing over the states of this site.
The number of distinct configurations for a column of height $L_h$ sites is then $L(L_h+1)$ for a
partially completed column, and $L(L_h)$ when the column has been completed.

 The transfer matrix algorithm described above is the same whether used in the
 calculation of partition function zeroes or eigenvalue crossings. Below we
 briefly outline how it is used in the two cases.
 
 \subsection{ \label{app:TMzero} Partition function zeros} 
 
To calculate partition function zeros we need the exact partition
function on an $L_v\times L_h$ lattice. This is simply a polynomial
in $z$ of degree $L_v\cdot L_h/3$ with integer coefficients. 
So for each state along the cut-line the partial partition function
is maintained as an array of integers of size  $L_v\cdot L_h/3+1$. The
coefficients become very large and in order to deal with this the calculations were 
performed using modular arithmetic. So the calculation for a given size lattice was performed
several times  modulo different prime numbers with the full integer coefficients  
reconstructed from the calculated remainders using the Chinese remainder theorem. 
Utilizing the standard 32-bit integers  we used primes of the form $p_{i} = 2^{30}-d_{i}$,
that is we used the set of largest primes smaller than $2^{30}$. 
Depending on the $L_v$ and $L_h$ the number of primes required to
reconstruct the exact integer coefficients can exceed 100. 
The zeros of the partition function can then be calculated numerically 
(to any desired accuracy)  using  root finders such as {\tt MPSolve} \cite{Bini00} 
or {\tt Eigensolve} \cite{Fortune02}. We used  {\tt MPSolve}  with a few
calculations checked by using {\tt Eigensolve}.

The transfer-matrix algorithm can readily be parallelised.  
One of the main ways of achieving a good parallel algorithm using 
data decomposition is to identify an invariant under the
operation of the updating rules. That is, we seek to find some property
of the configurations along the cut-line which does not alter in a single iteration.
As mentioned above only the `new' site can change occupation status. 
Thus,  any site not directly involved in the update cannot change from being 
empty to being occupied and vice versa. This invariant
allows us to parallelise the algorithm in such a way
that we can do the calculation completely independently on each
processor with just two redistributions of the 
data set each time an extra column is added to the lattice.   
This method for achieving a parallel algorithm has been used
extensively for other combinatorial problems and the interested reader can 
look at \cite{Jensen03} or \cite[Ch 7]{PolygonBook} for details.

Cylindrical boundary conditions are simply implemented by starting the transfer matrix calculation
with the all empty state having weight one (all other states having weight zero), iterating the
algorithm to add $L_h$ columns and then summing over all states. 

Toroidal boundary conditions are fairly easy to implement but they are computationally
expensive. The problem is that in order to include the interactions between sites in the first
and last columns we have to `remember' the state of the first column. Here we did
this by simply specifying the initial state of the first column $S_{\rm I}$, starting
with the initial weights $Z(S)=0$ when $S\neq S_{\rm I}$ and $Z(S_{\rm I})=z^m$,
where $m$ is number of occupied sites in $S_I$. For
each value of  $S_{\rm I}$ we then perform the transfer matrix calculation as described
above until the final column has been completed. Finally we put in the interactions between
the occupation numbers in the last column with state $S$ and those in the first column, sum over all
states $S$, and repeat for all   $S_{\rm I}$. 
The saving grace is that one does not have to do this calculation for all values of   $S_{\rm I}$. 
Indeed, any $S_{\rm I}$ related by translational and reflection symmetry give rise to the same result.  
In table~\ref{tab:PBnumstates} we have listed the number of distinct initial states $N_I$ one needs to
consider in a calculation of the partition function with toroidal boundary conditions on 
a lattice of  width $L_h$. The numbers $N_I$ are given by sequence A129526 in the OEIS \cite{oeis}.
Note that for large $L_h$ states which are invariant under the generators of the dihedral group
$D_{L_h}$ are exponentially rare. Therefore $L(L_h)/N_I \sim 2 L_h$ for $L_h \gg 1$, and this is
nicely brought out by the entries in table~\ref{tab:PBnumstates}.  
Naturally the calculations for different initial states can be done completely independently 
making it trivial to parallelise over $S_{\rm I}$  (one can also fairly easily combine this  with
the parallel algorithm over state space should this be required).

\begin{table}[htdp]
\begin{center}
\begin{tabular}{rrrcrrrc} \hline  \hline 
 $L_h$ &\quad  $N_I$   & \quad $L(L_h)$ &\quad   $L(L_h)/N_I$ & \qquad   $L_h$  
 &\qquad  $N_I$    & \quad $L(L_h)$ &\quad   $L(L_h)/N_I$  \\ \hline 
           3 &  2 &  4 &  2.00     & 6  &   5  & 18 & 3.60 \\ 
           9 &  9 &  76 &  8.44     & 12  &   26  & 18 & 12.38 \\ 
           15 &  64 &  1364 &  21.31     & 18  &   209  & 5778 & 27.64 \\ 
           21 &  657 &  24476 &  37.25     & 24 &   2359  & 103682 & 43.95 \\ 
           27 &  8442 &  439204 &  52.02     & 30 &   31836  & 1860498 & 58.44 \\ \hline \hline 
\end{tabular}
\caption{\label{tab:PBnumstates}
The number $N_I$ of distinct initial states required to calculate the partition function
with toroidal boundary conditions compared to the total number of states given by the
Lucas numbers.}
\end{center}

\end{table}%
 
The bulk of the large scale calculations for this part of the project were performed on the cluster 
of the NCI National Facility at ANU. The NCI peak facility is a Sun Constellation Cluster with 1492 nodes 
in Sun X6275 blades, each containing two quad-core 2.93GHz Intel Nehalem CPUs with 
most nodes having 3GB of memory per core (24GB per node). 
The largest size calculation we performed was for cylindrical boundaries where
we went up to $39\times 39$. This required the use of 256 processors (cores to be precise)  
taking around 140 CPU hours per prime with the calculation being repeated for 25 primes.
The largest calculation for toroidal boundaries was the $27\times 27$ lattice. This is
sufficiently small memory wise to fit on a single core so we only used a parallelisation 
over initial states. The calculation 
took around 550 CPU hours per prime with the calculation repeated for 14 primes.

 \subsection{ \label{app:TMev} Transfer matrix eigenvalues} 
 
The equimodular curves on $L_h\times \infty$ strips where eigenvalues of largest
modulus cross can be obtained from numerical studies using the
transfer matrix algorithm outlined above. As we have seen,
the dimension ${\rm dim}\, {\bf T}$ of the transfer matrix $\bf T$ for hard hexagons confined to a strip
of height $L_h$ grows exponentially fast with $L_h$.
Hence  the calculation of the eigenvalues of  $\bf T$ 
and the resulting equimodular curves quickly becomes a very demanding task.
We have however seen that the updating rules (\ref{updating}) do not require us to 
manipulate or store all the $({\rm dim}\, {\bf T})^2$ entries of $\bf T$, but rather
operate on two vectors of size ${\rm dim}\, {\bf T}$, namely $\bf v$ and $\bf T v$,
representing the set of conditional probabilities before and after a move of the cut-line.
In other words, the trick of adding the sites to the system one at a time has produced
a sparse-matrix factorization of $\bf T$.

It is possible to take further advantage of this gain to extract also the leading eigenvalues
of $\bf T$. Namely, we use a set of iterative diagonalisation methods in which the
object being manipulated is not $\bf T$ itself but rather its repeated action on a
suitable set of vectors. An iterative scheme that works well even in the presence of
complex and degenerate eigenvalues is known as Arnoldi's method \cite{Arnoldi51}.
This forms part of a class of algorithms  called Krylov subspace projection methods \cite{Saad92}.
These methods take full advantage of the intricate structure of the sequence of vectors $\bf T^n v$ naturally
produced by the power method. If one hopes to obtain additional information through various linear 
combinations of the power sequence, it is natural to formally consider the Krylov subspace
$$ \mathcal{K}_n({\bf T}, {\bf v}) = {\rm Span}\{{\bf v}, {\bf T v},  {\bf T^2 v}, ... ,  {\bf T^{n-1} v}\} $$
and to attempt to formulate the best possible approximations to eigenvectors from this subspace.
We make use of the public domain software package 
ARPACK \cite{ARPACK} implementing Arnoldi's method with suitable subtle stopping criteria.
The ARPACK package allows one to extract eigenvalues (and eigenvectors) based on various
criteria, including the one relevant to our calculations, namely the eigenvalues of largest modulus. 

The problem specific input for this type of calculation only consists in a user supplied subroutine
providing the action of $\bf T$ on an arbitrary complex vector $\bf v$. In our case this amounts to
iterating the update rules (\ref{updating}) until a complete column has been added to the lattice.
In particular, the sparse-matrix factorization and the subtleties having to do with the `inflation'
of the state space before the addition of the first site in a new row, as well as the `contraction'
after the addition of the last site in a completed row, are all hidden inside this subroutine and
not visible to Arnoldi's method. The iterations of $\bf T$ are thus computed for a fixed 
complex value of the fugacity $z$ until the ARPACK routines have converged.

Very briefly, we trace the equimodular curve
as follows. First we find a point on the equimodular curve; we can choose a point
on the negative real axis, say $z_0=-1$, that we know is on the curve.
To find a new point $z$ on the curve we start at the previous point and look
for a new point on a circle of radius $\epsilon$ (in general we use $\epsilon = 10^{-2}$) at an angle $\theta_0$ from
the previous point. In general this trial point will not lie on the equimodular curve.
Our algorithm then finds a new point by using the Newton-Raphson method to
converge in the angle $\theta$ towards a zero in the distance between 
leading eigenvalues. The pair $(z,\theta)$ is then used as the starting values
$(z_0,\theta_0)$ for a new iteration of the search algorithm.
This procedure in then iterated until the equimodular
curve has been traced. Points where the curve branches are detected by 
noting that the third leading eigenvalue becomes equal in modulus to the
leading eigenvalue. End points are detected by noting that the procedure
cannot find a new point (in fact it turns around and converges towards a point on the part of the curve
already traced). Many aspects of this search algorithm involve subtleties, in particular
automatizing the procedure in the case where the equimodular curve has a complicated
topology with many branchings; this will be described fully in a separate publication \cite{JJinprep}.

As for the partition function zeroes, we are interested in tracing the equimodular curves for both
toroidal and cylindrical boundary conditions. Since in both cases we use the same transfer
matrix $\bf T$ (i.e., with periodic boundary conditions in the $L_h$ direction) it might seem that
the curves would be identical. This is not the case. Indeed, in the cylindrical case the initial
condition imposed on the first column of the lattice is that all sites in the preceding column are empty.
In particular, this initial state is translational invariant and thus has momentum $P=0$. This momentum
constraint can be imposed by rewriting $\bf T$ in the translational and reflection symmetric subspace
of dimension $N_I$. Once again, an appropriate `inflation' and `contraction' of the state space has to
be performed at the beginning and the end of the user supplied subroutine, as the kink on the cut-line
describing the intermediate states breaks the dihedral symmetries explicitly. But since these intermediate
steps are hidden from Arnoldi's method, the end result amounts to diagonalising a transfer matrix of
smaller dimension, ${\rm dim}\, {\bf T} = N_I$. Meanwhile, the equimodular curve for toroidal boundary
conditions is obtained by diagonalising the original transfer matrix without the $P=0$ constraint,
i.e., with dimension ${\rm dim}\, {\bf T} = L(L_h)$.

The memory requirements of the algorithm up to the largest size $L_h=30$ that we attempted is quite
modest and the calculation can be performed on a  basic desktop or laptop computer. As an
example the calculation of the equimodular curve for  $L_h=30$ with cylindrical boundary conditions
took about  10 days on a MacBook Pro with a quad core I7 2.3GHZ processor.

\section{ \label{app:FSS}  Finite-size scaling analysis of  $z_c(L)$ and $z_d(L)$}

According to the theory of finite-size scaling (FSS) \cite{Cardybook},
the free energy per site corresponding to the $j$-th eigenvalue of
the transfer matrix has the scaling form
\begin{equation}
 \frac{1}{L} f_j \left( |z-z_c| L^y, u L^{-|y'|} \right) \,,
\end{equation}
where $z_c$ is the critical point, $y$ is the leading relevant eigenvalue under
the renormalization group (RG), and $u$ is the coupling to an RG irrelevant
operator with eigenvalue $y' < 0$. If more than one RG irrelevant coupling is
present there will be further arguments to the function, which we
here omit for clarity. The equimodularity condition $|f_1| = |f_2|$ can
obviously be written in the same scaling form as can the partition function zeros.
Moreover, FSS assumes that the functions $f_j$ are analytic in their arguments
for $z \neq z_c$, which implies at leading order that
\begin{equation}
 |z-z_c| = A L^{-y} + B u L^{-y-|y'|} + \ldots \,,
 \label{FSS_development}
\end{equation}
where $A$ and $B$ are non-universal constants. To higher orders, the terms
appearing on the right-hand side involve powers of $L^{-1}$ that can be
any non-zero linear combination of $y$ and $|y'|$ with non-negative
integer coefficients. There is obviously no guarantee that all such terms will
appear, since some of the multiplying constants ($A,B,\ldots$) may be zero.

When it is known that $z \to z_c$ as $L \to \infty$, with $z_c$ real, one can similarly analyze
distances other than $|z-z_c|$ to the critical point that vanish linearly with $z-z_c$.
Examples include $||z|-z_c|$,  ${\rm Re}(z)-z_c$, ${\rm Im}(z)$, and ${\rm Arg}(z)$.%
\footnote{Obviously we here exclude cases where the variable is identically zero,
such as when ${\rm Im}(z)=0$, or when $||z|-z_c|=0$ because of a circle theorem.}
According to the general principles of FSS \cite{Cardybook} these variables can be
developed on $|z-z_c|$ and the irrelevant RG couplings, and (\ref{FSS_development})
will follow, albeit necessarily with different values of the non-universal constants ($A,B,\ldots$).

\medskip

The critical point $z_c > 0$ in the hard hexagon model is known to be in
the same universality class as the three-state ferromagnetic Potts
model \cite{FYWu82}. The energy operator of the latter \cite{Dotsenko84}
provides the RG eigenvalue $y = 2 - 2 h_{2,1} = 6/5$, where we have
used the Kac table notation $h_{r,s}$, familiar from conformal field theory
(CFT), for the conformal weight of a primary operator $\phi_{(r,s)}$.
Subdominant energy operators, $\phi_{(3,1)}$ and $\phi_{(4,1)}$, follow
from CFT fusion rules and lead to RG eigenvalues $y' = -4/5$ and
$y''=-4$ respectively. Our numerical analysis of $|z_c(L)| - z_c$ for
$L$ up to $39$ (see table~\ref{tab:zeroendpoints}) gives good evidence for the FSS form
\begin{equation} 
 |z_c(L)|-z_c = a_0 L^{-6/5} + a_1 L^{-2} + a_2 L^{-14/5} + \ldots \,.
\end{equation}
\label{eq:zcasymp}
The powers of $L^{-1}$ appearing on the right-hand side can be
identified with $y$, $y+|y'|$ and $y+2|y'|$. This is compatible with
the above general result; note however that the power $2y = 12/5$,
which is a priori possible, is not observed numerically.

The CFT of the Lee-Yang point $z_d < 0$ is much simpler \cite{cardy},
since there is only one non-trivial primary operator $\phi_{(2,1)}$. It provides the RG
eigenvalue $y = 12/5$. Our numerical analysis of $|z_d(L) - z_d|$ 
(see table~\ref{tab:zeroendpoints}) gives strong evidence for the FSS form
\begin{equation} \label{eq:zdasymp}
 |z_d(L)-z_d| = b_0 L^{-12/5} + b_1 L^{-17/5} + b_2 L^{-22/5} + \ldots \,.
\end{equation}
The powers of $L^{-1}$ on the right-hand side can be identified with
$y$, $y+1$ and $y+2$.  The integer shifts in  (\ref{eq:zdasymp}) can be related 
to descendent operators in the CFT, since $|y'|$ is a positive integer for descendents of
the identity operator; note that some descendents are ruled out by symmetry arguments,
as in section~\ref{sec:zerodens}. Another source of corrections in powers of $L^{-1}$ is that the data
 of table~\ref{tab:zeroendpoints} are computed for 
$L \times L$ systems with cylindrical boundary conditions. Indeed, while the length $L$ along 
the periodic direction is unambiguous, the one along the free direction should possibly be interpreted 
as $L+a$ in the continuum limit, where $a$ is a constant of order unity. 
In any case, it is remarkable that $y+1 = 11/5$ does not occur in (\ref{eq:zdasymp}).
 
\medskip
 
To probe the finite-size scaling form, i.e., determine which terms actually occur in the 
asymptotic expansion,  we carry out a careful numerical analysis of how the endpoints 
$z_d(L)$ and $z_c(L)$ approach $z_d$ and $z_c$ in the thermodynamic limit. 
Since  our data for end-point positions is most extensive in the case of partition function
zeros with cylindrical boundary conditions we analyse the data of table~\ref{tab:zeroendpoints}.  
We also tried to analyse the data in table~\ref{tab:endpoints} obtained from equimodular calculations,
but we found that this data set suffers from numerical instability presumably because our
determination of the end-point position is not sufficiently accurate. Note that
the data in table~\ref{tab:zeroendpoints} can be calculated to any desired numerical 
accuracy since it is obtained from the zeros of polynomials.
Obviously the data for $z_d(L)$  is much closer to the thermodynamic limit $z_d$ than
is the corresponding data for $z_c(L)$ so it is no surprise that the analysis of $z_d(L)$
is `cleaner' than that for  $z_c(L)$ and hence we start our exposition with the former.

\begin{figure}[htbp] 
   \centering
   \begin{picture}(260,130)
   \includegraphics[height=4.5cm]{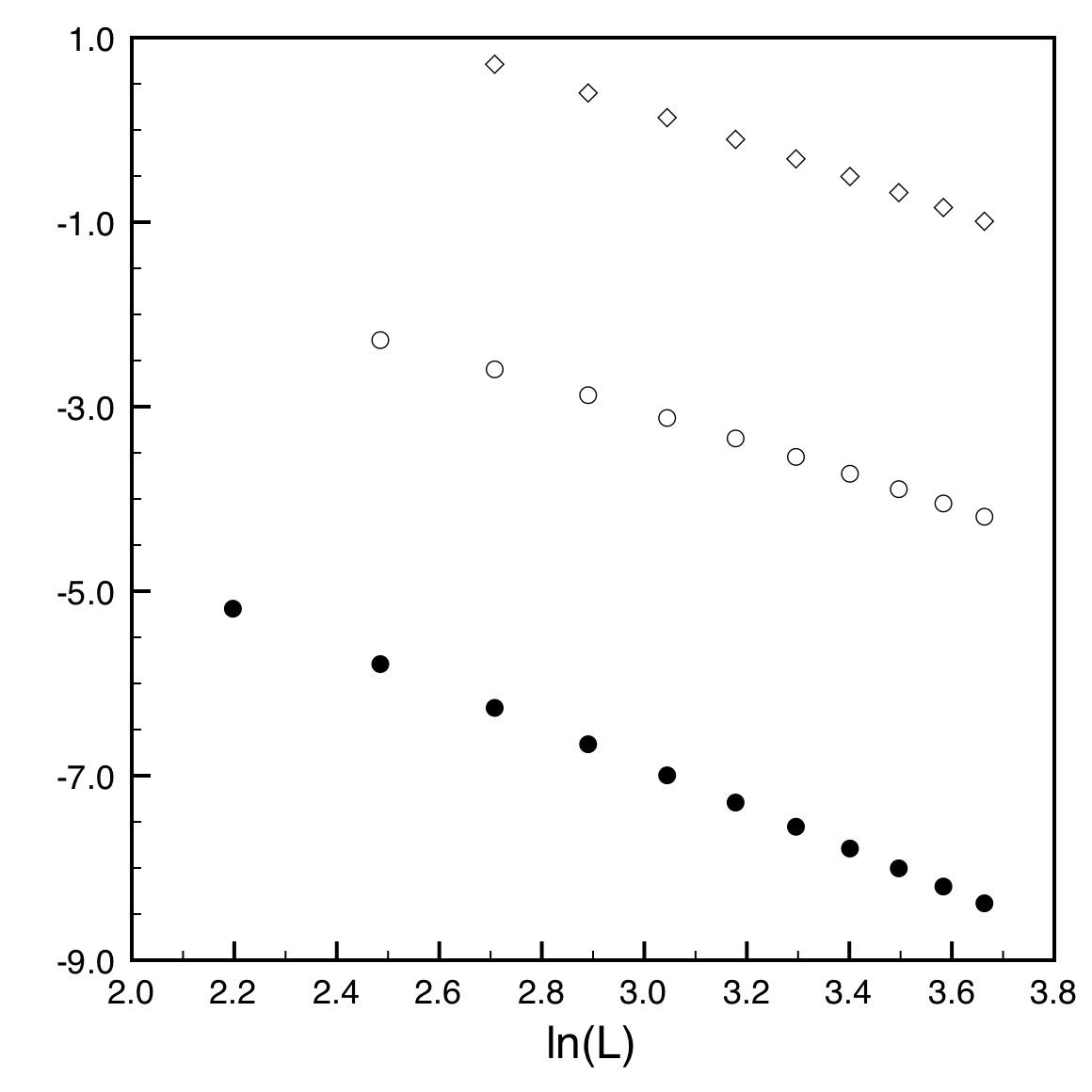} 
   \includegraphics[height=4.5cm]{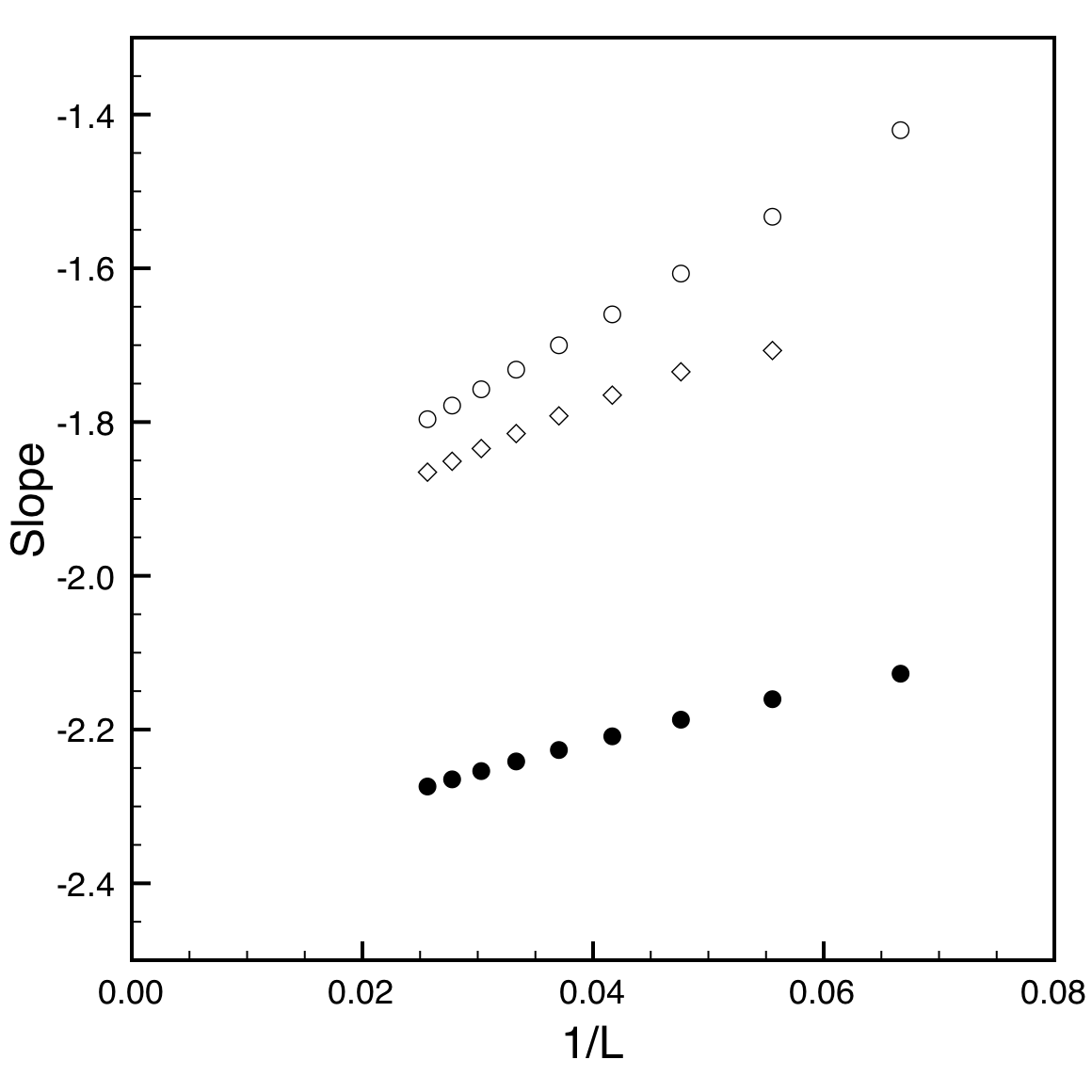} 
   \end{picture}
   \caption{The left panel are log-log plots of the data for $ |z_d(L)-z_d| $ (filled circles),
   $d(L)$ (open circles) and $L^2d(L)-(L-3)^2d(L-3)$ (diamonds). In the right panel we
   plot the corresponding local slopes versus $1/L$.
 }
   \label{fig:zdanalysis}
\end{figure}

Firstly, plotting $\ln |z_d(L)-z_d|$ versus $\ln L$ confirms a power-law relationship
(see left panel of figure~\ref{fig:zdanalysis}). To estimate
the exponent we take a pair of points at $L$ and $L-3$, calculate the resulting 
slope of a straight line through the data-points, and in figure~\ref{fig:zdanalysis} we 
plot the slope versus $1/L$. Clearly, the slope can be extrapolated to the predicted value, $12/5$, for the exponent.
We next look for sub-dominant exponents. Accepting the 12/5 exponent as exact we
form the scaled sequence, $s(L)= L^{12/5} |z_d(L)-z_d| \simeq a+b/L^{\alpha}$, and look
at the sequence of differences, $d(L)=s(L)-s(L-3) \propto 1/L^{\alpha+1}$, thus
eliminating the constant term. As before we calculate the slope of $\ln d(L)$ versus $\ln L$ 
and plot against $1/L$. From figure~\ref{fig:zdanalysis} the slope is seem to extrapolate to a value of $-2$,
so $\alpha =1$ and hence the sub-dominant exponent is $17/5$. We then repeat the analysis starting with
the $d(L)$ sequence which we scale by $L^2$. The estimates for the local slopes are shown
in  figure~\ref{fig:zdanalysis} and are again consistent with a slope of $-2$, indicating that the third exponent in the
asymptotic expansion is $22/5$.

\begin{figure}[htbp] 
   \centering
  \begin{picture}(260,260)
  \put(0,130){ \includegraphics[height=4.5cm]{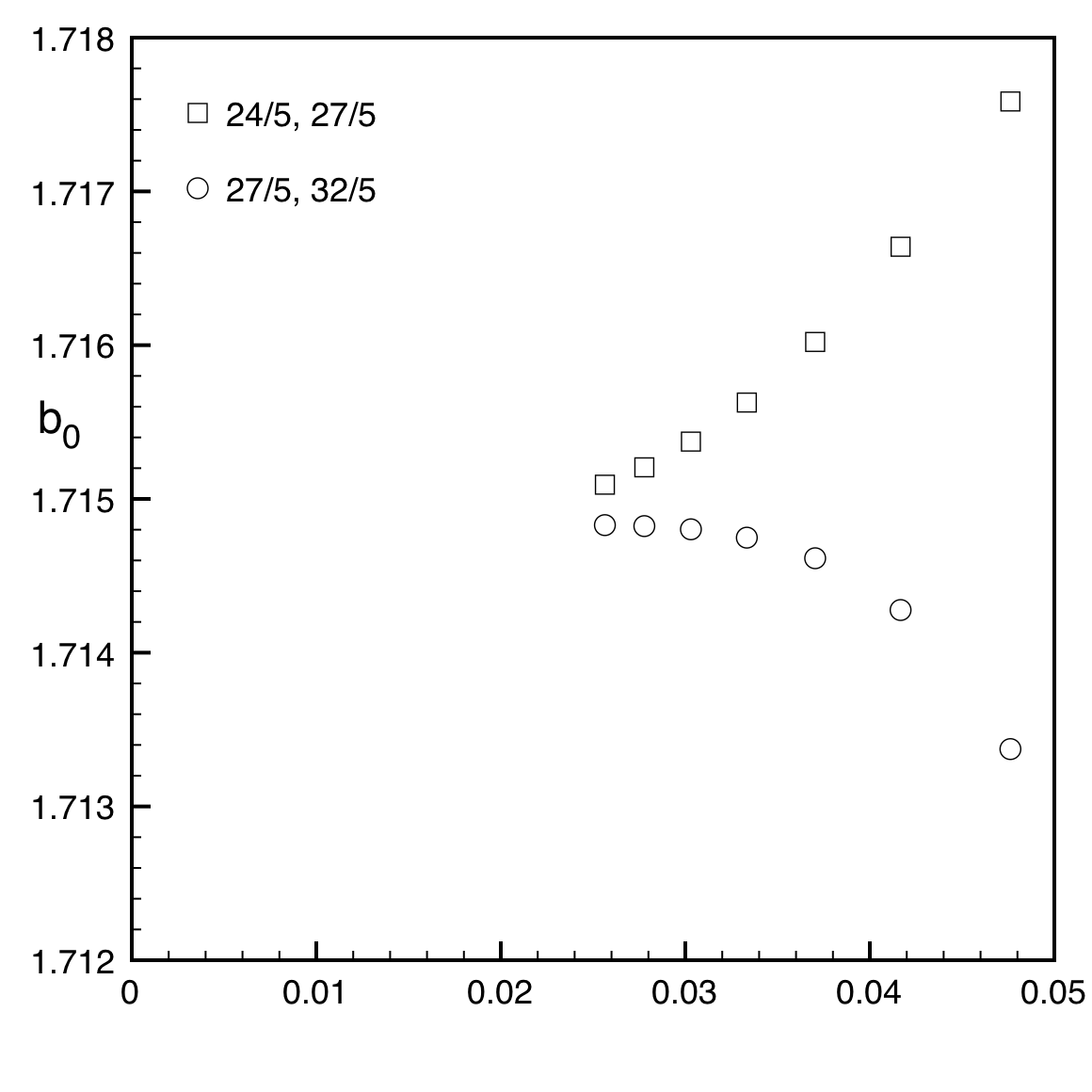}}
  \put(150,130){ \includegraphics[height=4.5cm]{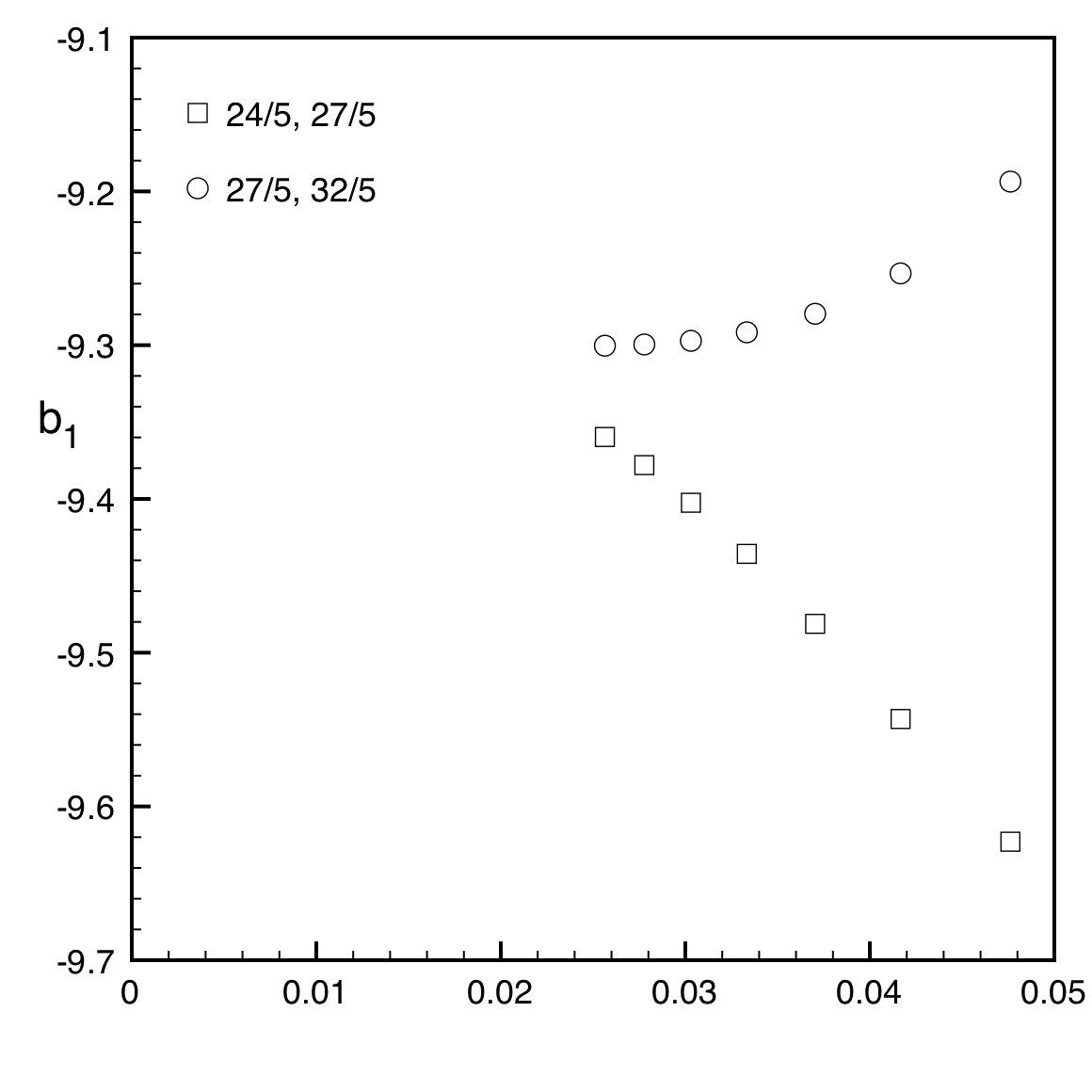}}
  \put(0,0){ \includegraphics[height=4.5cm]{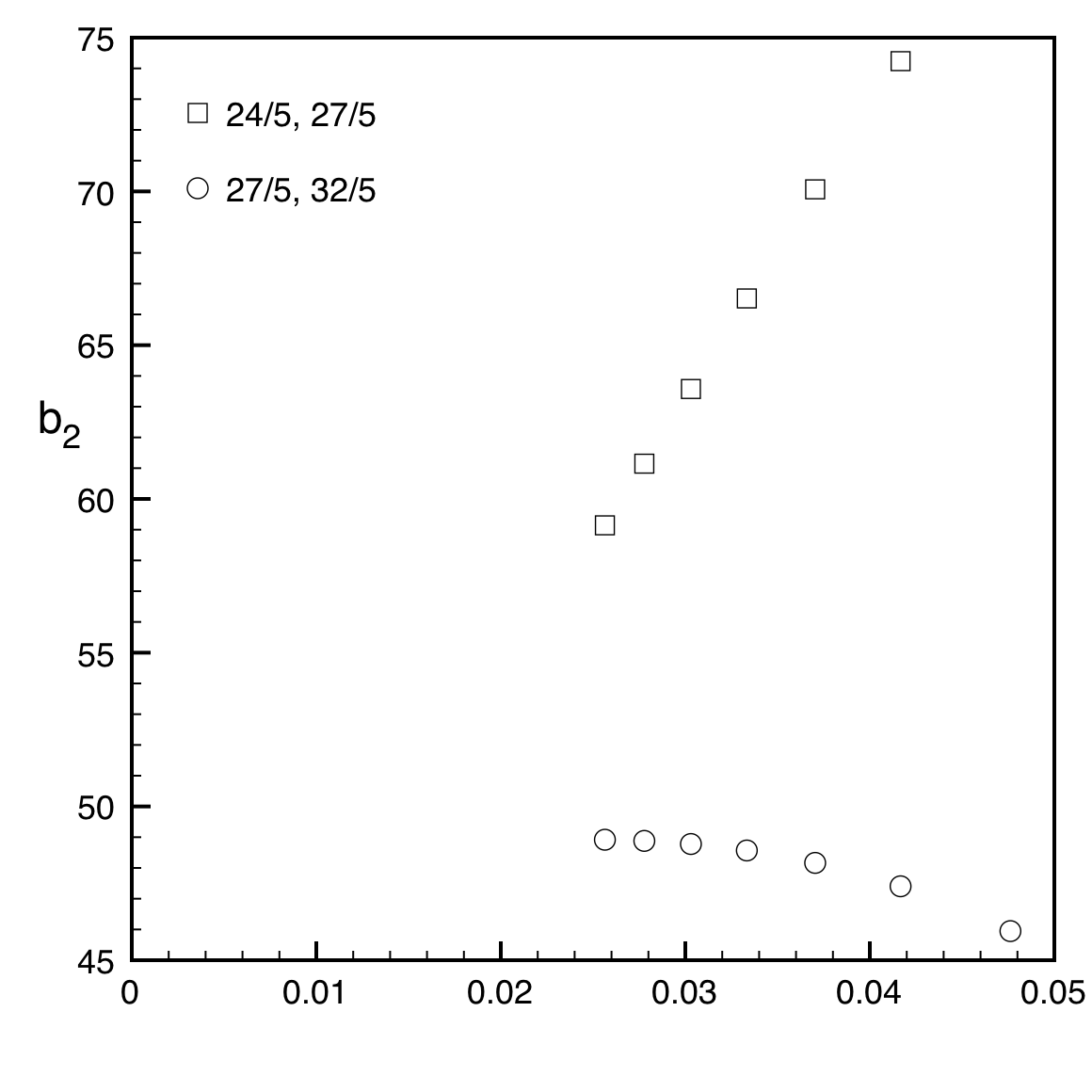}}
  \put(150,0){ \includegraphics[height=4.5cm]{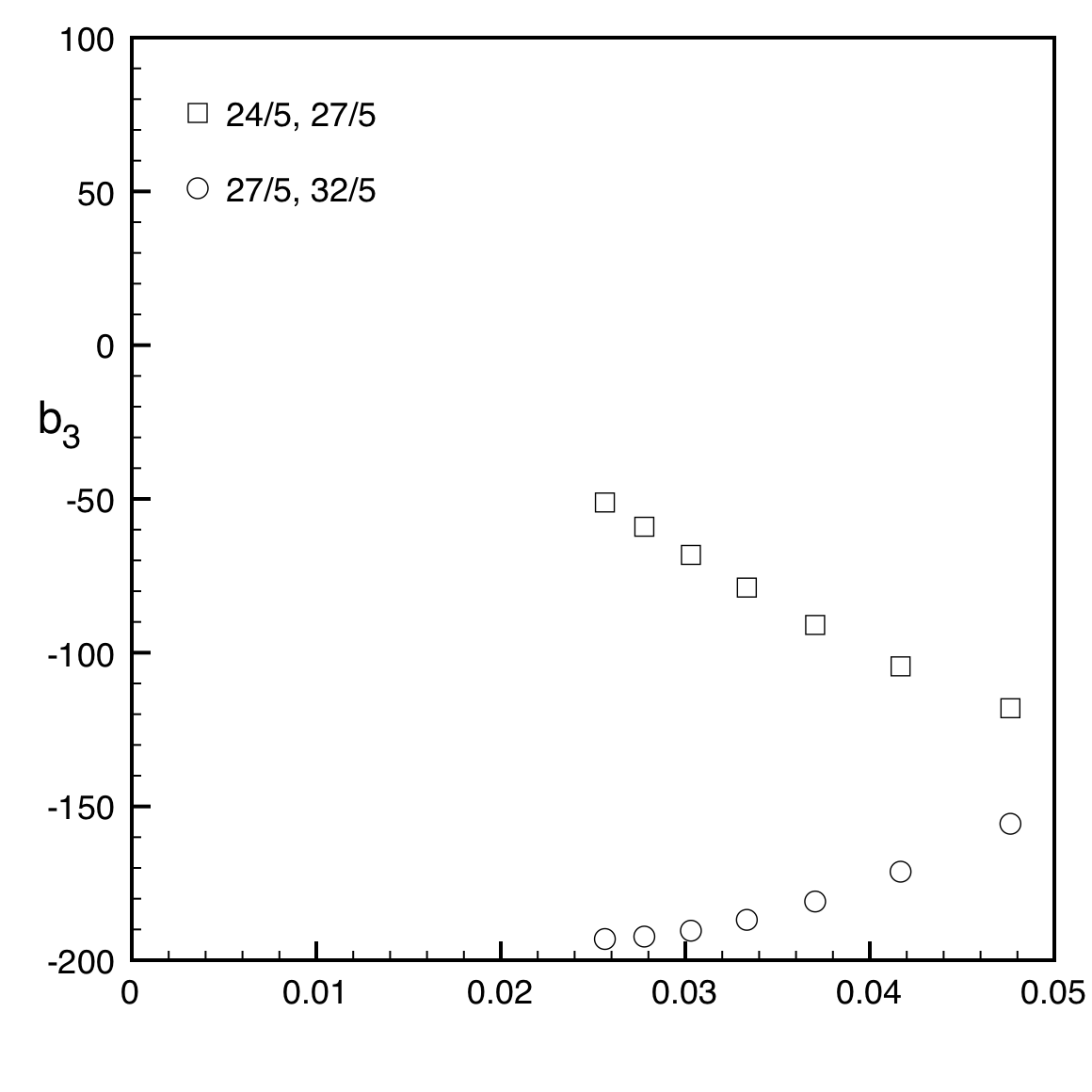}}
   \end{picture}
   \caption{Amplitude estimates versus $1/L$ when fitting to a five-term asymptotic form
   akin to (\ref{eq:zdasymp}), but with two additional exponents as indicated on the plots.}
   \label{fig:zd_5ampl}
\end{figure}

\begin{figure}[htbp] 
   \centering
  \begin{picture}(130,130)
 \includegraphics[height=4.5cm]{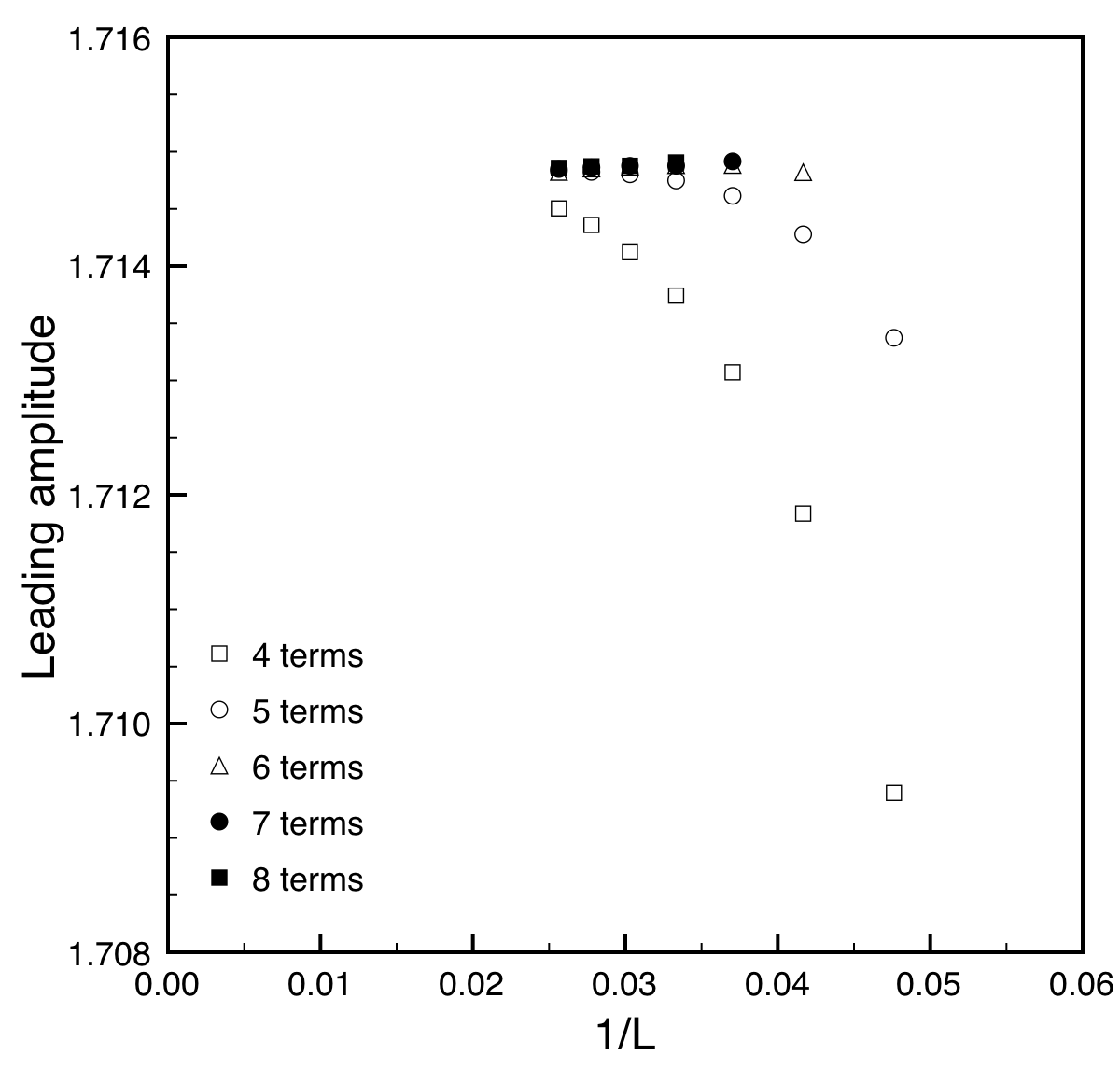} 
   \end{picture}
   \caption{Estimates for the leading amplitude $b_0$, plotted versus $1/L$ in the
   asymptotic expansion (\ref{eq:zdasymp}) when truncating after 4 to 8 terms and
   using only exponents $24/5+k$. }
   \label{fig:zdampl}
\end{figure}

From the above analysis we conclude that the correct asymptotic form is
(\ref{eq:zdasymp}).
 It is tempting to conjecture that the sequence of integer spaced
corrections $L^{-12/5-k}$ will continue indefinitely. However, we cannot completely rule
out the presence of  extra terms such as $L^{-n \cdot 12/5}$ for $n\geq 2$. 
This is borne out by a further analysis using five terms in the asymptotic expansion
namely the three exponents firmly established above and a further two terms with exponents
$2y=24/5, \, y+3=27/5$, and $y+3=27/5, \, y+4=32/5$, respectively. The resulting
amplitude estimates are shown in figure~\ref{fig:zd_5ampl}.
Clearly, the fits using only terms of the form $y+k$ display much less variation against $1/L$
and this could be an indication that only these types of terms are present in the asymptotic expansion.
However, the amplitude $b_3$ when fitting using an exponent $24/5$ does not appear to vanish
and hence we are not willing to claim with certainty that this term is absent. 
If we assume only terms of the form $y+k$, we can obtain refined amplitude estimates by truncating the expansion
after a fixed number of terms and fitting using sub-sequences of consecutive data points.
Results for the leading amplitude $b_0$ are displayed in  figure~\ref{fig:zdampl}.
Note that the estimates are quite accurate and that as more terms from the
asymptotic expansion are included the estimates have less variation.   
We estimate that
\begin{equation}\label{eq:zdampl}
b_0=1.7147(1), \quad b_1=-9.30(2), \quad b_2=48(2), \quad b_3=-180(30).
\end{equation}

\begin{figure}[htbp] 
   \centering
   \begin{picture}(260,130)
   \includegraphics[height=4.5cm]{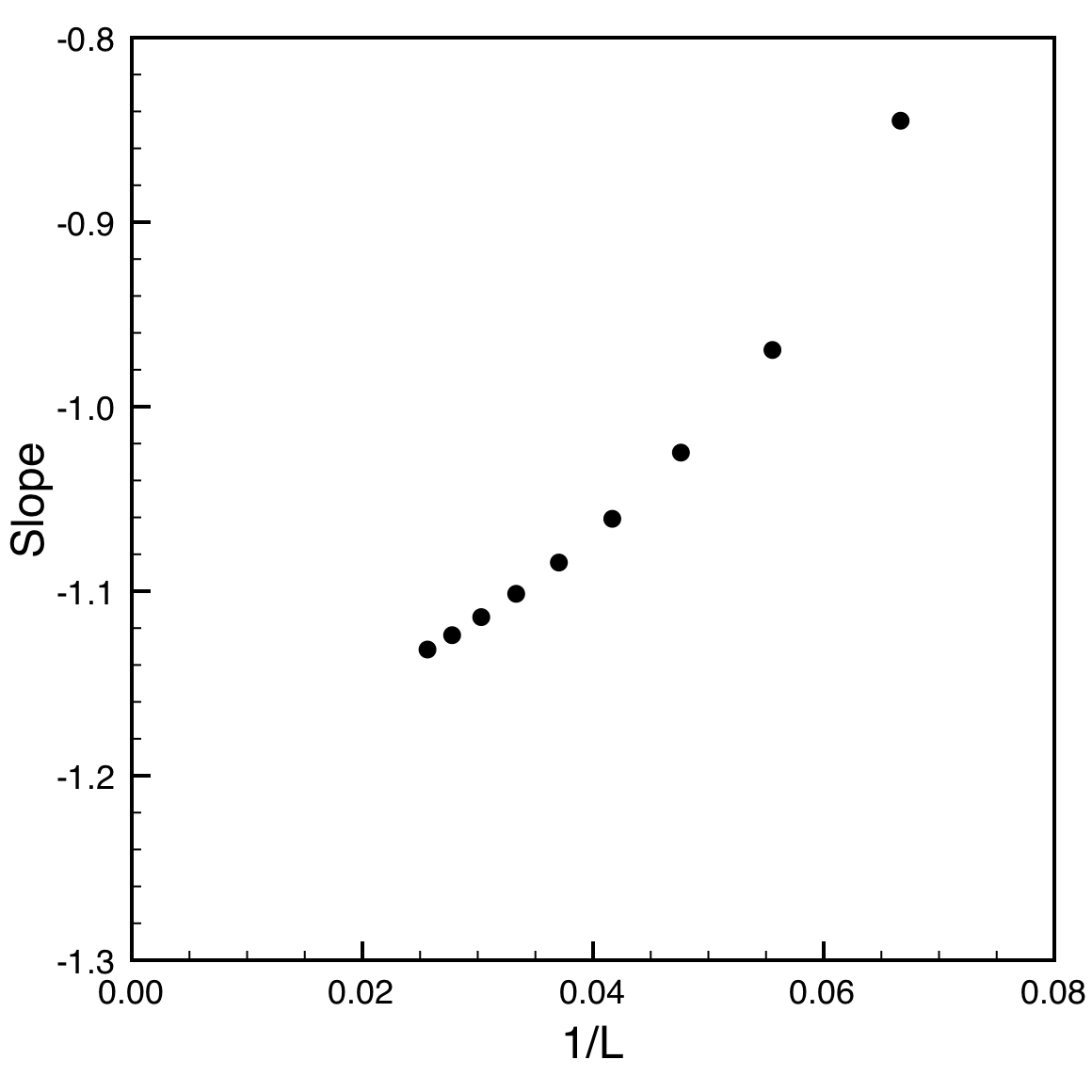} 
   \includegraphics[height=4.5cm]{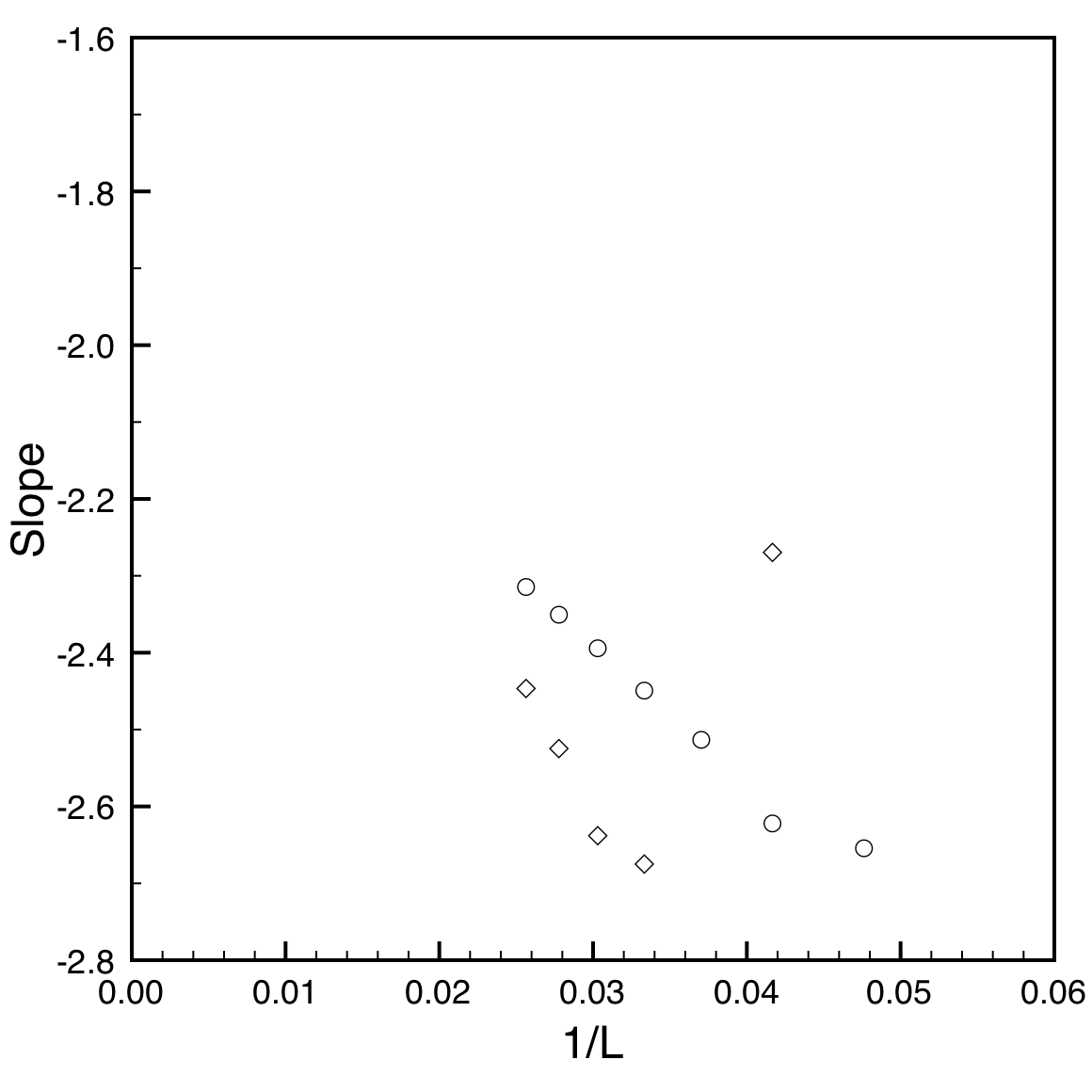} 
   \end{picture}
   \caption{Local slopes versus $1/L$ for  $ ||z_c(L)|-z_c|$ in the left panel and in the right panel
   the sequences $d(L)$ (open circles) and $L^{9/5}d(L)-(L-3)^{9/5}d(L-3)$ (diamonds).}
   \label{fig:zcanalysis}
\end{figure}

We now turn to $z_c(L)$ where we start by analysing the data
for $|z_c(L)|-z_c$.  Note that the modulus $|z_c(L)|$  can be viewed as a crude 
approximation to where the zeros intercept the real axis since it amounts to saying
that the zeros approach the real axis along a circle.   Many other measures of the
distance/approach to $z_c$ could be used but this one happens to be particularly
well-behaved so we start our exposition with this quantity.
As above we first look at the local log-log slope for
this data shown in the left panel of figure~\ref{fig:zcanalysis}. In this case
the data displays pronounced curvature but nevertheless it seems reasonable
that the slope can be extrapolated to the predicted value $-6/5$.
We next look for sub-dominant exponents. Accepting the 6/5 exponent as exact we
form the scaled sequence, $s(L)= L^{6/5} (|z_c(L)|-z_c) \simeq a+b/L^{\alpha}$.
We again look at the sequence $d(L)$ of differences and plot the local slopes in
the right panel of figure~\ref{fig:zcanalysis} using open circles.  In this case
the results are not as clear cut.  The data can be extrapolated to a value 
$>-2$ and it is consistent with the predicted exponent $y+|y'|=2$,
which would yield a slope of $-1.8$. We then repeated the analysis 
scaling $d(L)$ by $L^{9/5}$ and looking at the differences. The
local slopes are shown as diamonds in the right panel of figure~\ref{fig:zcanalysis}.
Clearly no meaningful extrapolation can be performed on this data
other than to say that a value of  $-1.8$ cannot be ruled out.

\begin{figure}[htbp] 
   \centering
  \begin{picture}(400,120)
    \includegraphics[height=4.5cm]{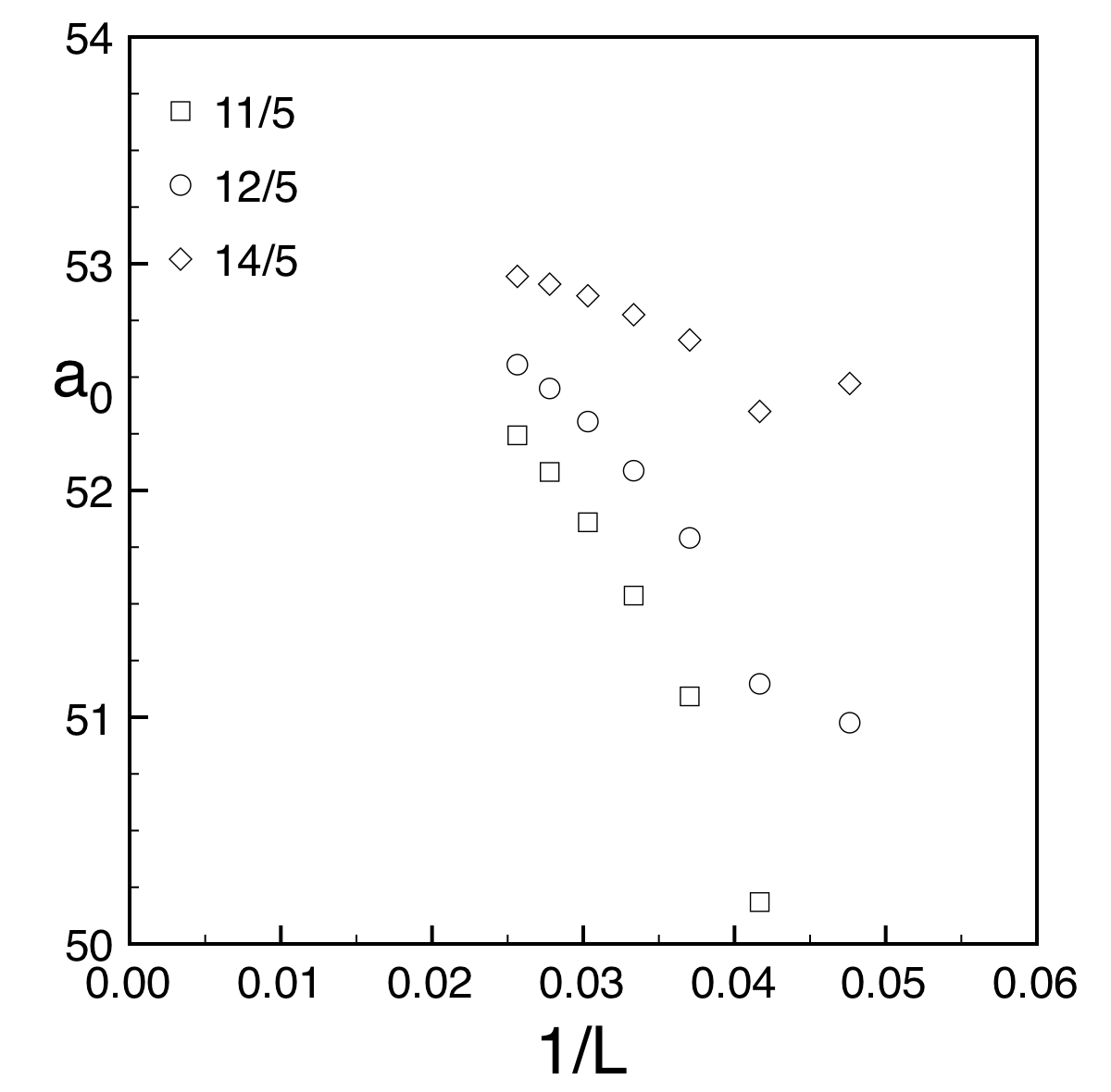} 
    \includegraphics[height=4.5cm]{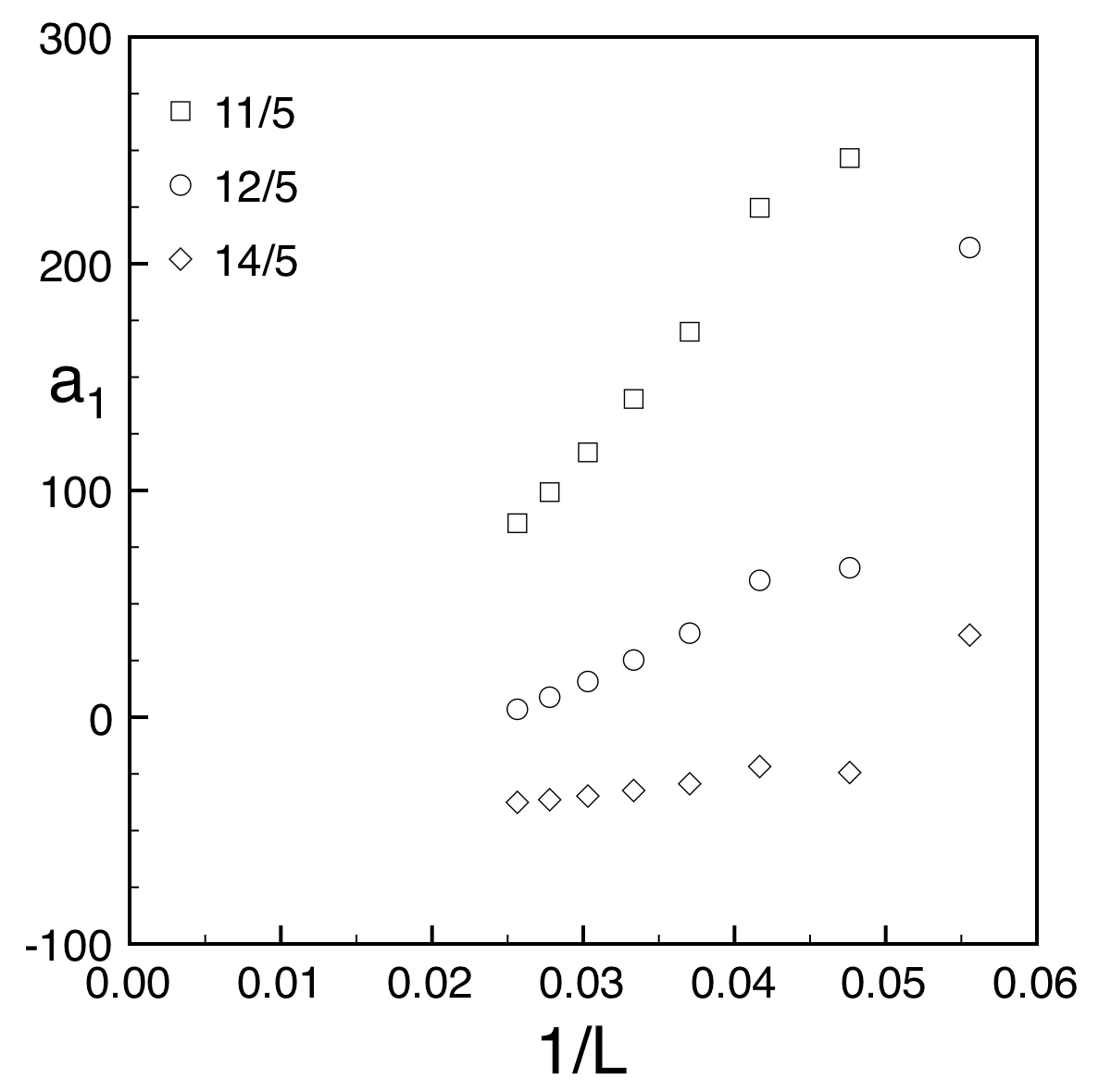} 
    \includegraphics[height=4.5cm]{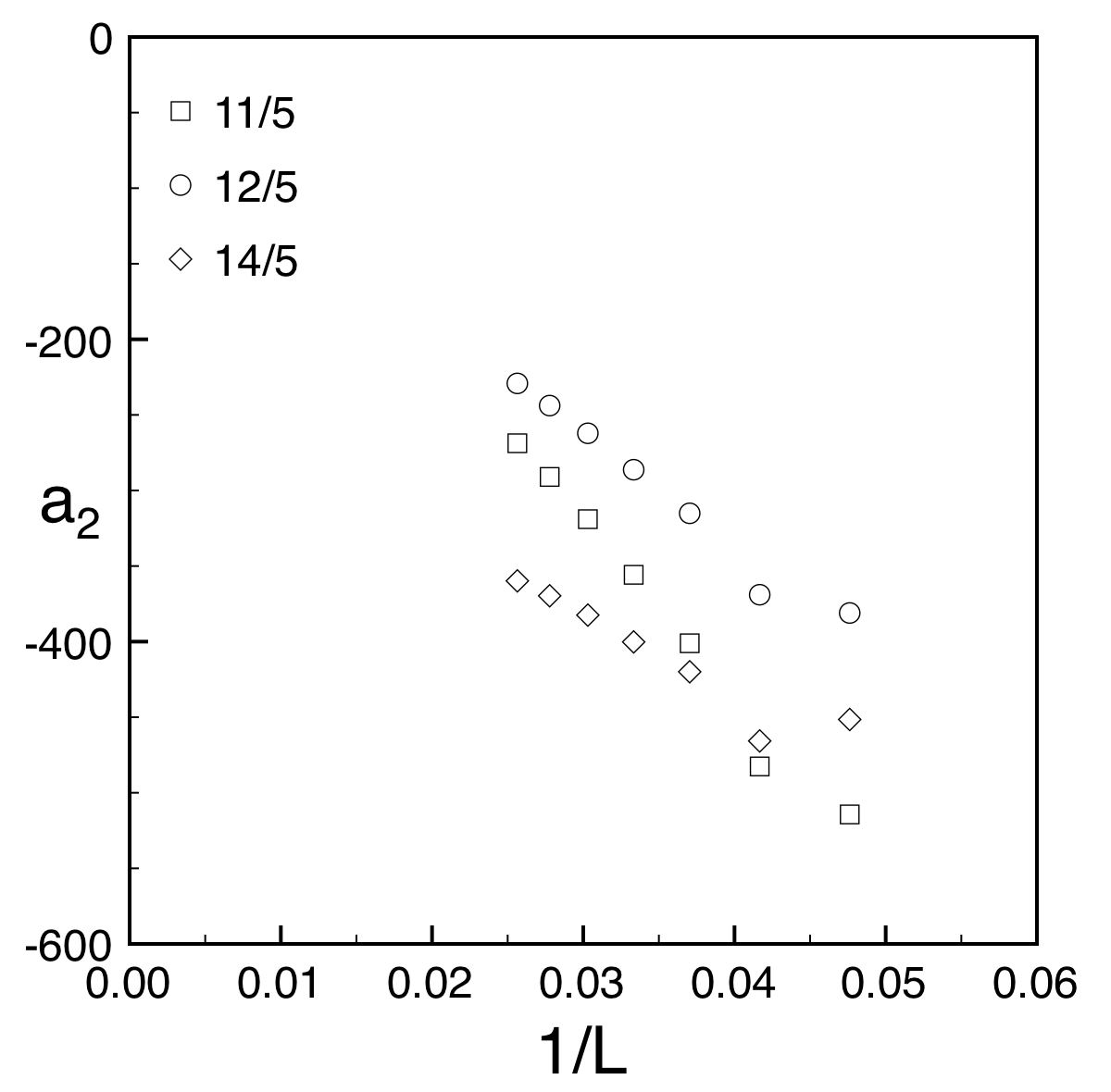} 
   \end{picture}
   \caption{Amplitude estimates versus $1/L$. The panels (from left to right) shows the estimates for
   the amplitudes $a_0$, $a_1$ and $a_2$ when fitting to the asymptotic form (\ref{eq:zcasymptest})
   while using three different values for the third exponent $\Delta$. }
   \label{fig:zcampl}
\end{figure}

To further investigate the asymptotic scaling form we turn to amplitude fitting using 
\begin{equation}\label{eq:zcasymptest}
 |z_c(L)|-z_c = a_0/L^{6/5}+a_1/L^{10/5}+a_2/L^{\Delta}.
\end{equation}
We look at three possible values for the third exponent $\Delta$, namely $y+1=11/5$, 
$2y=12/5$, or $y+2|y'|=14/5$. The results
are displayed in figure~\ref{fig:zcampl} where we plot the estimated values of the three
 amplitudes for the three different values of $\Delta$. Firstly, we note that the estimates
 for the amplitude $a_0$ (left panel) are quite stable though the estimates become more stable
 as the value of $\Delta$ is increased. Secondly, the data for the amplitude
 $a_1$ (middle panel) is very striking; for a $\Delta$ of 11/5 or 12/5 the estimates
 vary greatly with $L$ and even have the wrong sign from the extrapolated value;
 in sharp contrast for $\Delta=14/5$ the estimates are quite well converged with only
 a mild dependence on $L$. Finally, for the amplitude $a_2$ (right panel) we see
 that the amplitude estimates for  $\Delta=11/5$ or 12/5  may well extrapolate to
 a value of 0 while the estimates for $\Delta=14/5$ clearly extrapolate to a non-zero
 values around $-200$ or so. Taken together this is quite clear evidence that the
 correct value of the third exponent is $\Delta=y+2|y'|=14/5$. We estimate
 roughly that
 
 \begin{equation}\label{eq:zcampl}
a_0=53.0(1), \quad a_1=-50(5), \quad a_2=-200(50).
\end{equation}

\begin{figure}[htbp] 
   \centering
   \begin{picture}(260,130)
   \includegraphics[height=4.5cm]{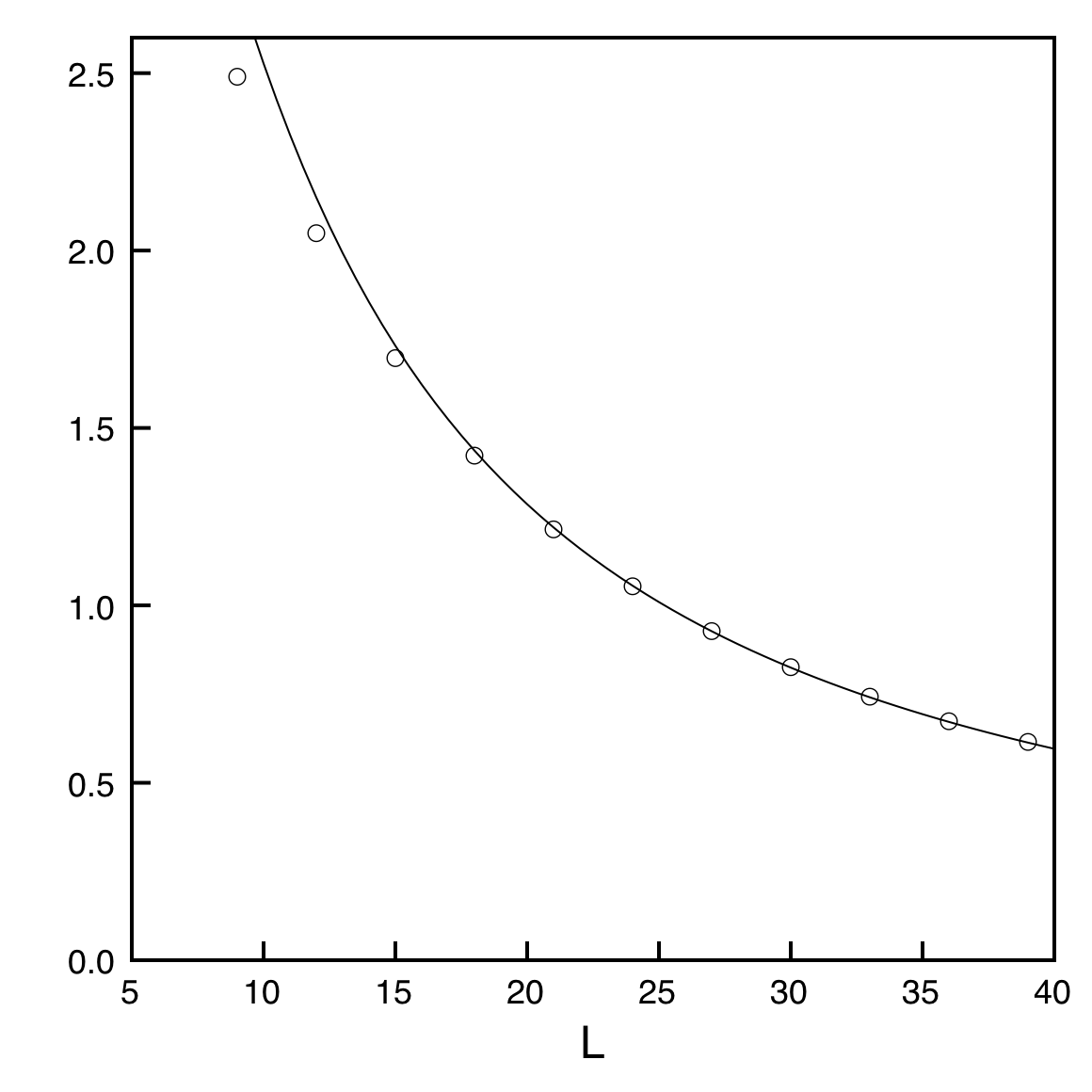} 
   \includegraphics[height=4.5cm]{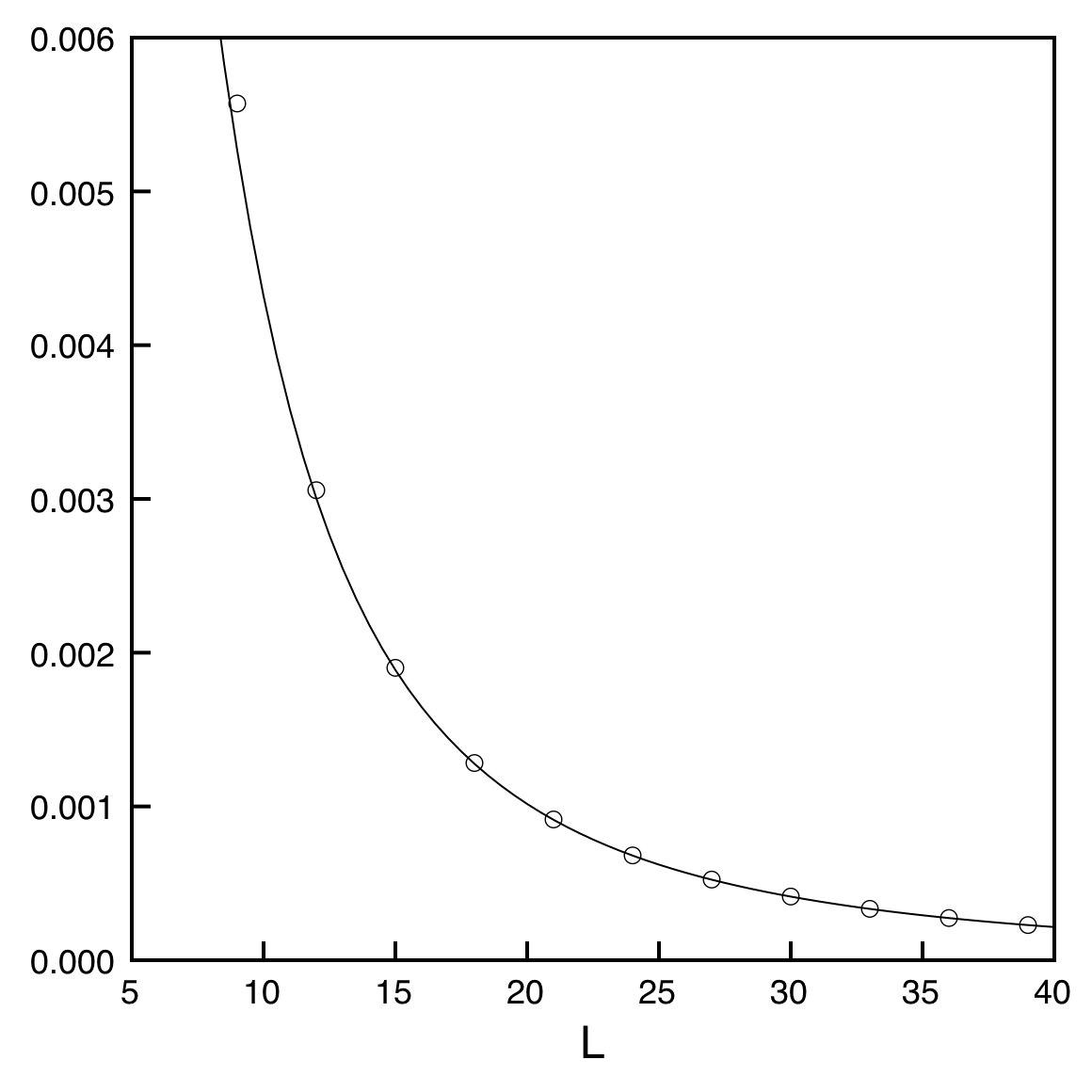} 
   \end{picture}
   \caption{Data for $|z_c(L)|-z_c$ (left panel) and  $|z_d(L)-z_d|$ (right panel) 
   and fitted curves to the asymptotic forms (F.3) and (\ref{eq:zdasymp}) 
   using the listed amplitude estimates (\ref{eq:zcampl}) and (\ref{eq:zdampl}).}
   \label{fig:zczdfit}
\end{figure}

In figure~\ref{fig:zczdfit} we plot the data for $|z_c(L)|-z_c$ (left panel) 
and  $|z_d(L)-z_d|$ (right panel) and the asymptotic fits obtained above. 

\begin{figure}[htbp] 
   \centering
  \begin{picture}(400,130)
    \includegraphics[height=4.5cm]{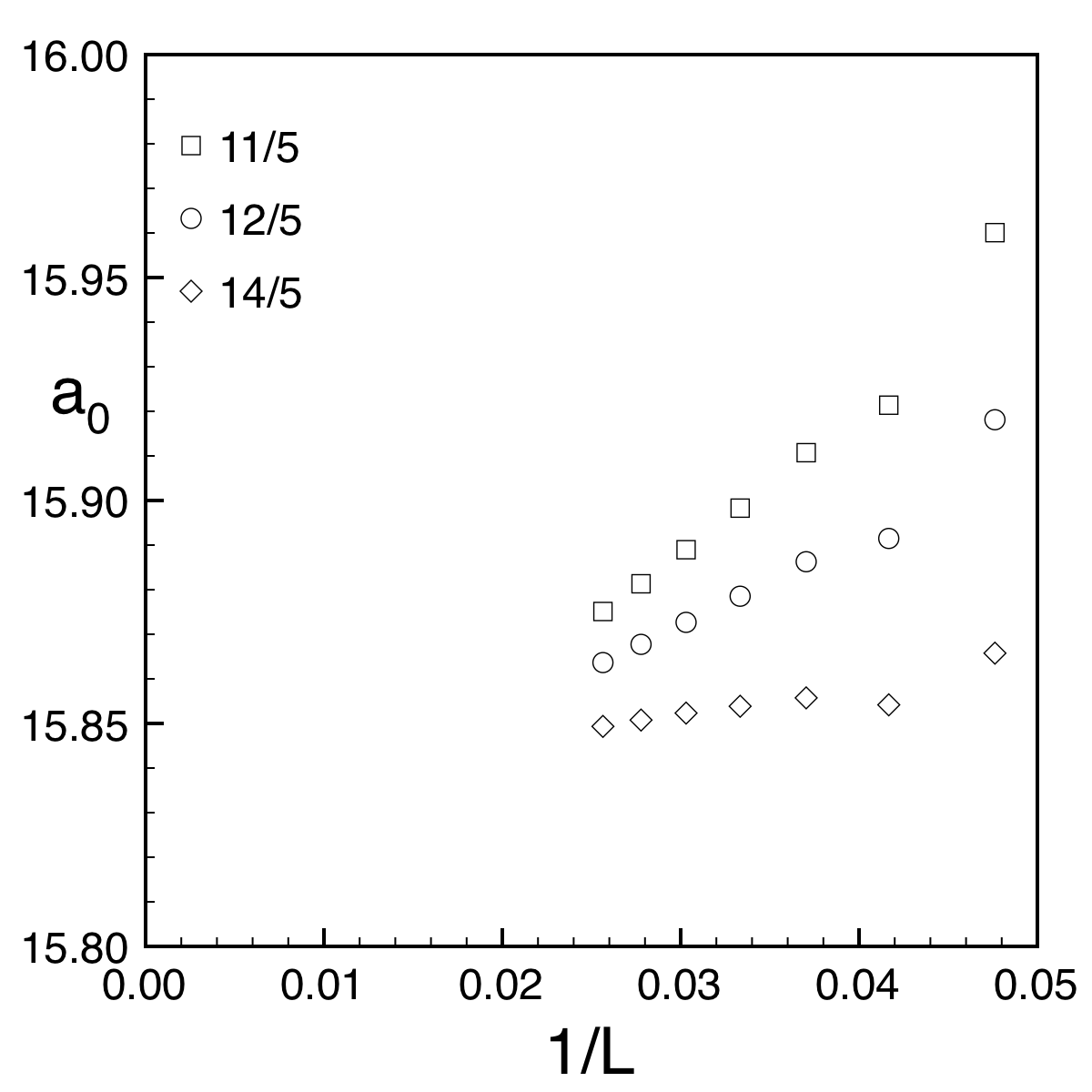} 
    \includegraphics[height=4.5cm]{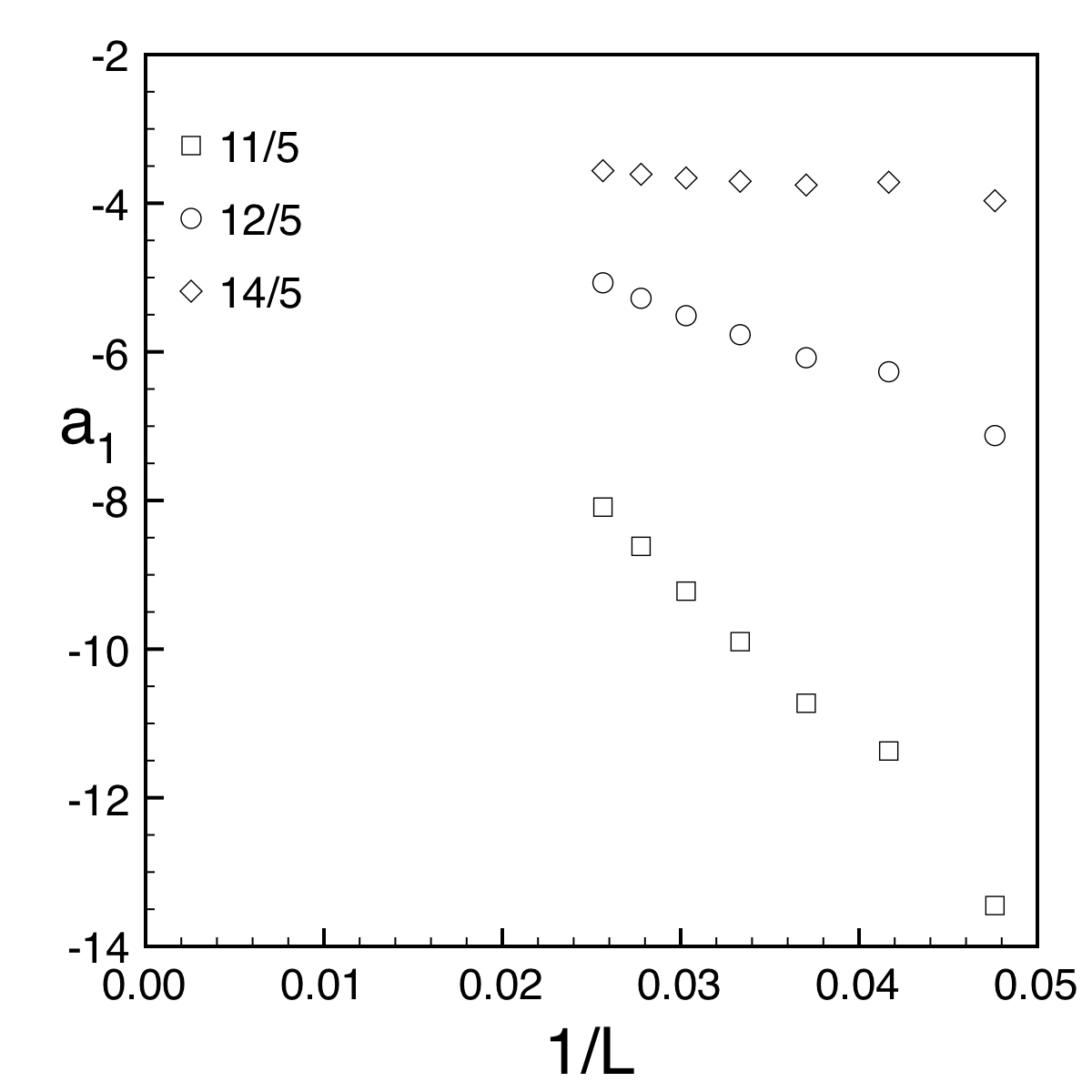} 
    \includegraphics[height=4.5cm]{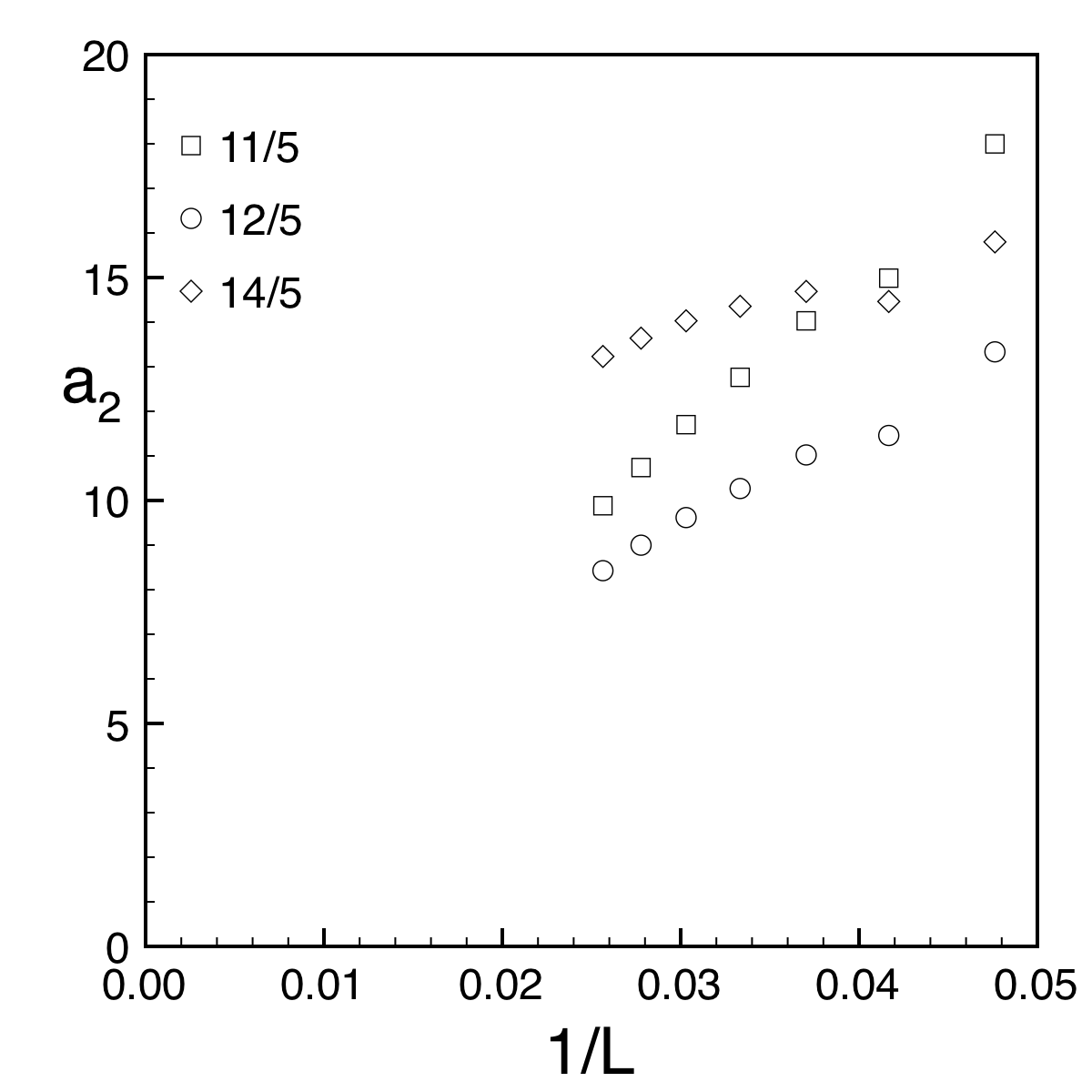} 
   \end{picture}
   \caption{ Amplitude estimates versus $1/L$. The panels (from left to right) shows the estimates for
   the amplitudes $a_0$, $a_1$ and $a_2$ when fitting ${\rm arg} (z_c(L))$ to the asymptotic form (\ref{eq:zcasymptest})
   while using three different values for the third exponent $\Delta$. }
   \label{fig:zc_argampl}
\end{figure}

It is universally expected that the end-point $z_c(L)$ converges towards $z_c$, as can
be confirmed by analyzing the behavior of ${\rm Re}(z_c(L))$ and ${\rm Im}(z_c(L))$ against $1/L$.
This obviously means that the imaginary part must vanish as $L\to \infty$. 
To examine this we repeat the above analysis for  ${\rm arg} (z_c(L))$. 
The `local-slope' analysis is not as clear-cut in this case but it is consistent
with the two leading terms in (\ref{eq:zcasymptest}). The amplitude analysis
is again very clean as can be seen in figure~\ref{fig:zc_argampl} and
from this we obtain the amplitude estimates  
\begin{equation}\label{eq:zc_argampl}
a_0=15.83(2), \quad a_1=-3.0(5), \quad a_2=8(2).
\end{equation}
These values of course differ from those in (\ref{eq:zcampl}) since we are
analyzing a different quantity.

\begin{figure}[t] 
   \centering
  \begin{picture}(400,130)
    \includegraphics[height=4.5cm]{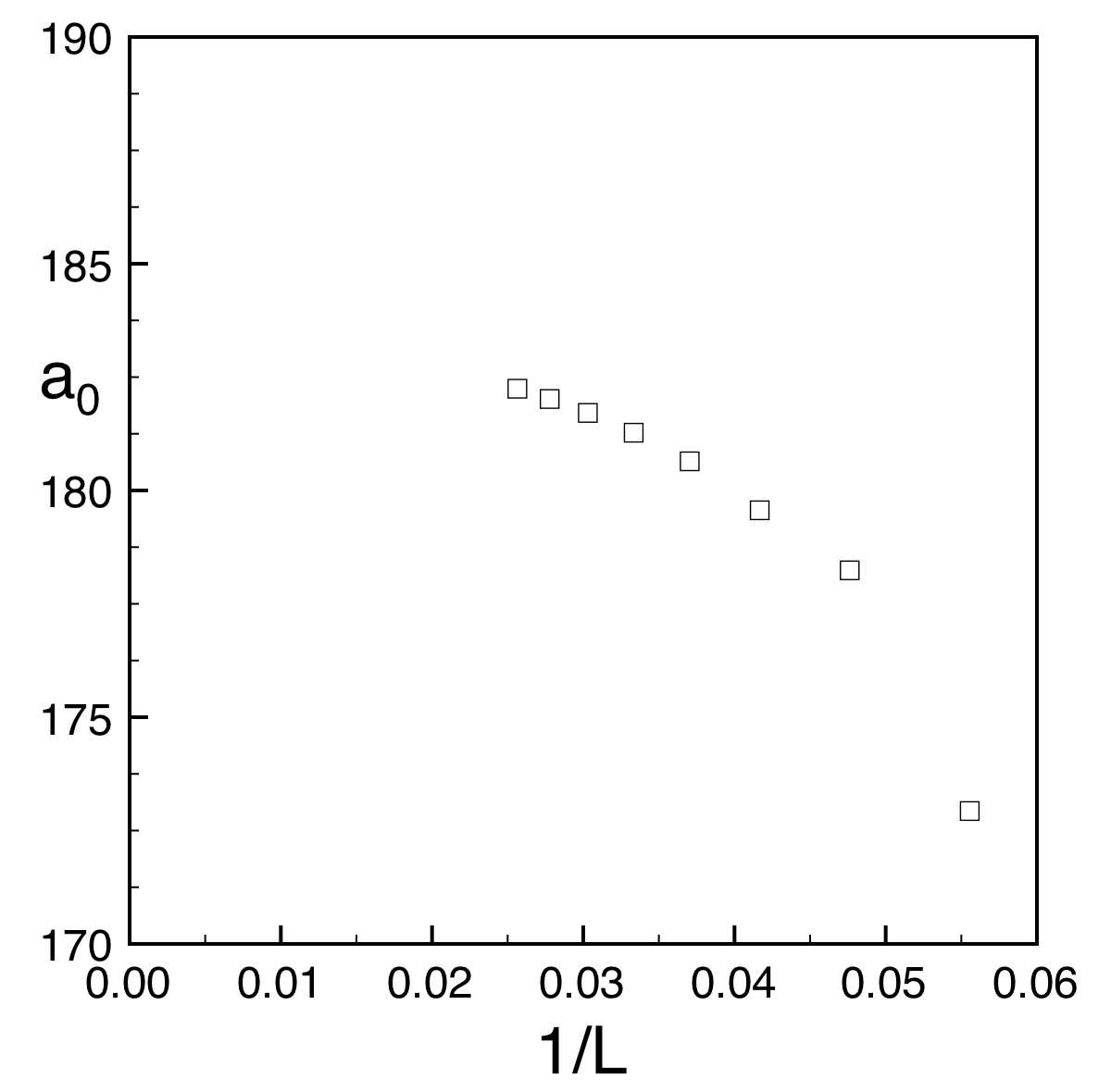} 
    \includegraphics[height=4.5cm]{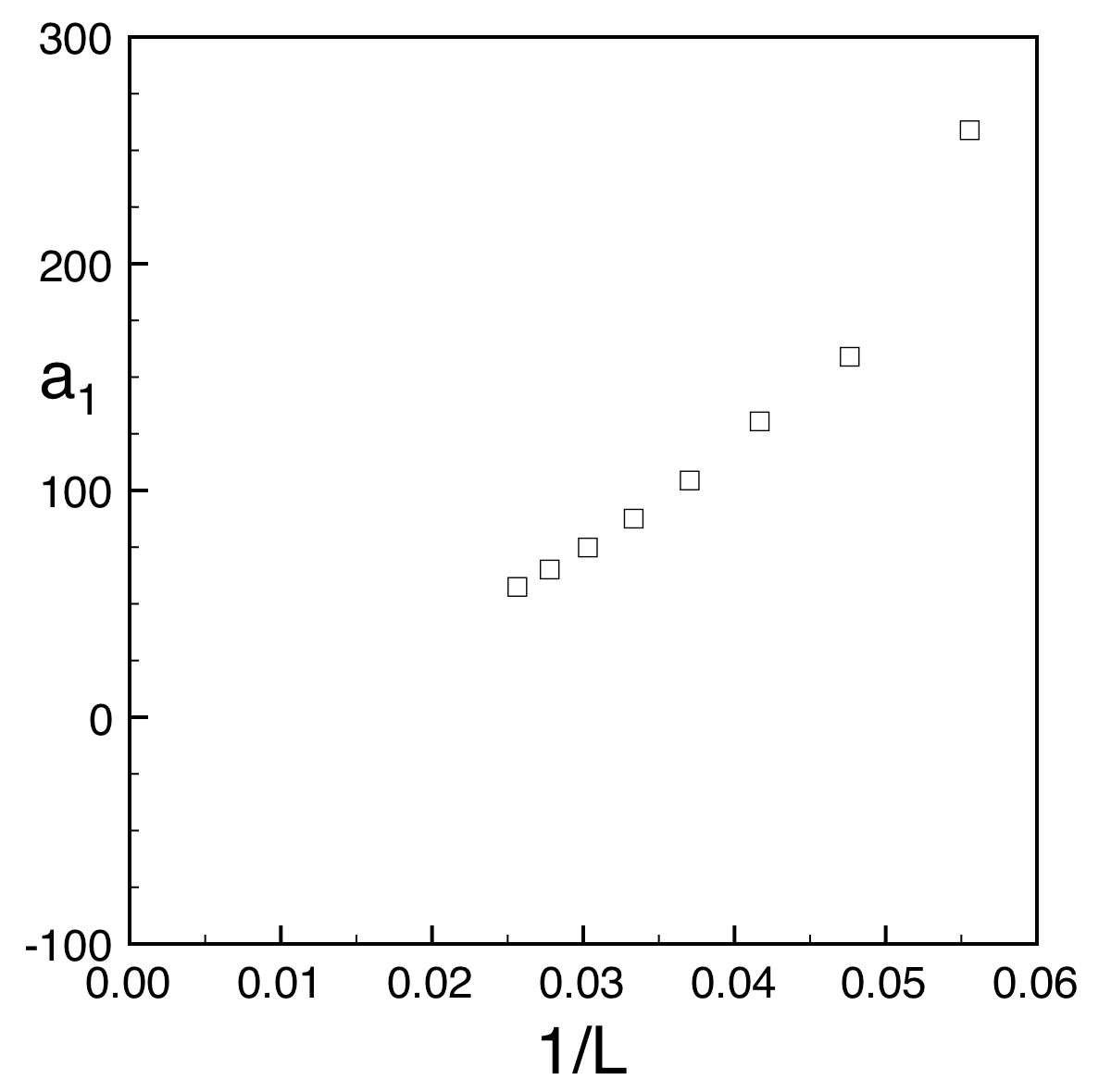} 
    \includegraphics[height=4.5cm]{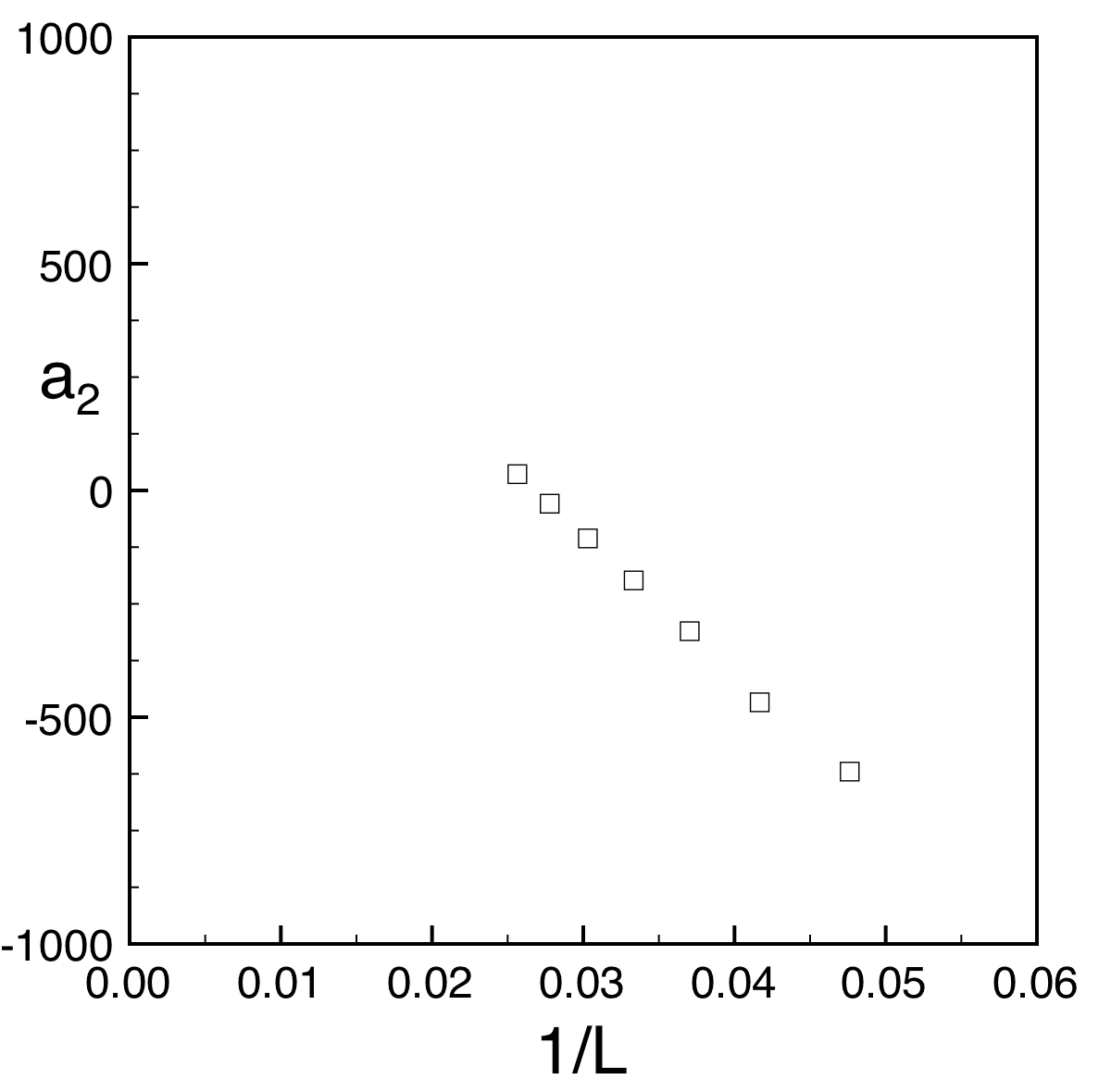} 
   \end{picture}
   \caption{ Amplitude estimates versus $1/L$. The panels 
(from left to right) shows the estimates for
   the amplitudes $a_0$, $a_1$ and $a_2$ when fitting to the 
asymptotic form (F.3)
   while using the data for $|z_c(L)-z_c|$.}
   \label{fig:zc_cdampl}
\end{figure}

Finally we carried out a similar analysis for the quantity $|z_c(L)-z_c|$. Again the
evidence for the leading amplitude being $-6/5$ was firm. However, the local-slope
analysis for the sub-dominant term was inconclusive. In figure~\ref{fig:zc_cdampl} we
plot the amplitude estimates  obtained when fitting to (F.3).
We observe a very strong variation in $a_1$ and $a_2$, but on the other hand
the amplitude estimates are nice  and monotonic, suggesting that the data might just
be really hard to fit. In particular we note that $a_1$ could extrapolate to 0.
We then tried a new fit using an additional fourth term with exponent $-18/5$,
thus assuming exponents of the form $y+k|y'| = 6/5+k\cdot 4/5$
In this case we found much more stable amplitude estimates with $a_0=183.5(5)$.
The other amplitudes displayed quite a bit of scatter so we will not quote error-bars,
but we found $a_1\simeq 6.5$, $a_2\simeq 930$ and $a_3\simeq -5200$.
Remarkably $a_1$ is quite small compared to the other quantities which may
well explain the numerical difficulties we had with the analysis.
Note that we make no claim that  $y+k|y'| $ exhausts the exponents and it is quite likely
that other exponents, such as $2y+|y'|= 16/5$, could occur, but our data sets
are too limited to answer such questions beyond the terms explicitly included
in (F.3).

\end{document}